\documentclass{article}
\usepackage[utf8]{inputenc}
\usepackage[legalpaper, margin=.8in]{geometry}
\usepackage[colorlinks,citecolor=red,urlcolor=blue,bookmarks=false,hypertexnames=true]{hyperref} 
\usepackage{hyperref}

\usepackage[acronym]{glossaries}
\usepackage{graphicx}
\usepackage{float}
\usepackage{caption}
\usepackage{subcaption}
\usepackage[parfill]{parskip}
\usepackage{siunitx}
\usepackage{graphicx}
\usepackage{ifthen}
\usepackage{fancyhdr}

\newcommand{\thesistitle}{Modelling of shattered pellet injection experiments on the ASDEX Upgrade tokamak}			
\newcommand{\yourname}{Anshkumar Himanshu Patel}				
\newcommand{\studentID}{1671545}					
\newcommand{\defensedate}{October 26, 2023}			
\newcommand{\capacitygroup}{Science and technology of Nuclear Fusion}			
\newcommand{\supervisor}{prof.dr. R.J.E. Jaspers}		
\newcommand{\extsupervisor}{dr. G. Papp}						
\newcommand{\studyload}{45}						

\newcommand{\memberone}{prof.dr. R.J.E. Jaspers}		
\newcommand{\membertwo}{dr. T.W. Morgan}		
\newcommand{\memberthree}{dr. X.C Mi}		
\newcommand{\memberfour}{}		
\newcommand{\firstadvisor}{dr. G. Papp}		
\newcommand{\secondadvisor}{}		

\newcommand{\external}{Max Planck Institute for Plasma Physics, Garching}				

\newcommand{\confidential}{false}	 				
\newcommand{\confidentialperiod}{1/2/3/4/5}			
\newcommand{\publicationdate}{January 1, 2025}			

\usepackage[
backend=biber,
style=nature,
clearlang=true
]{biblatex}
\AtEveryBibitem{
\clearlist{language}
\clearfield{urlyear}
\clearfield{urlmonth}
}
\DeclareUnicodeCharacter{03B2}{\ensuremath{\beta}}
\begin{document}
\begin{titlepage}

\parskip            \bigskipamount
\parindent         0mm
\oddsidemargin  0mm
\evensidemargin 0mm
\textwidth        15cm      
\textheight       25cm      
\topmargin      -16mm   
\pagestyle{empty} 

\includegraphics[width=6cm]{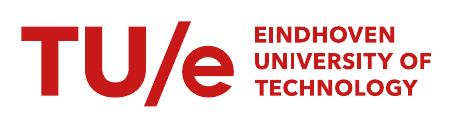}

\quad\textbf{Department of Applied Physics and Science Education} \\

\vspace*{4cm}

\begin{center}
    \LARGE{\textbf{\thesistitle}}   
    \vspace*{0.5cm}
    
    \large{by}
    \vspace*{0.5cm}
    
    \Large{\textbf{\yourname}} 
    \vspace*{2cm}
    
    \Large{MSC THESIS}
\end{center}
\vspace{2cm}

\begin{minipage}[t]{6cm}
\textbf{Assessment committee}\medskip

\begin{tabular}{@{}ll}
Member 1 (chair): & \memberone\\
Member 2: & \membertwo\\
Member 3: & \memberthree\\
\ifthenelse{\equal{\memberfour}{}}{}{Member 4: & \memberfour\\}\ifthenelse{\equal{\firstadvisor}{}}{}{Advisory member 1: & \firstadvisor\\}\ifthenelse{\equal{\secondadvisor}{}}{}{Advisory member 2: & \secondadvisor}
\end{tabular}
\end{minipage}
\qquad\qquad\qquad
\begin{minipage}[t]{5cm}
\textbf{Graduation}\medskip

\begin{tabular}{@{}ll}
Program: & Science and Technology of Nuclear Fusion\\
Capacity group: & \capacitygroup\\
Supervisor: & \supervisor\\
Date of defense: &  \defensedate\\
Student ID: & \studentID \\
Study load (ECTS): & \studyload\\
\ifthenelse{\equal{\extsupervisor}{}}{}{External supervisor(s): & \extsupervisor}
\end{tabular}
\end{minipage}
\vspace*{5mm}

\ifthenelse{\equal{\external}{}}{}{
The research of this thesis has been carried out in collaboration with \emph{\external}.\\}
\ifthenelse{\equal{\confidential}{true}}{This thesis is confidential for a period of \confidentialperiod~years until the publication date \publicationdate.}{This thesis is public and Open Access.}

This thesis has been realized in accordance with the regulations as stated in the TU/e Code of Scientific Conduct.\\\\\\
Disclaimer: the Department of Applied Physics and Science Education of Eindhoven University of\\
Technology accepts no responsibility for the contents of MSc theses or practical training reports.
\vfill
\end{titlepage}


\newpage
\section*{Abstract}
A disruption mitigation system (DMS) is necessary for fusion-grade tokamaks like ITER in order to ensure the preservation of machine components throughout their designated operational lifespan. To address the intense heat and electromagnetic loads that occur during a disruption, a shattered pellet injection (SPI) system will be employed. This SPI system involves injecting material into the plasma through the means of a cryogenic pellet that is shattered on a bent tube before entering the plasma. The penetration and assimilation (ionized material that stays inside the plasma volume) of the injected material is influenced by various SPI parameters, including the fragment sizes, speeds, and composition of the shattered fragments. An SPI system was installed on the ASDEX Upgrade tokamak to study the effect of the aforementioned parameters. In this thesis, 1.5D simulations with the INDEX code have been utilised to conduct parametric scans, thus examining the influence of fragment sizes, velocities, and pellet composition on the efficacy of disruption mitigation. \\\\
When injecting only deuterium, I found material assimilation to be limited to the edge of the plasma with larger and faster fragments leading to higher assimilation. For mixed deuterium/neon injections, again, larger and faster fragments enabled higher assimilation. The amount of assimilated neon increased with increasing injected neon amounts but saturated for larger neon fraction pellets. I also carried out comparisons with previous experimental results of penetration, material assimilation and pre-TQ duration. Previous experimental results for pure deuterium injections indicated that larger and faster fragments exhibit greater penetration, aligning with findings from the simulations. Simulated material assimilation trends for pure deuterium injections were also found to be qualitatively similar to the experiments. Nonetheless, a major difference in the quantitative assimilation values was identified, likely associated with the experimental assimilation criterion. By assuming a semi-empirical TQ onset condition, pre-TQ duration dependence on fragment sizes and speeds for mixed deuterium/neon injections is also studied. I found that faster fragments lead to a shorter pre-TQ duration in the experiments, similar to the simulations. In the case of fragment size variation, larger fragments had a longer pre-TQ duration in the simulations, in contrast to the experiments. 
\newpage
\tableofcontents
\newpage
\section*{Acknowledgement}
The successful completion of this thesis was made possible through the support and guidance of numerous individuals. Working under Gergely Papp's (Geri) supervision was a truly enriching experience. His trust in my abilities and his encouragement for independent thinking allowed me to take ownership of my research and develop critical thinking skills. Geri consistently offered valuable insights and support across various research aspects, leading to significant personal growth and research achievements. I am also deeply grateful for Akionbu Matsuyama's mentorship, not only in regards to getting familiar and using the INDEX code, but also for assisting with countless invaluable insights with interpretation of results. \\\\
Roger, my university supervisor, provided consistent support and oversight during the thesis, always ready to assist as needed. My ITER colleagues, Michael Lehnen, Stefan Jachmich, Javier Artola regularly provided me with inputs on my results throughout the duration of my thesis. Their guidance really helped in shaping the results obtained in this thesis. Colleagues at ASDEX-Upgrade, IPP, including Paul Heinrich, Peter Halldestam, Weikang Tang, Emiliano, and Matthias Hoelzl, contributed constructive feedback and readily shared AUG data and analysis results, facilitating the research process greatly.  \\\\
Finally, I extend my heartfelt gratitude to all who contributed to my personal and professional development during this thesis journey, with special thanks to my family for their unwavering support, encouragement, and guidance.

\newpage
\section{Introduction}
\begin{figure}[h]
    \centering
    \includegraphics[width = 0.8\linewidth]{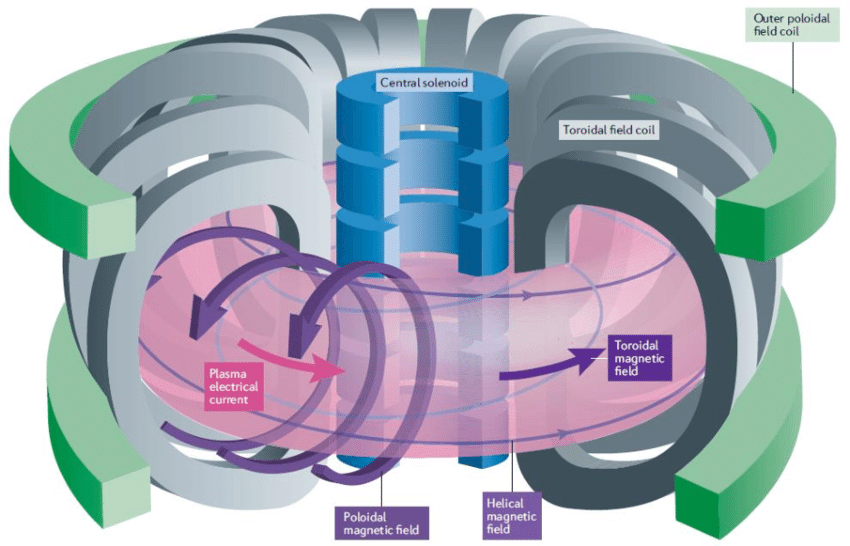}
    \caption{Magnetic fields in a tokamak. The toroidal magnetic field is generated by toroidal field coils shown by grey color. The central solenoid induces current in the plasma which creates a poloidal magnetic field. The net magnetic field is helical. Image source: \href{https://euro-fusion.org/}{EUROfusion}}
    \label{fig:tokamak_sch}
\end{figure}

To tackle the issue of growing energy demand \cite{birol_world_nodate} while mitigating the emission of greenhouse gases, nuclear fusion is a promising solution. The remainder of this section closely follows the discussion from 'Plasma Physics and Fusion Energy' by Jeffrey P. Freidberg  \cite{freidberg_plasma_2008}. Fusion reactions involve merging two light nuclei, resulting in a combined mass that is smaller than the original nuclei. The excess mass is converted into usable energy for electricity generation. To obtain net electricity from fusion reactions, the fusing particles must be confined such that fusion power can overcome scattering losses \cite{harms_principles_2002}. Various confinement schemes have been developed and explored for fusion reactors. These schemes aim to achieve the extreme conditions of temperature and pressure necessary for initiating and sustaining fusion reactions. A widely researched form of confinement, magnetic confinement, utilizes magnetic fields to confine a mixture of hot charged particles in a plasma state. Magnetic confinement fusion reactors plan to confine a mixture of deuterium (D) and tritium (T).
\begin{equation}
    \text{D}+ \text{T} \rightarrow ^4_2{}\text{H}_\text{e}(3.5 \text{MeV}) + \text{n}(14.1 \text{MeV}). 
    \label{eq:DTfusion}
\end{equation}
The D-T mixture will be utilized as fuel due its highest reaction rate \cite{glasstone_controlled_1960} among other fuels such as D-D, D-$\text{He}^3$. A single D-T fusion reaction emits an alpha particle and a neutron as shown in \autoref{eq:DTfusion}. 80\% of the total energy generated in fusion reactions is carried by neutrons while 20\% is transferred to alpha particles that would be used to self-heat the plasma to sustain the fusion reaction. The energy of the emitted neutrons can be converted into heat to generate electricity through a turbine. Among various magnetic confinement approaches, tokamaks are the most extensively studied. The tokamak, such as ITER, utilizes magnetic coils arranged around a torus-shaped vessel, which generate a toroidal magnetic field to confine the plasma. A central solenoid induces current in the plasma which generates a secondary magnetic field along the poloidal direction. The two field components result in a helical magnetic field as shown in \autoref{fig:tokamak_sch}. The helical magnetic field is required to confine the plasma without being affected by grad-\textit{B} drifts that would otherwise push the plasma towards the plasma facing components. In an axisymmetric tokamak, the magnetic field lines form nested toroidal flux surfaces, where the magnetic flux through the surfaces is constant \cite{wesson_tokamaks_2011}. The plasma temperature and pressure on the flux surface is also constant as the energy and particle transport along the magnetic field lines is significantly faster than transport between the surfaces. The flux surfaces can also be denoted with a $q$ value which corresponds to the number of toroidal turns a field line on the flux surface makes to complete one poloidal turn.

Another magnetic confinement concept is the stellarator, which, unlike the tokamak, uses magnetic coils around the plasma vessel to generate the helical magnetic field directly without externally driven plasma current. All eyes are currently on ITER \cite{editors_chapter_1999} which is designed to confine a D-T plasma in which $\alpha$ particle heating will dominate all other forms of plasma heating. It is foreseen to obtain inductive plasmas with an energy gain $Q$ = 10 beyond break-even where $Q$ is the ratio of the output fusion power to the input heating power provided to the plasma. It is also aimed to demonstrate the integrated operation of the technologies for a fusion power plant, test the components required as well as the concepts for a tritium breeding module for the tokamak. After that, DEMO \cite{ongena_magnetic-confinement_2016} will be deployed at which stage nuclear fusion energy will be considered a viable energy source. However, the commercial realization of nuclear fusion power faces various engineering and technological challenges. \\\\
One challenge faced by tokamaks is that they are susceptible to disruptions \cite{wesson_tokamaks_2011}, which are unwanted events that occur on a milliseconds timescale leading to a loss of thermal and magnetic confinement of the plasma. These events are a result of strong magnetohydrodynamic instabilities in the plasma. Disruptions in ITER can release substantial thermal and magnetic plasma energy, potentially damaging the machine components and reducing their lifespan \cite{lehnen_disruptions_2015}. Additionally, electrons with relativistic energies known as runaway electrons (RE) \cite{breizman_physics_2019}, are generated and accelerated in this phase and can also cause severe damage to the plasma-facing components (PFC) and even penetrate to deeper structures. To ensure the longevity of machine components, a disruption mitigation system (DMS) is essential as an investment protection system. In the case of ITER, a shattered pellet injection (SPI) system will be employed \cite{lehnen_rd_2018}. Cryogenic pellets composed of a hydrogen-neon mixture will be launched toward the tokamak and will shatter in a bent tube before entering the plasma. The resulting fragments will ablate, assimilate in the plasma which then radiates its energy uniformly, reduces heat loads on the tokamak components, increases plasma density and suppresses runaway electrons \cite{lehnen_rd_2018}. \\\\
The penetration and assimilation of the injected material in the hot plasma depends on a few key parameters: the shattered fragment sizes, fragment speeds and the composition of the pellet. Here, the injected material that is deposited and stays within the plasma volume is considered to be assimilated. To study the effect of these parameters, the ASDEX Upgrade (AUG) tokamak recently installed a highly flexible SPI system. In this thesis, I carry out parametric scan simulations of fragment sizes, speeds and pellet composition and study the plasma response using the INDEX disruption simulation code. Furthermore, I also present comparisons of the simulation results with AUG experimental SPI campaign results. The rest of thesis is structured as follows: In chapter 2, the theoretical background and available literature regarding disruptions and SPI is summarized and a research question is defined. Chapter 3 describes the INDEX code, relevant AUG inputs for INDEX and introduces two example SPI simulations in AUG. With this knowledge, simulation results of the parametric scans are presented in chapter 4. Chapter 5 compares the simulation trends of penetration and assimilation with the experimental measurements and chapter 6 summarizes the main results in the thesis. 
\section{Theoretical background}
\label{sec:theoreticalback}

\subsection{Disruptions in tokamaks}
\begin{figure}[H]
    \centering
    \includegraphics[width = 0.8\textwidth]{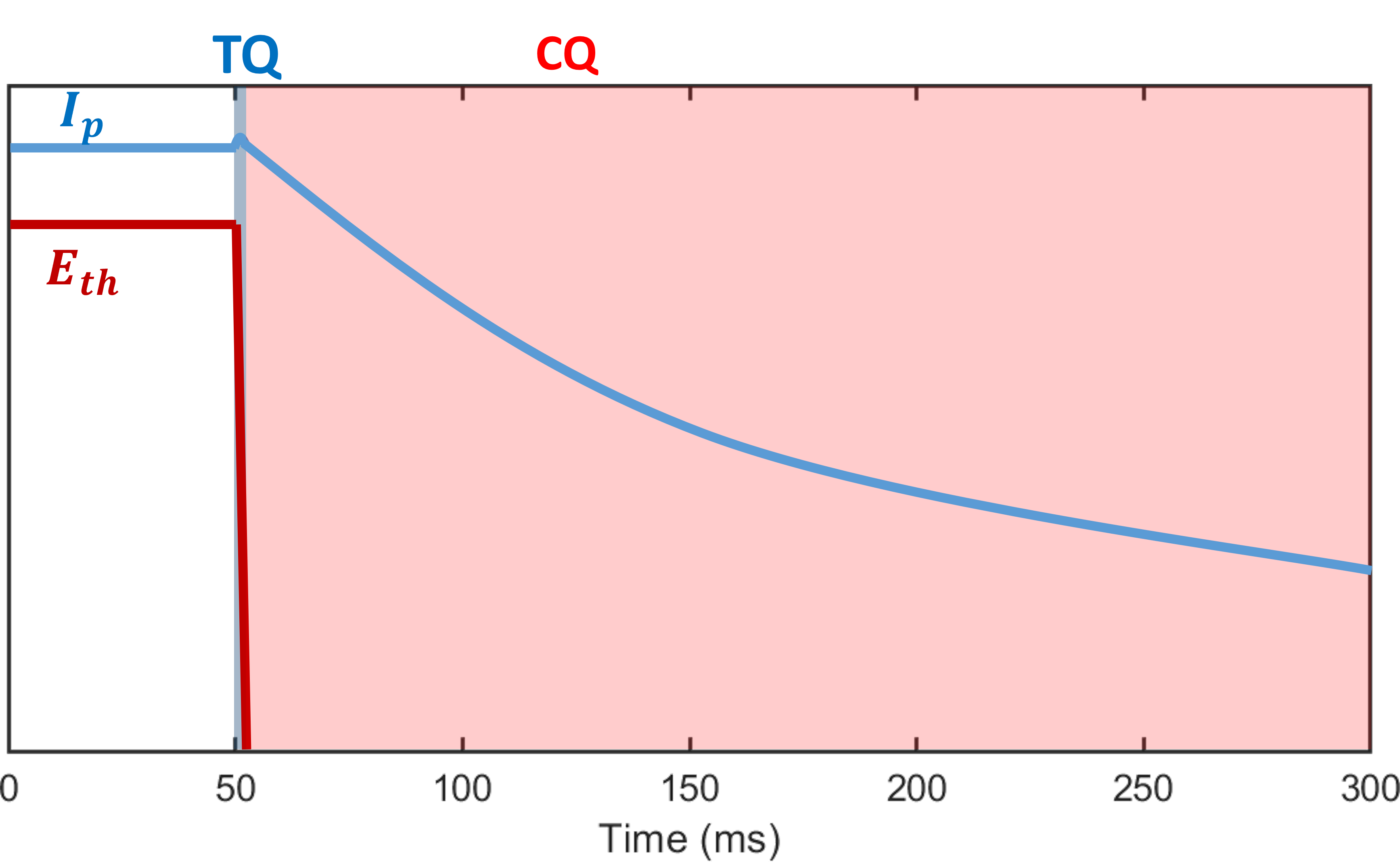}
    \caption{Schematic diagram of plasma current ($I_p$) and plasma thermal energy ($E_{th}$) during a disruption event in ITER and the different phases of disruptions.}
    \label{fig:Ip_unmitigated}
\end{figure}

The ITER tokamak, like other tokamaks, will be prone to disruptions. These events can result from e.g. a control failure when the magnetohydrodynamic stability limits are exceeded. These instabilities create perturbations to the magnetic field structure. If these perturbations become large enough, they can overlap with each other, causing reconnection of the magnetic field lines and driving significant heat and particle transport across the plasma. During the reconnection event, the plasma current increases slightly (around $10\%$) due to the redistribution of the plasma current profile \cite{hender_chapter_2007}. The perturbation can lead to an influx of impurities into the plasma that can radiate energy through impurity line radiation. 
A disruption is usually characterized by two phases, the thermal quench (TQ) and the current Quench (CQ) phase. These different phases of an unmitigated disruption are shown in a schematic diagram in \autoref{fig:Ip_unmitigated}. In the TQ phase, the large-scale perturbations drive heat transport across the plasma, cooling it down rapidly (of sub-milliseconds to a few milliseconds) from several keV (kiloelectronvolts) down to a few eVs or tens of eVs. At the end of the TQ phase, the magnetic flux surfaces may start to re-form and cross-field transport losses may be reduced. \\\\
Post-TQ plasma temperature is determined by the power balance between ohmic heating by the plasma current and impurity radiation loss \cite{lehnen_disruptions_2015}. The drop in plasma temperature increases the resistivity of the plasma, causing the plasma current to start to decay. The sudden increase in resistivity also gives rise to a high in-plasma electric field \cite{hender_chapter_2007}. The decay of the plasma current depends on the resistivity and inductance of the plasma and the surrounding vessel elements. In unmitigated disruptions, the radiated power can be less due to the lower amount of impurities entering the plasma. As a result, the plasma current decays slowly, and the plasma column is displaced vertically following the loss of vertical stability control, resulting in the generation of halo currents \cite{lehnen_disruptions_2015} and causing high electromagnetic (EM) loads on the vacuum vessel. Alternatively, if the radiated power is too high, the plasma current can decay faster, generating eddy currents in the surrounding structures and intense EM loads between components \cite{lehnen_disruptions_2015}. The induced electric field is large and can enable electrons to overcome the Coulomb collisional drag force in the plasma, causing them to gain relativistic energies (MeV range), which are then termed runaway electrons (REs) \cite{breizman_physics_2019}. Depending on the induced electric field and the background plasma parameters, the generated REs can carry a significant fraction of the plasma current. The high energy RE beam can melt plasma facing components and even penetrate to deeper vessel structures. 

Disruptions are already of concern in present-day large tokamaks such as JET and will be more detrimental in ITER because the stored plasma energy will be much higher than that of present tokamaks \cite{lehnen_disruptions_2015}. For reference, in JET, the thermal energy ($\le 12$ MJ) and the poloidal magnetic energy ($\sim 10$ MJ) are dissipated in unmitigated TQ and CQ phases in the form of heat to the plasma-facing components (PFCs) on time scales of $\sim 1$ ms and $>20$ ms, respectively \cite{riccardo_timescale_2005}. In ITER, a significant fraction of the thermal energy ($\le 350$ MJ) and the poloidal magnetic energy ($\le 1300$ MJ) can be deposited on the PFCs \cite{loarte_transient_2007} during unmitigated disruptions. Taking into account the increase in the wetted divertor area that will receive the heat loads and also the linear increase in the time scale of the TQ energy deposition, the expected peak power density will still be higher in ITER than in JET by a factor of 4 during the TQ \cite{riccardo_jet_2010}. Considering that JET already runs close to the material limits, the concern for PFC lifetime is even higher for ITER due to the shorter deposition timescales and higher asymmetries in deposition than JET. \\\\
Another concern is the avalanche multiplication of runaway electrons that occurs when an existing seed of REs generates more REs in the presence of the sufficiently high electric field after the TQ. The avalanche multiplication rate increases roughly by a factor of 10 with an increase in 1 MA of plasma current \cite{breizman_physics_2019}. Therefore, the multiplication rate for ITER which will have a maximum 15 MA plasma current will be higher than that of JET, where the highest currents are $\sim 4$ MA, by a factor of $10^{11}$. This can lead to the formation of runaway beams that carry up to 10 MA current with a beam energy of 20 MJ \cite{lehnen_disruptions_2015, hender_chapter_2007, pusztai_runaway_2022, pusztai_runaway_2023}. 
The disruption mitigation system (DMS) at ITER is responsible for addressing threats posed by the different phases of disruption that could affect the operation of tokamak components until the end of their intended lifespan. The basic mechanism of the DMS involves injecting impurities in the plasma to radiate energy uniformly and to increase the plasma density for RE avoidance. To summarize, the main concerns to be addressed by the DMS are the following:

\begin{itemize}
    \item In the TQ phase: Loss of thermal energy of the plasma on milliseconds timescale that can cause surface melting of PFCs due to high heat loads and avoid generation of runaway electrons after TQ,
    \item In the CQ phase: Intense electromagnetic forces arising from halo and eddy currents in the vacuum vessel and the blanket module, respectively,
    \item If a RE beam is formed: Generation of runaway electron beams that can carry a substantial amount of plasma current, which deposits energy on the PFCs that is highly localized and can cause localized melting and even damage to deeper components such as cooling channels \cite{lehnen_disruptions_2015}.
\end{itemize}

\subsection{Disruption mitigation} 
\label{sec:DMS_ITER}
\begin{figure}[h]
    \centering
    \includegraphics[width = 0.8\linewidth]{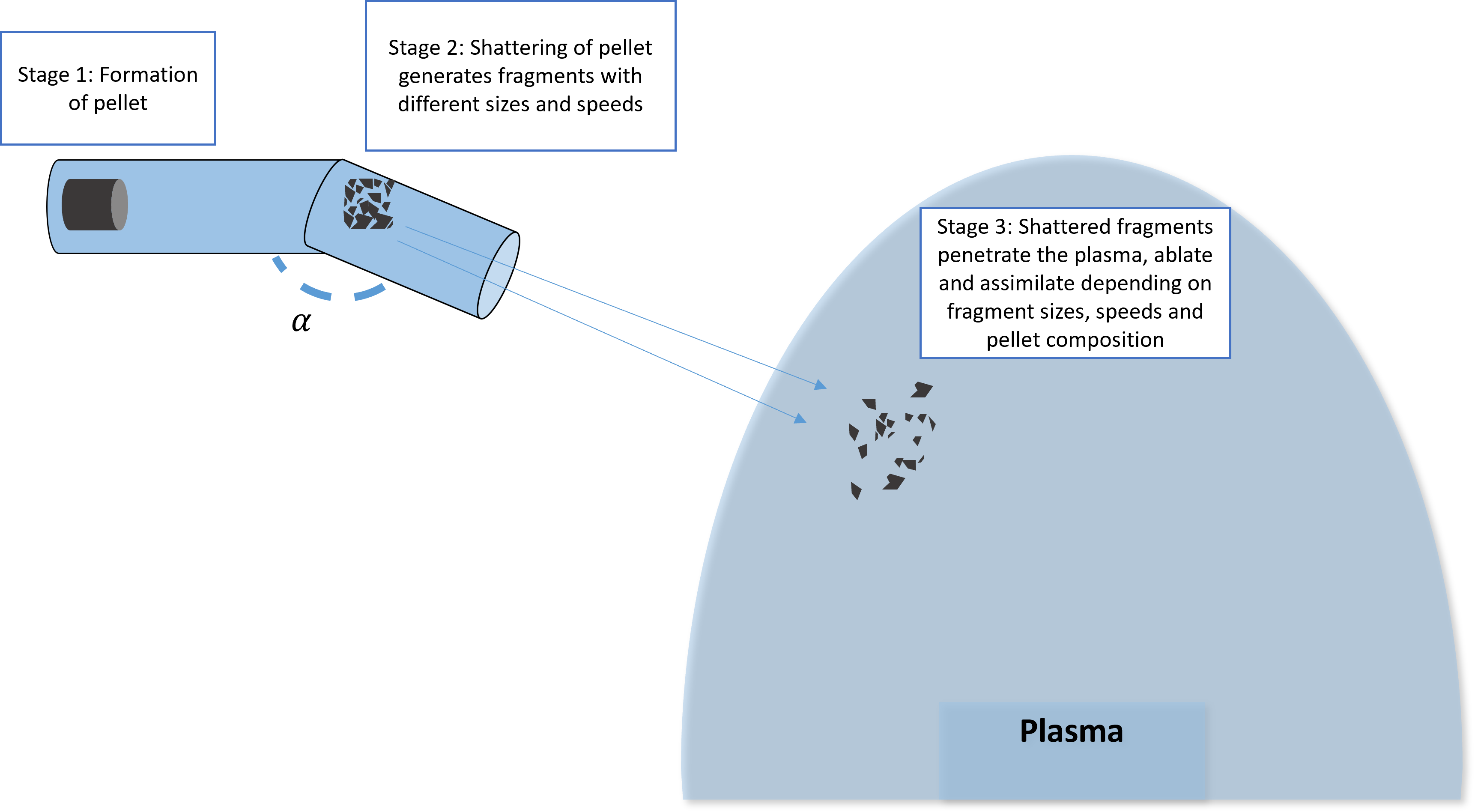}
    \caption{Schematic diagram of the shattered pellet injection system and its different stages. $\alpha$ denotes the shattering angle.}
    \label{fig:SPI_schematic}
\end{figure}

\begin{figure}[h]
\centering
    \begin{subfigure}[b]{0.9\textwidth}
    \includegraphics[width=\textwidth]{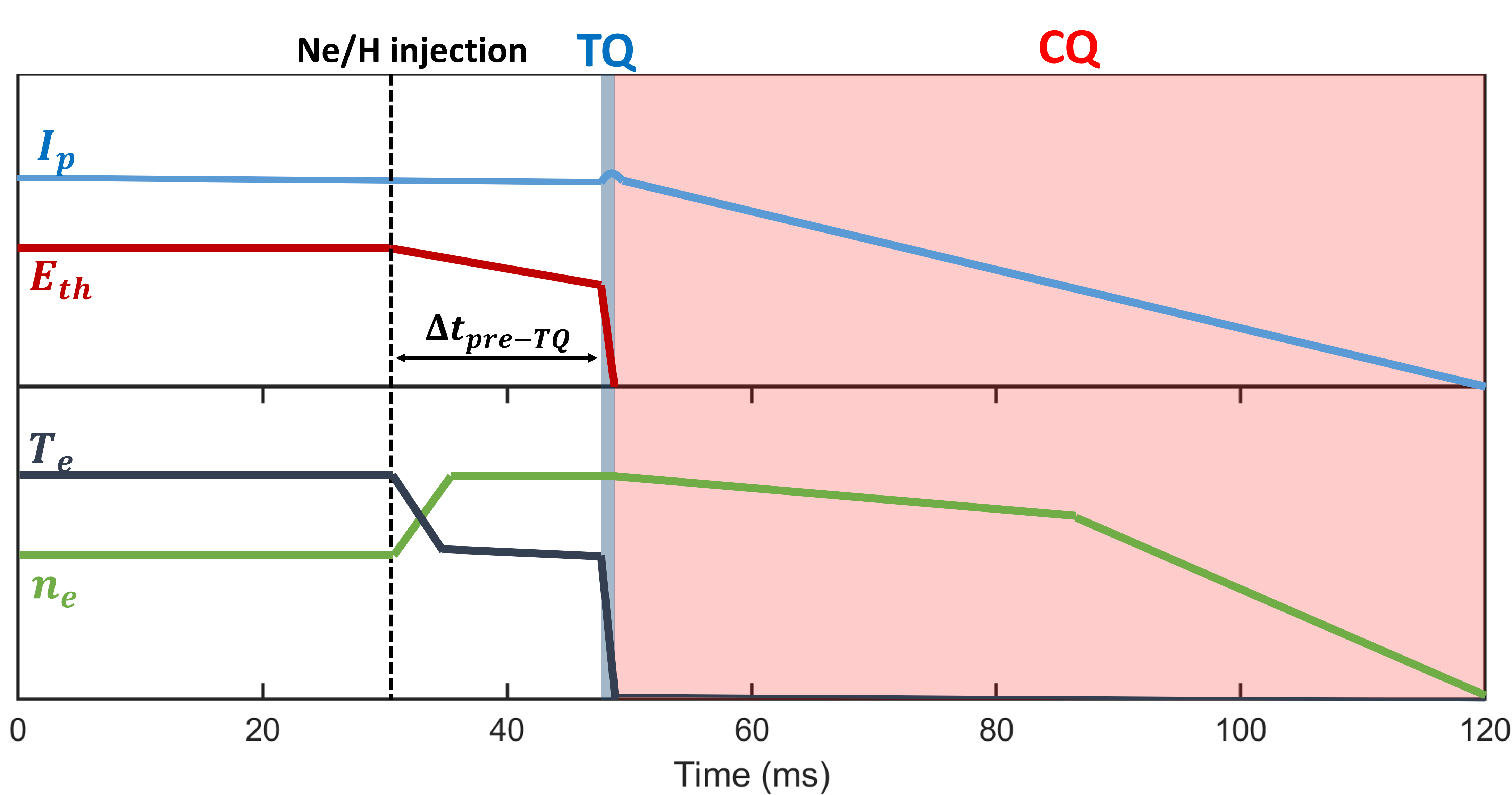}
    \caption{Single mixed Ne/H injection.}
    \label{fig:single_schematic}
    \end{subfigure}
    
    \begin{subfigure}[b]{0.9\textwidth}
    \includegraphics[width=\textwidth]{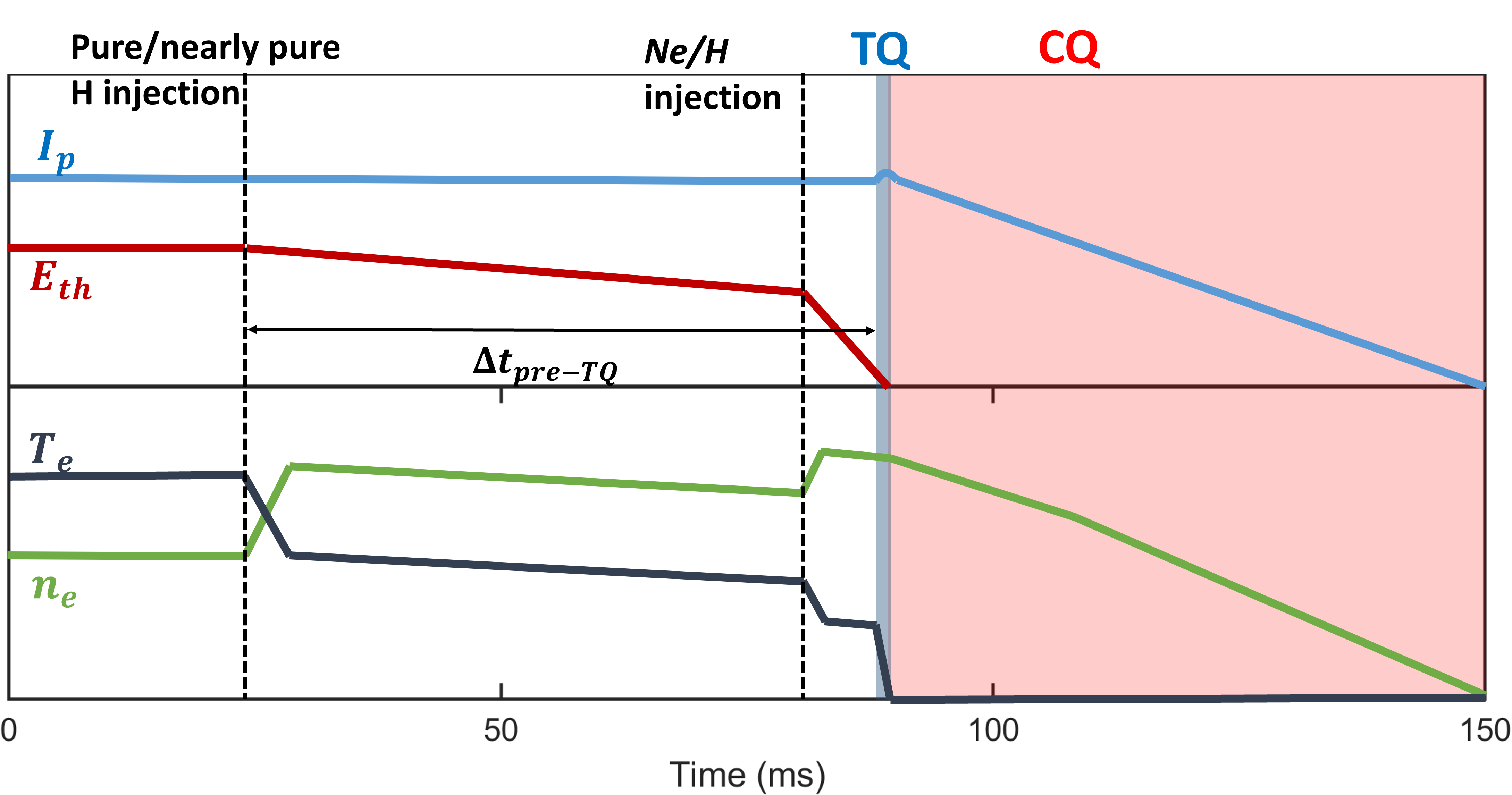}
    \caption{Staggered injection using pure H injection followed by mixed Ne/H injection.}
    \label{fig:staggered_schematic}
    \end{subfigure}
   \caption{Schematic diagram of plasma parameters during a mitigated disruption event in ITER and for two different schemes: [Top] Single Ne/H injection, [Bottom] Staggered injection. $I_p$ is plasma current, $E_{th}$ is plasma thermal energy, $T_e$ is plasma electron temperature and $n_e$ is the plasma electron density.}
    \label{fig:mitigated_schematics}
\end{figure}

The DMS in ITER is a final resort option to be utilized when disruption avoidance through plasma control is no longer possible. The requirements for reducing heat and electromagnetic loads to prevent PFC melting, prevent RE formation, and suppress a runaway electron beam, if it is formed in the CQ phase, are described in Lehnen et. al. \cite{lehnen_disruptions_2015} and Hollmann et. al. \cite{hollmann_status_2014}. The DMS must reliably dissipate 350 MJ of thermal energy and up to 1 GJ of magnetic energy, while maintaining a CQ duration between 50 and 150 ms to avoid extreme heat and electromagnetic loads. Additionally, the RE current must not exceed 150 kA when it strikes the ITER wall \cite{lehnen_iter_2021}. \\\\
While there are multiple methods of disruption mitigation \cite{hollmann_status_2014}, the more researched methods used for disruption mitigation in present tokamaks are Massive Gas injection (MGI) and Shattered Pellet Injection (SPI). Both methods involve injecting large amounts of material in the plasma to dissipate thermal and magnetic energy through line radiation, increasing the plasma density to prevent RE formation, and dissipating RE energy by collisional dissipation and synchrotron radiation if a RE beam is formed \cite{lehnen_disruptions_2015}. The ITER DMS concept relies on SPI technology, which was chosen due to the deeper and faster material assimilation in the plasma core compared to MGI before the TQ, as discussed in Lehnen et. al. \cite{lehnen_rd_2018}. \\\\
The SPI system involves the formation and acceleration of cryogenic pellets that are shattered into smaller fragments against a tilted surface before entering the plasma. The shattered fragments ablate into neutral gas particles because of the hot background plasma, which then ionize and assimilate within the plasma. Here, assimilation refers to the ablated and ionized particles that are deposited within the plasma volume. The efficiency of the SPI system for disruption mitigation depends on multiple parameters of the SPI system. These parameters include elements like pellet injection schemes (discussed later), optimum fragment sizes and velocities, injection location, pellet compositions and required quantities. These elements are being experimentally evaluated on different tokamaks \cite{jachmich_shattered_2022, park_deployment_2020, papp_asdex_2020} and are also being studied using simulations \cite{hu_radiation_2021, matsuyama_transport_2022, kim_shattered_2019}. A schematic of the SPI injection system with its various stages is shown in \autoref{fig:SPI_schematic}. There are two different injection schemes envisioned for the DMS before the TQ \cite{lehnen_iter_2020}: single and staggered injections. A schematic diagram of the plasma parameters during mitigated disruption for the two injection schemes is shown in \autoref{fig:mitigated_schematics}. 

Single injections involve injecting a mixed H/Ne pellet. ITER will utilize hydrogen rather than deuterium, which is utilized in present devices \cite{herfindal_injection_2019, papp_asdex_2020} due to the classification of deuterium as a nuclear material \cite{baylor_design_2021}. The neon will radiate the plasma energy through line radiation i.e. radiative cooling, while hydrogen assimilates in the plasma to increase the plasma density before the disruption for RE avoidance as shown in \autoref{fig:single_schematic}. However, mixed injection of H/Ne pellets has been shown to lead to shorter delay to the TQ that can restrict material assimilation \cite{sheikh_disruption_2021}. It is possible that multiple pellets of the same composition will be injected using different injectors, depending on the material assimilation requirements. The second injection scheme, known as the staggered injection scheme, is shown in \autoref{fig:staggered_schematic}. In this scheme, a nearly or completely pure H pellet is injected first that reduces the plasma temperature and increases plasma density by diluting the plasma at constant pressure i.e. dilution cooling, which suppresses RE generation without a radiative collapse \cite{nardon_theory_2021, vallhagen_effect_2022}. For the second injection, mixed H/Ne pellet(s) can be injected before the TQ to mitigate heat and electromagnetic loads \cite{lehnen_iter_2020}. This scheme has the advantage of allowing more time for the hydrogen to assimilate compared to mixed H/Ne pellets. However, it has to be assessed whether the second mixed pellet can ablate and assimilate in a cooled plasma after the first injection. The second injection then radiates plasma thermal and magnetic energy before the disruption to avoid excessive loads. For both injection schemes, a pre-TQ phase with duration $\Delta \text{t}_\text{pre-TQ}$ is defined as the phase from when the first pellet fragments appear in the plasma until the TQ as indicated in \autoref{fig:mitigated_schematics}.\\\\
The pellets generated by the ITER SPI system will have a diameter of 28.5 mm, a length-to-diameter ratio of 2 which contains $\sim 2\times10^{24}$ hydrogen atoms per pellet \cite{luce_progress_2021}. There will be 27 injectors at different toroidal and poloidal locations in six different ports. 24 of these injectors in three large equatorial ports will serve the function of mitigating thermal loads, increasing plasma density for RE avoidance, and dissipating RE energy in the pre-TQ phase. The remaining 3 injectors will be housed in three upper ports that are $120^\circ$ apart and serve the purpose of mitigation of the CQ in the event that the DMS has not been triggered before the TQ. The upper injectors can inject smaller fragments and even gaseous material compared to the equatorial launchers to ensure sufficient assimilation in the colder post-thermal plasma. The design of the system allows to implement shattering angles in the range of $12.5^\circ \text{ to } 30^\circ$.  The pellet velocity will be determined by the desired fragment size and velocity distribution and is expected to be between 200 and 600 m/s \cite{luce_progress_2021}.  

\subsection{Progress on the disruption mitigation system}
\label{ssec:dms_progress}
Apart from engineering studies relevant for development, testing and commissioning of the DMS elements, there are various physics relevant studies required to assess the following aspects of the DMS:  
\begin{enumerate}
    \item Adequacy of injection locations,
    \item Efficiency of multiple simultaneous injections, 
    \item Efficacy of runaway electron avoidance and mitigation schemes,
    \item Optimum fragment sizes and velocities for highest assimilation,
    \item Required quantities, pellet composition, injection sequences.
\end{enumerate}
Experiments are being carried out on present tokamaks with SPI systems which include KSTAR, JET, DIII-D, J-TEXT and ASDEX Upgrade (AUG) \cite{baylor_design_2021, herfindal_injection_2019, papp_asdex_2020}. Combined experimental results from these tokamaks and modelling studies will be utilized to assess the aforementioned aspects of the DMS. While ITER has multiple injections at different locations described in the previous section, the injection location(s) can affect the final assimilation of the material and also affect the asymmetries in radiation associated with SPI which arise because of spatially concentrated line radiation from the fragments near the PFCs.  For the sake of completeness, the information that is expected to be obtained from different tokamaks is briefly described. \\\\
KSTAR's SPI system has two identical injectors which are 180 degrees toroidally separated \cite{baylor_design_2021}. JET's SPI system has two injectors at the same injection point. DIII-D has two injectors which are 120 degrees toroidally separated \cite{herfindal_injection_2019}. These experiments with different injection locations and number of injectors are being used to address SPI aspects such as the adequacy of injection location to avoid first wall melting due to localised line radiation. The effectiveness of the mitigation process on the difference in the arrival timings of multiple pellets can also be tested. Additionally, owing to the different plasma size and energy content of these devices, an extrapolation to ITER SPI can be made in terms of material assimilation, 0D radiation fraction (fraction of heat dissipated before disruptions by SPI) and 3-D radiation asymmetries. Another important set of parameters of the SPI system are the fragment sizes and speeds. ASDEX Upgrade plays an important role here due to its SPI system being capable of adjusting fragment sizes and speed distributions independently. This is possible due to its unique combination of shattering tubes with different shattering angles \cite{dibon_design_2023} (discussed in \autoref{sec:methodology}). Along with the adequacy of injection locations and the number of injections, optimal fragment sizes and speeds must be determined for maximum penetration, assimilation, radiation fraction and distribution of radiation load.

Apart from fragment sizes and speeds, the composition of the pellet also affects the penetration and assimilation of the injected material. The pellets at ITER will be formed by a mixture of Z=1 hydrogen (H) and Z=10 neon (Ne) depending on the injection scheme \cite{bandyopadhyay_summary_2021}. It is well known from pellet fuelling experiments that pellets injected from the low field side (LFS) of the toroidal magnetic field, as will be the case in ITER, experience a drift that causes a shift of the deposition profile relative to the ablation profile towards the LFS \cite{muller_high-_1999, vallhagen_drift_2023}. This drift is known as the plasmoid drift. Ideally, pellet assimilation should happen in the core of the plasma to radiate maximum energy, however, the plasmoid drift can prevent this from happening \cite{vallhagen_disruption_2023}. It was recently shown \cite{matsuyama_enhanced_2022} that adding a small fraction of neon to the pellet reduces outward drift and facilitates deeper deposition than pure H pellets. The drift effect is minimized due to the line radiation from neon and is discussed in further detail in \autoref{ssec:backavgTS}. However, larger fractions of neon can also disrupt the plasma before the material can assimilate and also cause a very fast CQ leading to intolerable electromagnetic loads \cite{matsuyama_transport_2022}. The shorter pre-TQ duration for a mixed neon injection is due to the formation of a radiative cold front in the edge plasma by the injected impurities that destabilise tearing modes at low-order rational $q$ surfaces. When the radiative cold front reaches the q=2 surface, the modes can grow near the edge, causing enhanced radial transport and the collapse of the hot core plasma, similar to MGI-triggered disruptions \cite{hollmann_measurements_2005, granetz_gas_2007, bozhenkov_generation_2008, pautasso_disruption_2009, reux_experimental_2010, fable_transport_2016}. 

\subsection{Shattered pellet injection at ASDEX Upgrade}
\label{ssec:AUGSPI}
ASDEX Upgrade (AUG) is a medium-sized tokamak with a divertor configuration and a full tungsten wall located at the Max Planck Institute for Plasma Physics, Garching, Germany \cite{stroth_progress_2022}. AUG installed a highly flexible SPI system in late 2021 to provide inputs for the design and optimisation of the ITER DMS \cite{dibon_design_2023}. The AUG SPI system is a triple-barrel system where each independent guide tube can be equipped with different shatter heads. The pellets in AUG were made with pure deuterium, pure neon and mixed deuterium/neon injections with varying compositions. After extensive laboratory testing \cite{peherstorfer_fragmentation_2022}, the three shatter heads (shown in \autoref{fig:shatter_tubes_zoom}) selected included a 46 mm long, circular cross-section with shattering angle $25^\circ$ which generate a fragment plume with an increased spatial spread. The remaining two shatter heads had a similar 78 mm long, rectangular cross section but with two different shattering angles $12.5^\circ$ and $25^\circ$ respectively. The fragment size distribution was studied to be strongly dependent on the impact velocity $v_\perp$ of the pellet with respect to the shatter head where $v_{\perp} = v_\text{pellet}\cdot \sin \alpha$. Lab experiments\cite{peherstorfer_fragmentation_2022} showed that the mean fragment size $\langle d_\text{frag} \rangle$ decreases for increasing normal velocity. The mean fragment velocity was found to be close to the pellet velocity and initial analysis shows that it lies between the pellet velocity and the parallel component of pellet velocity with respect to the shattering angle i.e. $\langle v_{\text{frag}} \rangle \in [v_\parallel, v_\text{pellet}]$ where $v_\parallel = v_{\text{pellet}} \cdot \cos \alpha$. Injecting pellets with similar pre-shatter speed, one can generate varying fragment size distributions by using shatter heads with different shattering angles. Similarly, by comparing injections with similar impact velocity from different shattering angles, different fragment velocity distributions can be created. An experimental campaign was conducted in 2022 where $\sim$ 240 tokamak discharges were allocated to the SPI experiments. Some of the primary machine, plasma, and SPI parameters are shown in \autoref{table:AUGparams}.

\begin{table}[h]
\caption{Machine and plasma parameters of AUG for discharges in this report.}
\rule[0.5ex]{15 cm}{1.5pt}
\centering
\begin{tabular}{p{5.2cm} p{2cm} p{5cm}}
Plasma major radius & $R_0$ & 1.65 m \\
Plasma minor radius & $a$ & 0.5 m \\
Toroidal magnetic field strength & $B_T$ & $1.8$ T \\
Plasma current & $I_p$ & $\le 800 $ kA \\
Heating power & $P_\text{total}$ & 27 MW \\
H-mode plasma energy content & $W_{th}$ & $500 \text{~kJ} < W_{th}< 800$ kJ \\
Pellet length & $L_\text{pellet}$ & 4.5 mm $\le$ $L_\text{pellet} \le$ 10.5 mm \\
Pellet diameter & $d_\text{pellet}$ & 4 mm / 8 mm \\
\end{tabular}
\label{table:AUGparams}
\end{table}

\subsection{Modelling of shattered pellet injections}
In support of the experimental activities carried out at various tokamaks, additional simulation activities are also required for the physics-based extrapolation from the experimental results to ITER. The primary focus of the simulation activities include three-dimensional (3D) MHD simulations performed using codes JOREK \cite{hu_collisional-radiative_2023}, M3D-C1 and NIMROD \cite{kim_shattered_2019} to understand the plasma dynamics during mitigated disruptions and development of models to improve understanding of generation and mitigation of runaway electrons. These activities are ongoing and have been successful in interpreting present experimental results and extrapolating to future tokamaks \cite{hu_3d_2018, kim_shattered_2019, nardon_fast_2020}. JOREK simulations for AUG are also being carried out presently \cite{w_tang_non-linear_2023}. However, as 3D modelling is computationally expensive, only a limited number of simulations to explore the design parameters can be carried out before the design completion of the DMS. On the opposite end of the spectrum in terms of simplicity, zero dimensional (0D) modelling using particle and energy balance can be used to assess key trends of material assimilation and radiation loss \cite{shiraki_particle_2020}. While the simplicity allows for assessing key trends of the SPI assimilation, it does not take into account the plasma size information which is relevant for assessing material penetration, radial distribution of assimilated material and extrapolating to other present devices and larger devices such as ITER. To complement the aforementioned modelling approaches, SPI simulations based on one-dimensional (1D) transport simulations can also provide inputs for parameter spaces that can be studied in more detail with more complex simulations. The simplified modelling permits extensive parametric scans taking into account plasma size and SPI configuration information. This thesis focuses on such simulations using the disruption simulation code INDEX \cite{matsuyama_transport_2022} to study and match the experimental trends of penetration, and assimilation of the injected material for different SPI parameters, these being different fragment sizes, speeds, and pellet composition. 

\subsection{Literature review}
Previous modelling and experimental activities provide some indications about the effect of fragment sizes, speeds and pellet composition on the efficiency of the DMS which are discussed in this section. Previous modelling activities reported the following:
\begin{itemize}
    \item ITER simulations with the INDEX code have shown that larger fragment sizes, higher velocity, and higher velocity dispersion lead to deeper penetration and higher assimilation of the material \cite{matsuyama_transport_2022}. Additionally, benchmarking of INDEX with axisymmetric JOREK simulations for ITER SPI scenario was carried out and agreement between total ablation rate, radiated power and plasma profile evolution was found when a similar SPI setup (of fragment sizes, speeds, and injection timing) is used. 
    \item Simulations with the 3D MHD code NIMROD indicate that a mixed D/Ne injection might be more benign than pure deuterium injection in terms of conducted heat loads on the plasma facing components after pellet injection \cite{kim_shattered_2019}. 
    \item 3D JOREK simulations \cite{hu_collisional-radiative_2023} studied the MHD response for ITER SPI. An important 'size-effect' was pointed out which indicates that even for similar plasma thermal energy density, electron temperature, and ratio of injected atoms to plasma volume, devices with larger major radius and stronger magnetic field would cool down slower after SPI.    
\end{itemize}
Analysis of the SPI experiments at AUG is currently ongoing, which includes studies on the effect of fragment size, speed and composition on pre-TQ duration, material assimilation, toroidal radiation asymmetry after injection, and radiated energy fraction \cite{ansh_patel_internship_2023, paul_heinrich_analysis_2023, s_jachmich_shattered_2023}. Further studies of the laboratory shattering experiments \cite{peherstorfer_fragmentation_2022, illerhaus_machine_2022} are also being carried out to determine fragment size and speed distribution for different pellet velocities and shattering angles and also different shatter heads. The results of the experimental data analysis relevant for this thesis is discussed in \autoref{ssec:sim_exp_comp}. Additional modelling activities involving JOREK \cite{w_tang_non-linear_2023} and DREAM \cite{peter_halldestam_modeling_2023} simulations for AUG SPI discharges are also being carried out.

\subsection{Objective of this thesis}
\label{ssec:objective}
The research question that this report aims to answer is the following:
\textbf{To what extent can the experimental trends observed in penetration, assimilation of shattered fragments in AUG be corroborated with disruption simulations?} \\\\
This thesis focuses on validating the relationship between injected material penetration and assimilation in the ASDEX Upgrade (AUG) tokamak with three key SPI parameters: fragment sizes, speeds, and pellet composition. The simulations carried out in the thesis will focus on the pre-TQ phase when the majority of material penetration and assimilation takes place \cite{matsuyama_transport_2022}. In this thesis, assimilation is linked to the ablated and ionized material that remains within the plasma volume. \\\\
The INDEX code presents a viable approach for conducting extensive parametric scans of SPI parameters, taking into account material penetration and plasma size information, while remaining computationally efficient compared to 3D codes. As ITER simulations with INDEX progress \cite{matsuyama_transport_2022}, it is crucial to validate the SPI models within INDEX using existing devices. \\\\
By successfully comparing INDEX simulations with experimental trends, the simulation results can be used to identify ideal pellet and fragment parameters for deepest penetration, highest and uniform assimilation. Successful matching can be followed by identifying relevant quantities that should be used to extrapolate to ITER SPI in future work. Unsatisfactory matching with experimental results can also provide insights on missing aspects of experiments not being included in the simulations which can significantly affect plasma dynamics that require new or improved models. 
\section{Methodology}
\label{sec:methodology}

\subsection{Setup}
\subsubsection{INDEX}
\label{ssec:INDEXintro}
To simulate mitigated disruptions, I have used the INDEX (Integrated Numerical Disruption EXperiment) \cite{matsuyama_enhanced_2022} code. INDEX assumes that the plasma temperature and pressure depend on only a single spatial coordinate, the toroidal magnetic flux ($\rho$) which is a proxy for the plasma radius. The plasma properties are assumed to be axisymmetric and hence INDEX is referred to as a 1.5D code. \\\\
The INDEX code solves a magnetic diffusion equation on 1D flux coordinates and Grad-Shafranov (G-S) equilibrium on the 2D space with the boundary conditions given by solving a series of circuit equations of toroidally continuous filaments \cite{matsuyama_requirements_2020}.The magnetic diffusion equation (refer to \autoref{eq:magnetic_diffusion}) describes the time evolution of the magnetic fields in the plasma while the Grad-Shafranov equation describes the magnetic field and pressure configuration in the poloidal plane of the tokamak. The plasma equilibrium is evolved on a resistive timescale. This timescale corresponds to the temporal evolution of the equilibrium on the order of the time associated with the diffusion and dissipation of magnetic fields due to finite plasma resistivity. In essence, the G-S equation indirectly accounts for the ideal Magnetohydrodynamics (MHD) effects required for plasma stability.  \\\\
Hence, the INDEX code can be used to simulate the pre-TQ phase all the way to towards the end of the CQ phase, usually till fast vertical displacement events \cite{gruber_vertical_1993}. The transport module models the plasma ion species, and additional impurity ion species for different charge states based on OpenADAS ionisation and recombination rates. The particle and energy transport of the ion species $\alpha$ with the $i$th charge state and energy transport for electrons that are used are: 

Particle balance
\begin{equation}
\frac{1}{V^{\prime}} \frac{\partial}{\partial t}\left(n_\alpha^{i+} V^{\prime}\right)=\frac{1}{V^{\prime}} \frac{\partial}{\partial \rho} V^{\prime}\left\langle|\nabla \rho|^2\right\rangle D_\alpha \frac{\partial n_\alpha^{i+}}{\partial \rho}+S_\alpha^{i+},
    \label{eq:transport_1}
\end{equation}

Ion energy balance
\begin{equation}
 \frac{3}{2}\left(V^{\prime}\right)^{-5 / 3} \frac{\partial}{\partial t}\left(p_\alpha V^{\prime 5 / 3}\right) =\frac{1}{V^{\prime}} \frac{\partial}{\partial \rho} V^{\prime}\left\langle|\nabla \rho|^2\right\rangle \times\left[\kappa_\alpha \frac{\partial p_\alpha}{\partial \rho}+T_\alpha\left(\frac{3}{2} D_\alpha-\kappa_\alpha\right) \frac{\partial n_\alpha}{\partial \rho}\right] 
 \quad+P_{\mathrm{ex}}^{\alpha e}+\sum_\beta P_{\mathrm{ex}}^{\alpha \beta},
    \label{eq:transport_2}
\end{equation}

Electron energy balance
\begin{equation}
 \frac{3}{2}\left(V^{\prime}\right)^{-5 / 3} \frac{\partial}{\partial t}\left(p_e V^{\prime 5 / 3}\right) \\
 =\frac{1}{V^{\prime}} \frac{\partial}{\partial \rho} V^{\prime}\left\langle|\nabla \rho|^2\right\rangle \times\left[\kappa_e \frac{\partial p_e}{\partial \rho}+T_e\left(\frac{3}{2} D_e-\kappa_e\right) \frac{\partial n_e}{\partial \rho}\right] \\
 \quad+P_{\text {ohm }}-P_{\text {rad }}-P_{\text {ion }}+\sum_\alpha P_{\mathrm{ex}}^{e \alpha},
    \label{eq:transport_3}
\end{equation}

where $\rho$ is the magnetic surface label, $V(\rho)$ denotes the plasma volume enclosed by the magnetic surface, $V^\prime  = \frac{dV}{d\rho}$, and the bracket $\langle \bullet \rangle$ denotes the surface averaging process. $n_\alpha^{i+}$ is the the density of the ion species $\alpha$ charge state with charge state $i$. $p_e$ is the electron pressure. A single scalar pressure for all the ion species is considered $p_\alpha = \left( \Sigma_{i=1}^Z n_{\alpha}^{i+} \right) T_\alpha$ is considered where $T_\alpha$ is the temperature of each ion species $\alpha$. In INDEX, quasineutrality is assumed and hence the electron density is calculated at each time step from the ion density. $P_\text{ohm}$ is the ohmic heating power and $P_{ex}^{\alpha \beta}$ is the energy transfer between different ion species $\alpha$ and $\beta$. The ionisation and recombination rates used to calculate the particle source and sink terms $S_\alpha^{i+}$ and the radiative cooling rates used to calculate the radiation and ionisation losses, $P_\text{rad}$ and $P_\text{ion}$, are extracted from the OpenADAS database \cite{summers_ionization_2006}. $D_\alpha$ and $\kappa_\alpha$ are the particle diffusion and heat diffusivity coefficients. In addition to the 1D transport equations above, a neutral particle source equation is also solved:

\begin{equation}
\frac{1}{V^{\prime}} \frac{\partial}{\partial t}\left(n_\alpha^0 V^{\prime}\right)=\frac{1}{V^{\prime}} \frac{\partial}{\partial \rho} V^{\prime}\left\langle|\nabla \rho|^2\right\rangle D_\alpha^0 \frac{\partial n_\alpha^0}{\partial \rho}+S_\alpha^0+S_\alpha^{\mathrm{SPI}}
\label{eq:neutral_1D}
\end{equation}

where $S_\alpha^0$ and $S_\alpha^{SPI}$ are the neutral particle source/sink terms associated with ionisation/recombination and the particle source due to the SPI respectively. Similar to recent pre-TQ SPI simulations with INDEX \cite{matsuyama_transport_2022}, neutral particle sources introduced in the pre-disruption hot plasma ionize instantaneously and do not influence the plasma significantly in the pre-TQ phase. Hence, the neutral particle transport $D_\alpha^0$ is set to be equal to the ion diffusion coefficient of the same species. The 1D coordinate $\rho$ is related to the toroidal magnetic flux $\chi=2 \pi \oint \mathbf{B} \cdot d \mathbf{S}_{\varphi}$ as $\rho = \sqrt{\frac{\chi}{\chi_s}}$ where the subscript \textbf{s} denotes the plasma surface and $\phi$ is the toroidal angle. The current density profile is coupled with the plasma density and temperature through the magnetic diffusion equation which is expressed in terms of the rotational transform $\nu = \frac{d\psi}{d\chi}$ as

\begin{equation}
\left.\frac{\partial \nu}{\partial t}\right|_\chi=\frac{\partial}{\partial \chi}\left[\frac{\eta}{\mu_0} \frac{(d \chi / d V)^2}{A^2} \frac{\partial}{\partial \chi}(A K \nu)-\frac{\eta}{d \chi / d V}\left\langle\mathbf{j}_{\mathrm{ex}} \cdot \mathbf{B}\right\rangle\right]
\label{eq:magnetic_diffusion}
\end{equation}
where $K(\psi) = \left< \frac{|\nabla V|^2}{R^2} \right>$ and $A(\psi) = \left< \frac{1}{R^2} \right>$ are the surface averaged metric coefficients and $\eta(\psi)$ denotes neoclassical conductivity \cite{hirshman_neoclassical_1977}. The last term in \autoref{eq:magnetic_diffusion} is used to model externally or runaway driven current densities. While runaway electrons can play a significant role in the CQ phase, they are inconsequential in the pre-TQ phase and hence are not considered in the simulations carried out in this thesis. 
By coupling \autoref{eq:transport_1}, \autoref{eq:transport_2}, \autoref{eq:transport_3}, \autoref{eq:neutral_1D} with \autoref{eq:magnetic_diffusion} through $\eta(\psi)$, the evolution of the current density profile can be self-consistently solved taking into account changes in the plasma temperature and density due to the assimilation of the injected material. The INDEX free-boundary equilibrium solver is similar to that of the DINA code \cite{khayrutdinov_studies_1993, miyamoto_inter-code_2014}. 

\subsubsection{SPI in INDEX}
The INDEX code has a SPI module which has been applied to various tokamaks such as DIII-D \cite{lvovskiy_evolution_2022}, KSTAR, ITER \cite{matsuyama_transport_2022} and now AUG. To generate a fragment plume, first INDEX generates an initial fragment size distribution ($r_p (t=0)$) calculated using a Bessel function based on a statistical fragmentation model \cite{bosviel_near-field_2021} with the probability distribution: 

\begin{equation}
    P(r_p) = \frac{r_p K_0 (\kappa_p r_p)}{I}, \quad I = \int_0^\infty r_p K_0 (\kappa_p r_p) dr = \kappa_p^{-2} 
    \label{eq:fragmentation_size}
\end{equation}

where $K_0$ is the modified Bessel function of the second kind, $\kappa_p$ is the inverse of the characteristic fragment size and is related to the total number of molecules in the pre-shattered pellet and to the number of fragments generated after shattering \cite{hu_3d_2018}. INDEX pseudo-randomly samples $N_f$ number of different fragment sizes $r_i (i = 1,2...N_f)$ from the probability distribution in \autoref{eq:fragmentation_size}. Assuming each fragment has a spherical shape, it then calculates the number of molecules in each fragment $N_i = \frac{4}{3} n_m \pi r_i^3$ where $n_m$ is the molecular density of the pellet material. Since only a finite number of fragments are sampled, the sum of the molecules in the fragments do not match with $N_p$. To compensate for this, the fragment size distribution is renormalized to match the pellet material amount resulting in the number of molecules in the new fragments being $N_i^\prime * \frac{N_p}{\Sigma_i N_i} N_i$ and the new sizes being $r_i^\prime = (3N_i^\prime / 4 \pi n_m)^{1/3}$. Since the fragments sizes are sampled using pseudo-random numbers, different realisations of the same probability distribution can be generated. Experimentally, the pellet shattering process is also a stochastic one so the shattered fragments will have slightly different sizes even if the pellet injection parameters are the same. The effect of different realisations of fragment size distributions from the same probability distribution is only briefly discussed in this thesis and is being further investigated in DREAM simulations \cite{peter_halldestam_modeling_2023}. However, to ensure that trends in penetration and assimilation are due to the changes in the shattering parameters and not due to a particular outlier realization of the fragment size and speed probability distribution, 3-5 different realizations are used for simulation sets of parametric scans. However, to systematically eliminate the effect of outlier realizations, substantially more realizations should be utilized. \\\\
The initial velocity distribution of the fragments is modelled as a Gaussian distribution with a user-defined standard deviation and the pellet speed as the mean. The standard deviation of the fragment speed distribution for all the simulations in this thesis was set to 40\% based on JET SPI experiments \cite{jachmich_shattered_2021}. It is expected that different pellet speeds and shattering angles might modify the standard deviation \cite{peherstorfer_fragmentation_2022}, however additional experimental inputs are required for taking this into account. After the generation of fragments with corresponding sizes and speeds, the fragments travel towards the plasma with a toroidal and poloidal spread which was set to $\pm 20^\circ$ for all the simulations in this thesis \cite{peherstorfer_fragmentation_2022}. INDEX maps the 3D location of the fragments at each time step to the 1D flux coordinates used in the particle and energy transport equations taking into account the toroidal geometry. The ablation of shattered fragments in the plasma is modelled using the Neutral Gas Shielding (NGS) model which has been developed and used previously for fuelling pellets \cite{parks_effect_1978}. Experimentally, the dependence of the penetration depth of non-shattered pellets for fuelling the core plasma \cite{milora_pellet_1995,pegourie_review_2007} follows a scaling close to that obtained from the NGS model. The recession rate of the fragment radii used in INDEX has been proposed in Parks et. al. \cite{parks_theoretical_2017} and can be calculated for neon-deuterium mixed pellets with molar ratio $X$ of Deuterium as 

\begin{equation}
    \begin{aligned}
\frac{d N_{\text {mix }}}{d t} & =\frac{C \lambda(X)}{f_W(1-X)+X} n_e^{1 / 3} T_e^{5 / 3} r_p^{4 / 3} \text {   particles s }^{-1}, \\
\lambda(X) & =A+\tan (B X)
\end{aligned}
    \label{eq:ablation_recession}
\end{equation}

where $f_W$ is the mass ratio between the neon atom and the deuterium molecule, $n_e$ and $T_e$ are the electron density (in $m^{-3}$) and electron temperature (in eV), respectively at the location of the fragment with radius $r_p$ (in \unit{m}). Using \autoref{eq:ablation_recession}, the number of ablated deuterium and neon atoms are deposited in the plasma are estimated using: 
\begin{equation}
    \frac{dN_D}{dt} = 2X \frac{dN_{mix}}{dt}; \quad \frac{dN_{Ne}}{dt} = (1-X) \frac{dN_{mix}}{dt}.
\end{equation} 
The numerical coefficients in \autoref{eq:ablation_recession} are \cite{matsuyama_transport_2022, parks_theoretical_2017} $f_W \equiv W_{Ne}/W_{D_2} = 20.183/4.0282$, $A = 27.0837$, $B = 1.48709$, $C = 4.062 \times 10^{14}$. 
During a simulation, INDEX traces the motion of multiple pellet markers with given fragment velocities, calculates the recession in fragment sizes using the NGS model and assumes the ablated particles are deposited as a flux-surface-averaged neutral particle source. The particle source is convoluted into 1D radial flux coordinates using a Gaussian shape centred at the particle source location and with a width of a few centimetres in real space based on previous measurements of fuelling pellet ablation \cite{kocsis_fast_2004}.

\subsubsection{AUG inputs for INDEX}
\begin{figure}[H]
    \centering
    \includegraphics[width = 0.6\linewidth]{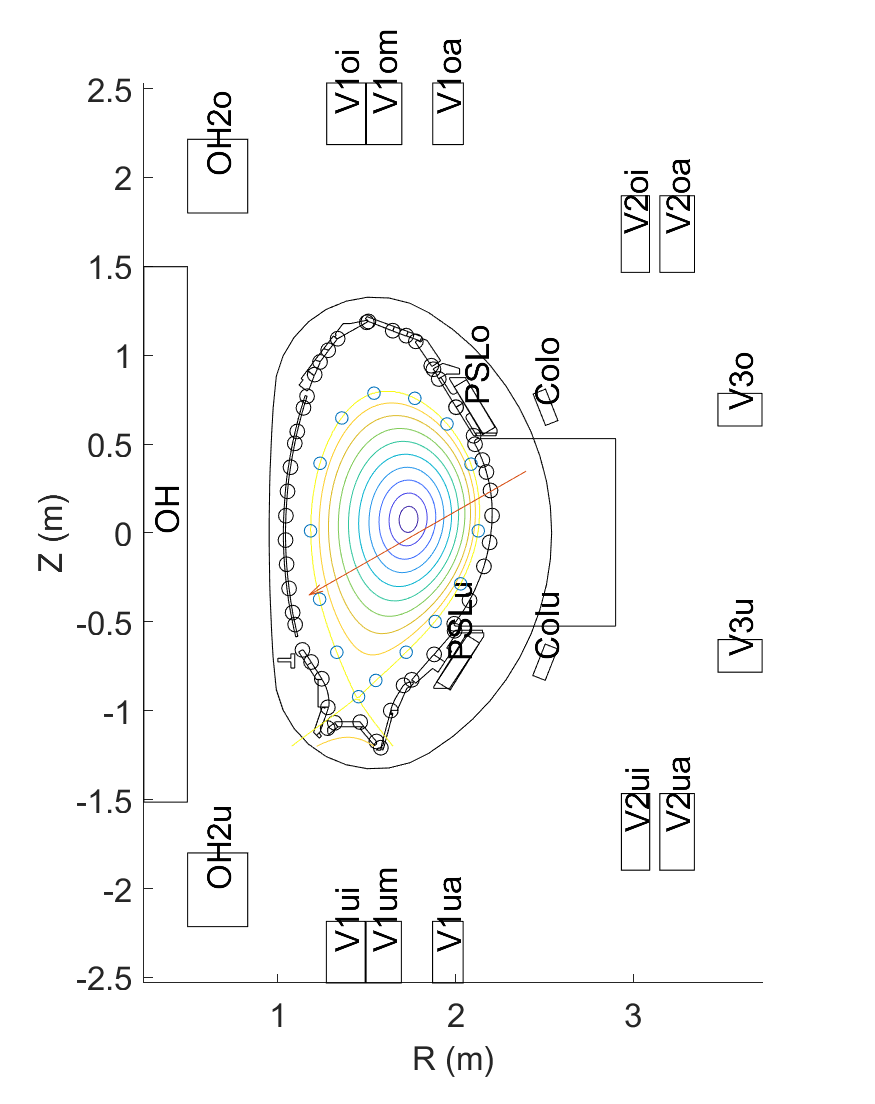}
    \caption{AUG geometrical components implemented in the INDEX code shown in a poloidal cross section.}
    \label{fig:AUG_geo_elements}
\end{figure}

Before running INDEX simulations for AUG SPI, I set up various input parameters and profiles. These inputs include geometric and electrical properties of various ohmic, poloidal field coils, and vessel elements that influence the plasma equilibrium convergence in INDEX. A set of initial conditions of electron, temperature and current density radial profiles are also required as input. A corresponding plasma equilibrium for the above inputs is also required. Coordinates of plasma facing components (PFCs) act as boundaries which when crossed by the plasma cause the simulation to stop. For the SPI, coordinates of the beginning and the end of the shattering tubes is also required. I implemented the geometrical and electrical properties of various AUG coil and vessel elements in the INDEX code, which are shown in \autoref{fig:AUG_geo_elements}. The rectangular elements outside the vacuum vessel in \autoref{fig:AUG_geo_elements} are the ohmic and the poloidal field coils. The coils with prefix "OH" refer to ohmic heating coils, prefix "V" are shaping coils, prefix "CoI" are position control coils and prefix "PSL" are passive stabilising loop coils \cite{koppendorfer_asdex_1986}. The plasma equilibrium surfaces of constant poloidal flux are shown in colours ranging from blue to yellow. Blue scatter points on the last closed flux surface show a few of the points provided as input for the plasma equilibrium in INDEX. Black scatter points on the plasma facing components are the boundary points provided to terminate the simulation if crossed. The orange arrow line shows the injection vector for one of the $25^\circ$ injection tubes in AUG. 

\begin{figure}[H]
   \centering
     \begin{subfigure}[t]{0.45\textwidth}
       \centering
       \includegraphics[width = \linewidth]{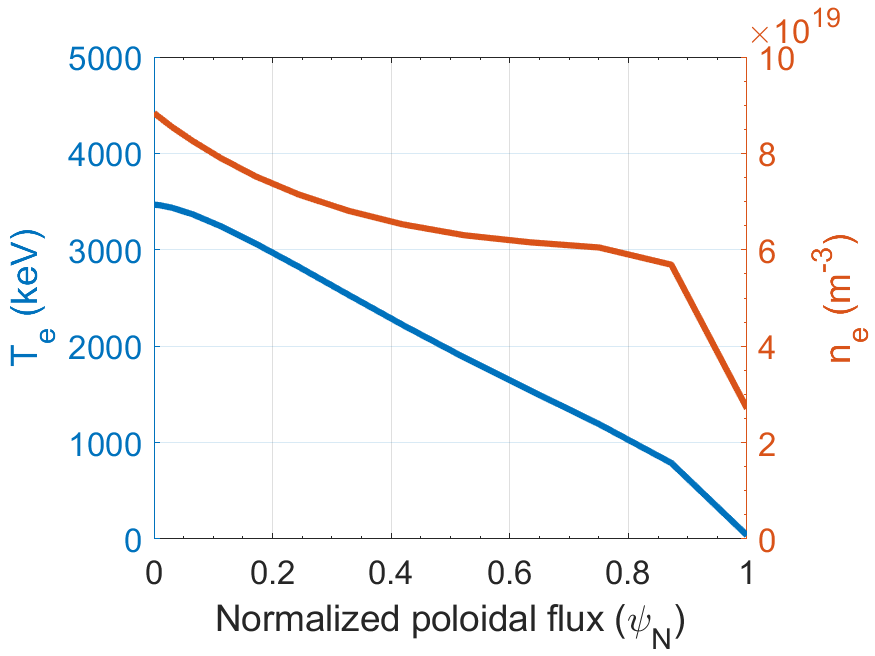}
       \caption{Electron temperature (left axis) and electron density (right axis) profiles}
       \label{fig:oneD_profs1}      
     \end{subfigure}
     \begin{subfigure}[t]{0.45\textwidth}
        \centering
        \includegraphics[width = \linewidth]{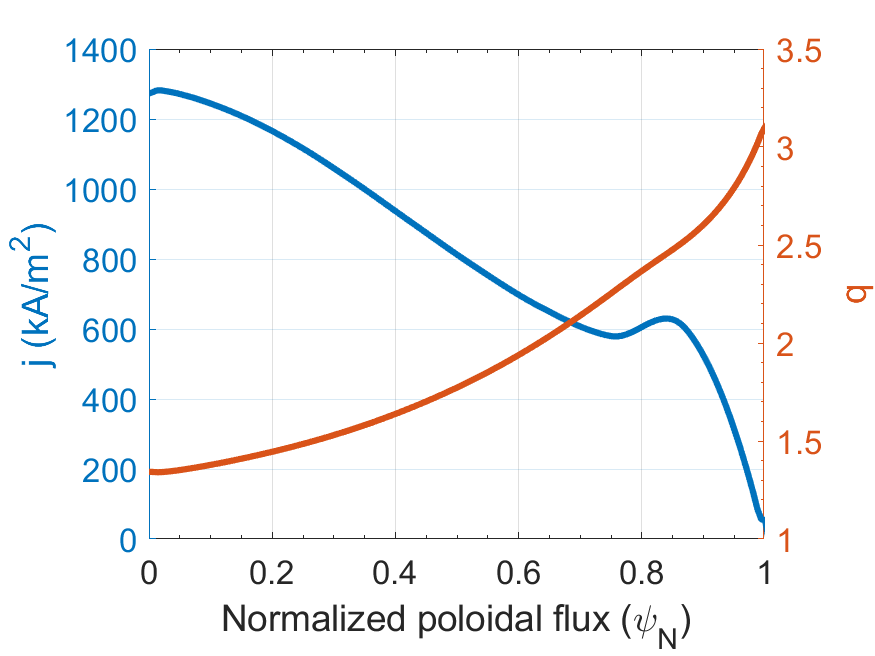}
        \caption{Current density (left axis) and resultant safety factor q (right axis)}
        \label{fig:oneD_profs2}                 
     \end{subfigure}
     \caption{Radial plasma profiles used for initial convergence of plasma equilibrium in INDEX.}
     \label{fig:oneD_profs}
\end{figure}

The 1D profiles of electron temperature, density and current density used for initial equilibrium convergence in INDEX are shown in \autoref{fig:oneD_profs}. These profiles were taken from \#40355, which was a typical H-mode discharge, at 2.25s. The same profiles are used as inputs for JOREK simulations of AUG SPI \cite{w_tang_non-linear_2023}. 

\begin{figure}
    \centering
    \includegraphics[width=0.6\linewidth]{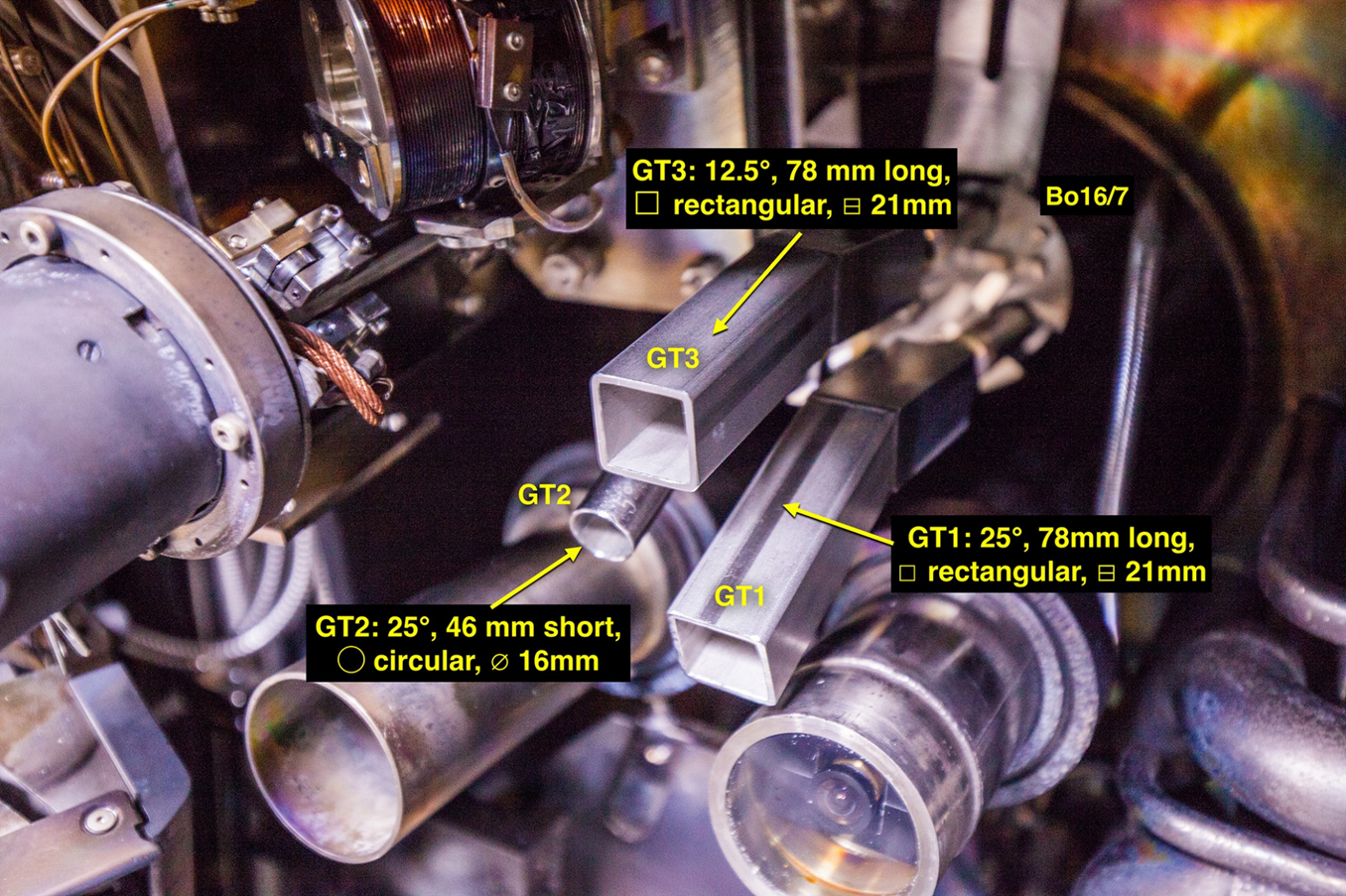}
    \caption{The three shatter tubes inside the vessel. Figure taken from Heinrich et. al. \cite{heinrich_characterization_2022}.}
    \label{fig:shatter_tubes_zoom}
\end{figure}

As mentioned before, AUG has 3 different shatter tubes whose start and end coordinates (X,Y,Z in machine coordinates) are provided in \autoref{table:AUGshattertubes}. An image of the shattering tubes inside the machine is also shown in \autoref{fig:shatter_tubes_zoom}. For all the simulations carried out in this thesis, the particle diffusion coefficient was set to $D_\alpha = 2 $ \unit{m^2/s} and the thermal diffusivity for ions and electrons was set to $\kappa_\alpha = 4.5$ \unit{m^2/s}. It should be noted that the diffusion coefficients were assumed to be radially uniform. In this thesis, the simulations were conducted without considering any inherent machine impurities (like tungsten) to prioritize the exploration of parametric variations in the SPI parameters.

\begin{table}[h]
\caption{Start and end coordinates of the shattering tubes at AUG (in machine XYZ coordinates with the geometric centre of the machine as the origin)}
\centering
\begin{tabular}{p{6cm} || p{5cm} | p{5cm}}
\hline
Shattering tube (shattering angle, cross-section shape) & Shattering tube start coordinate (X,Y,Z)  m& Shattering tube end coordinate (X,Y,Z) m\\
\hline
GT1 $(25^\circ$, rect.) & (2.340, -0.510, 0.347) & (2.227, -0.489, 0.308)\\
GT2 $(25^\circ$, circ.) & (2.233, -0.458, 0.343) & (2.295, -0.452, 0.324)\\
GT3 $(12.5^\circ$,  rect.) & (2.341, -0.490, 0.385) & (2.271, -0.469, 0.361)\\
\end{tabular}
\label{table:AUGshattertubes}
\end{table}

\begin{figure}[h]
    \centering
    \includegraphics[width = 0.4\linewidth]{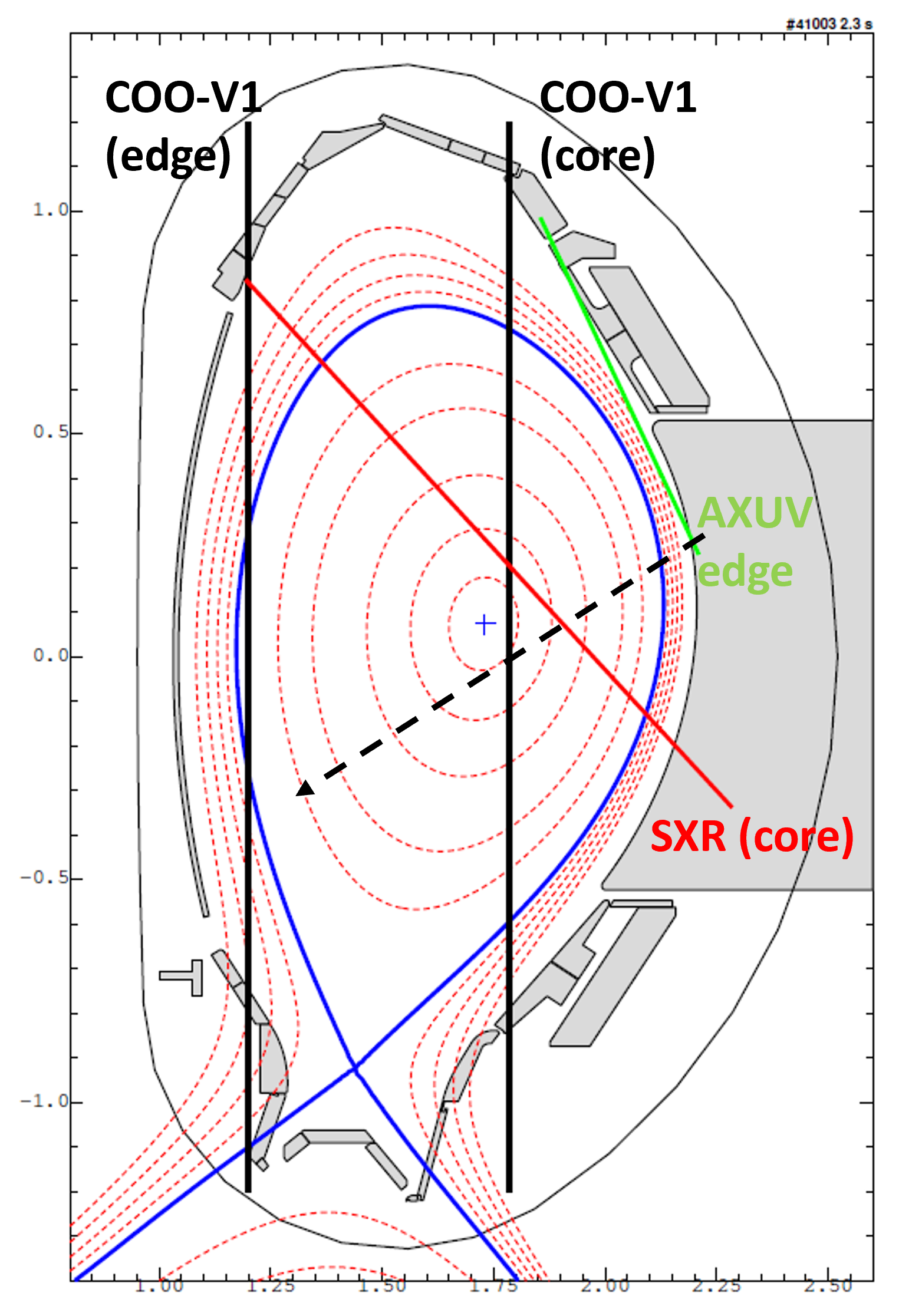}
    \caption{Relevant line of sight diagnostics for SPI in the AUG poloidal cross section. Black dashed arrow shows the injection vector for the GT1 shatter tube.}
    \label{fig:AUGDiagnosticLOS}
\end{figure}

For comparison of assimilation measurements, Thomson scattering \cite{kurzan_edge_2011, murmann_thomson_1992} and CO\textsubscript{2} laser interferometer \cite{mlynek_infrared_2012, mlynek_simple_2017} diagnostics were used in AUG. Initial assessment of the validity of the diagnostic signals and their limitations during SPI was reported in my ITER internship report \cite{ansh_patel_internship_2023}. To compare the INDEX simulations to the experiments, I implemented synthetic diagnostics of these two diagnostics in the INDEX code. The line of sights of these diagnostics along with some other relevant diagnostics are shown in the poloidal cross section in \autoref{fig:AUGDiagnosticLOS}. The COO interferometers have a vertical line of sight shown by the black solid lines. The Thomson scattering diagnostic has the same poloidal location as the core channel of the interferometer. 

\subsection{Example: mixed deuterium/neon injection}
\label{ssec:exampleSim}
\begin{figure}[H]
    \centering
    \includegraphics[width = 0.6\linewidth]{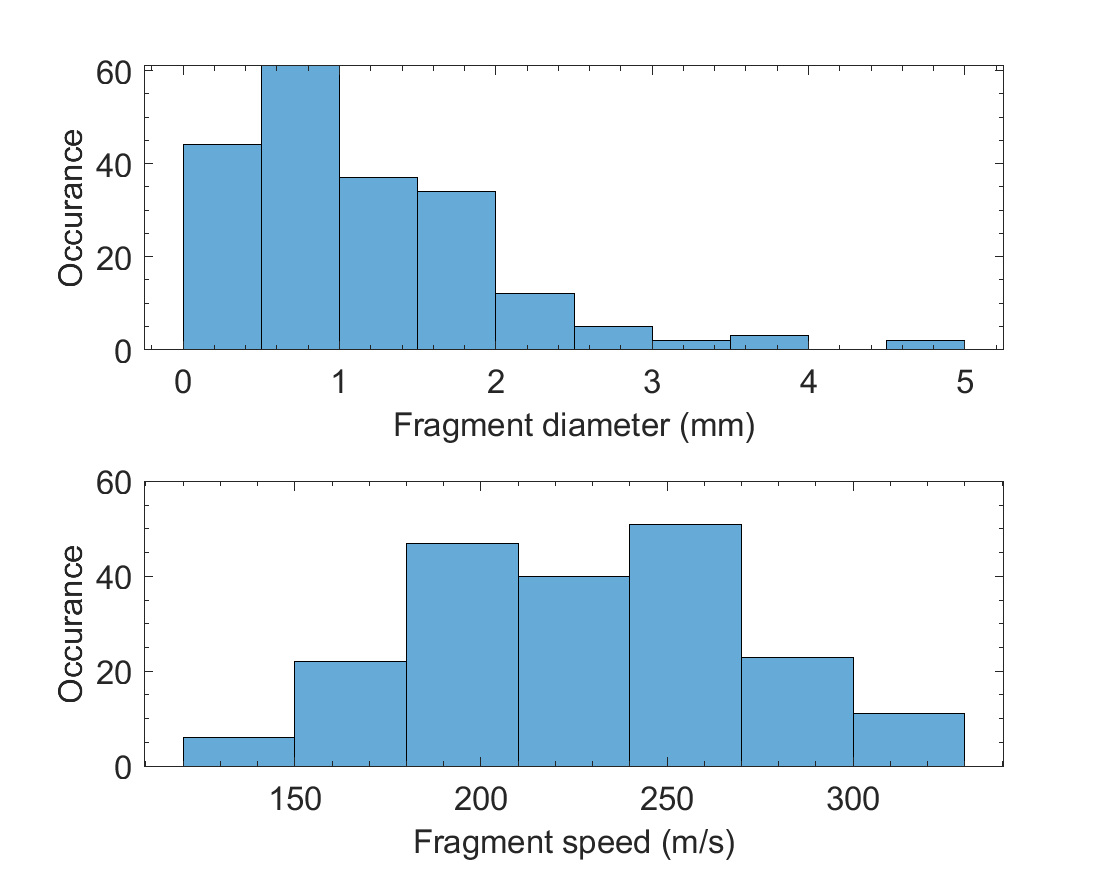}
    \caption{Binned fragment size and speed distributions for the example simulation.}
    \label{fig:size_speed_distribution_example}
\end{figure}

In order to introduce the simulation procedure, an example pre-TQ simulation with an injection of 10\% neon / 90\% deuterium (molecular number composition) is discussed. It starts at $t_\text{sim}=0$ ms when the fragments are assumed to be at the shattering point. In the present simulation, 200 fragments were sampled for a pellet of length of 10 mm and diameter of 8mm. The fragment velocities were sampled from a Gaussian distribution centred at $ \langle v_{\text{frag}} \rangle = 230$ m/s with $\Delta v/v = 40\%$. The resultant fragment size and speed distributions are shown in \autoref{fig:size_speed_distribution_example}. The average size from the fragment size distribution was 1.14 mm. In the remainder of this thesis, fragment size and speed distributions for multiple distributions and realizations will be shown as continuous distributions for plotting purposes (example: \autoref{fig:size_var_frag_size_speed_distributions}) however it should be kept in mind that the distributions have been discretized.

\begin{figure}[H]
         \centering
         \includegraphics[width = 0.7\linewidth,clip,trim={0 15.2cm 0 0}]{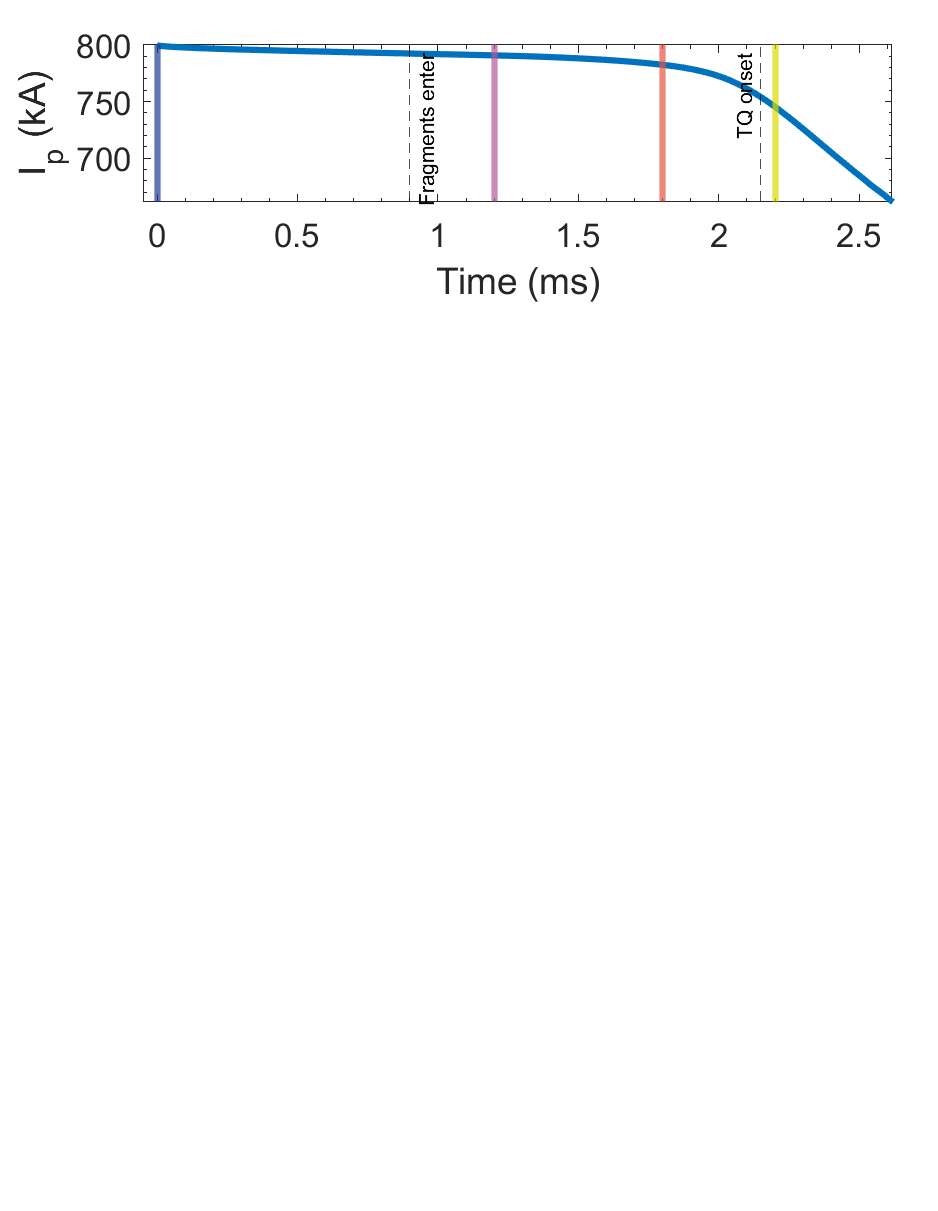}
         \caption{Time trace of plasma current evolution for a mixed deuterium/neon injection.}
         \label{fig:exampIp}    
         \centering
         \vspace{1cm}
         \includegraphics[width = 0.7\linewidth]{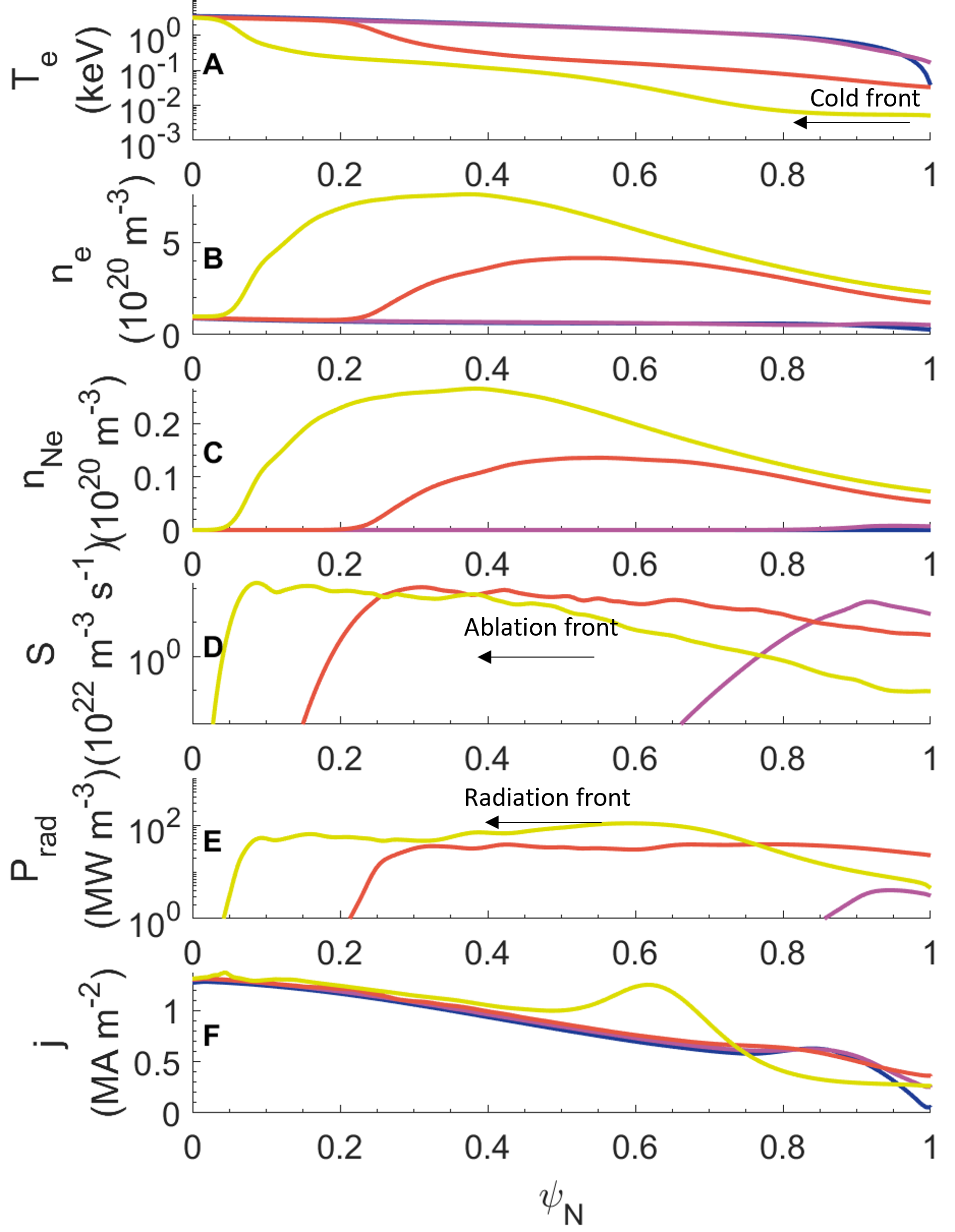}
         \caption{Profiles for various SPI relevant parameters against normalized poloidal flux: (A) electron temperature, (B) electron density, (C) neon density, (D) neutral source function, (E) radiated power, (F) Current density. Different colors indicate different time points of the simulation.}        
         \label{fig:exampRemainingsubplots}               
\end{figure}

The time trace of plasma current evolution is shown in \autoref{fig:exampIp}. An initial decay in the plasma current is caused by two reasons. First, the current profile is modified as the simulation progresses due to the absence of a bootstrap current model in INDEX which plays a significant role in H-mode discharges. Second, while the loop voltage (which drives the current) is taken into account in the circuit equations (refer to \autoref{ssec:INDEXintro}), the present simulation does not include "realistic" AUG coil currents actions to maintain constant $I_p$. To compare the pre-TQ duration in the experiments to the simulation, a condition similar to the experiment was considered. Experimentally, the arrival of the fragments in the plasma is expected when a significant rise in an edge AXUV channel (\autoref{fig:AUGDiagnosticLOS}) is observed. Similarly, in the simulation, the arrival of the fragments in the plasma is considered when the first fragment crosses a circle centred at the plasma magnetic axis and a radius equal to the distance between the plasma centre and the point of intersection of the edge AXUV channel and the injection vector, which is $\sim$ 50 cm. The time of fragments entering the plasma is marked in \autoref{fig:exampIp} at 0.8 ms with a black dashed line. Several vertical lines are marked as well which correspond to time stamps whose plasma profiles are plotted in \autoref{fig:exampRemainingsubplots}. The radial profiles are plotted against the normalized poloidal flux label $\psi_N = (\psi - \psi_a) / (\psi_s - \psi_a)$ where $s$ and $a$ denote the magnetic axis and the last-closed flux surface, respectively. Hence $\psi(0)$ corresponds to the magnetic axis while $\psi(1)$ corresponds to the last closed flux surface. The choice of this label was motivated by previous INDEX simulations \cite{matsuyama_transport_2022} that were used for benchmarking with JOREK results. \\\\
As discussed in \autoref{sec:DMS_ITER}, the pre-TQ duration is an important parameter that can limit the amount of material assimilation in the plasma before the onset of the disruption. Based on the discussion in \autoref{ssec:dms_progress}, the TQ onset is expected when the plasma temperature becomes less than 10 eV inside the q=2 surface. Such a semi-empirical TQ onset condition is required as the 1D transport simulations do not include any description of linear or nonlinear MHD instabilities. However, it is expected that the axisymmetric current contraction that can lead to the TQ can be captured by simulating the particle energy balance and the resulting current density profiles. The onset of the TQ condition in this example is marked in \autoref{fig:exampIp} at $\sim$ 2.15 ms. 

In \autoref{fig:exampRemainingsubplots}, the plasma response to the SPI is shown. The physical events are given in chronological order:
\begin{enumerate}
    \item As the fragments penetrate the plasma, initial rise in the neutral particle source function at the edge is observed in \autoref{fig:exampRemainingsubplots} (D) in the purple profile corresponding to 1.2 ms in the simulation. An increase in the radiated power is also observed in \autoref{fig:exampRemainingsubplots} (E) due to line radiation associated with neon.
    \item At 1.8 ms, corresponding to red colored profiles, a significant neutral particle source is deposited towards the core as shown in \autoref{fig:exampRemainingsubplots} (D). Following behind the neutral particle source in space, an increase in the electron and neon density can be observed in \autoref{fig:exampRemainingsubplots} (B), (C) respectively. A further rise in intensity in the radiated power can be observed in \autoref{fig:exampRemainingsubplots} (E) which only lags slightly behind the ablation profile in space. It should be noted that I have observed similar movement of the ablation front and radiation front in space in preliminary analysis of experimental SPI discharges \cite{ansh_patel_internship_2023}. Because of the radiated power, a drop in the plasma temperature can also be observed in \autoref{fig:exampRemainingsubplots} (A). However, the formation of a cold front ($<$ 10 eV) does not occur yet.  
    \item At 2.2 ms (dark yellow color profiles), the penetration of material continues which is ablated in the still sufficiently hot plasma. This can be observed in the ablation profile in \autoref{fig:exampRemainingsubplots} (D). More material is deposited which can be observed in \autoref{fig:exampRemainingsubplots} (B,C). Further radiative cooling of the plasma leads to the formation of a cold front towards the edge of the plasma. A small bump can be observed in the radiated power profile in \autoref{fig:exampRemainingsubplots} (E) which occurs at the forefront of the cold front. The small bump in the current density profile in \autoref{fig:exampRemainingsubplots} (F), arises due to the contraction of the current profile from the edge due to the movement of the cold front. As the TQ onset condition is satisfied at $\sim$ 2.15 ms, strong MHD activity can be expected at this time which would enhance transport significantly. As discussed in \autoref{sec:theoreticalback}, the onset of the TQ also leads to a small spike in the plasma current (also seen in experiments, refer \autoref{fig:40743_discharge_time_trace}) which is not observed in the present simulations. This discrepancy arises as the present simulations do not utilize a hyper-resistivity model \cite{nardon_origin_2023} that can emulate the enhanced plasma transport.       
\end{enumerate}

\begin{figure}[H]
   \centering
     \begin{subfigure}[b]{0.7\textwidth}
         \centering
         \includegraphics[width=\textwidth]{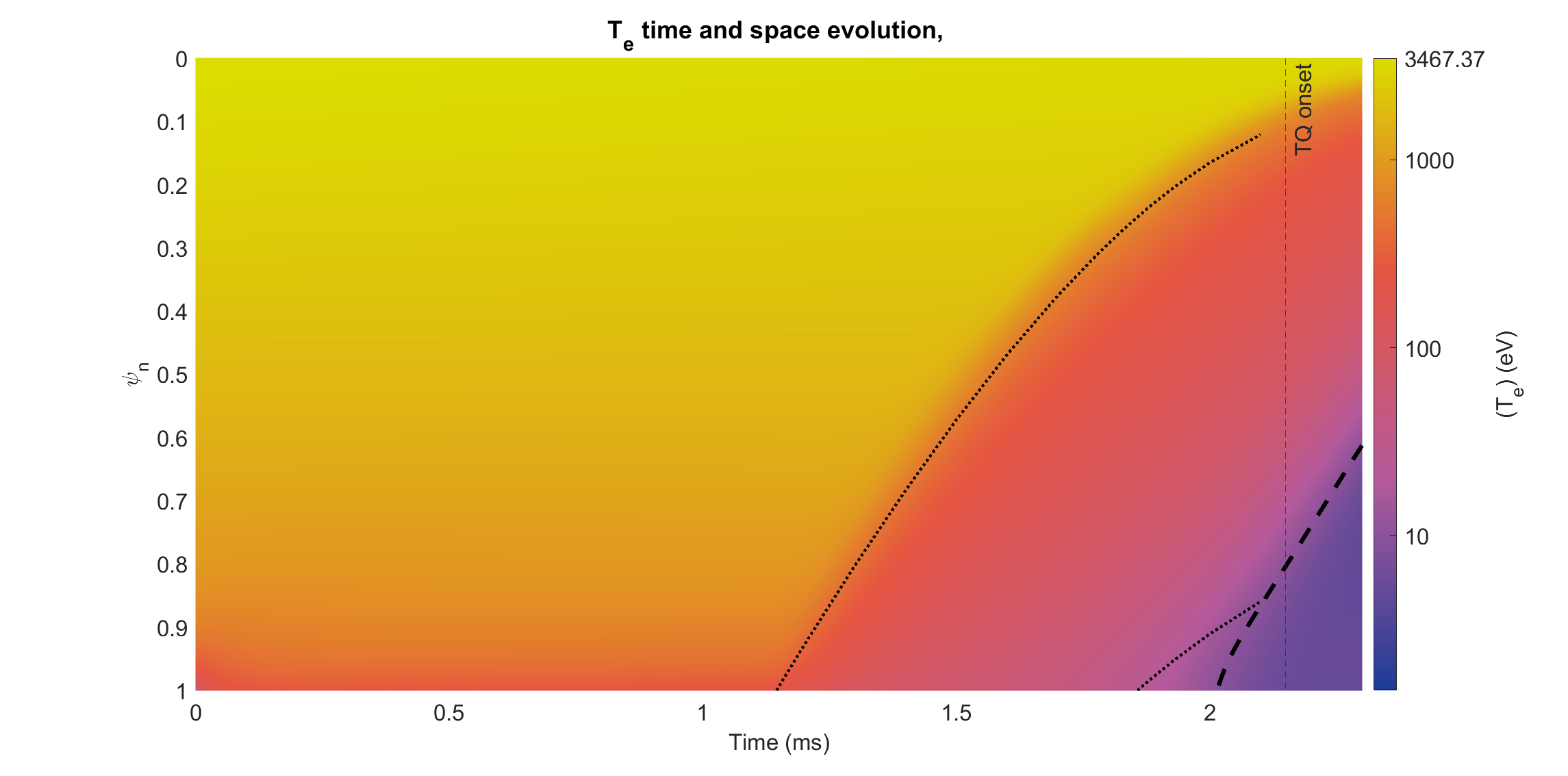}
         \caption{Electron temperature space and time evolution}
         \label{fig:examp_2D_Te}         
     \end{subfigure}
     \begin{subfigure}[b]{0.7\textwidth}
         \centering
         \includegraphics[width=\textwidth]{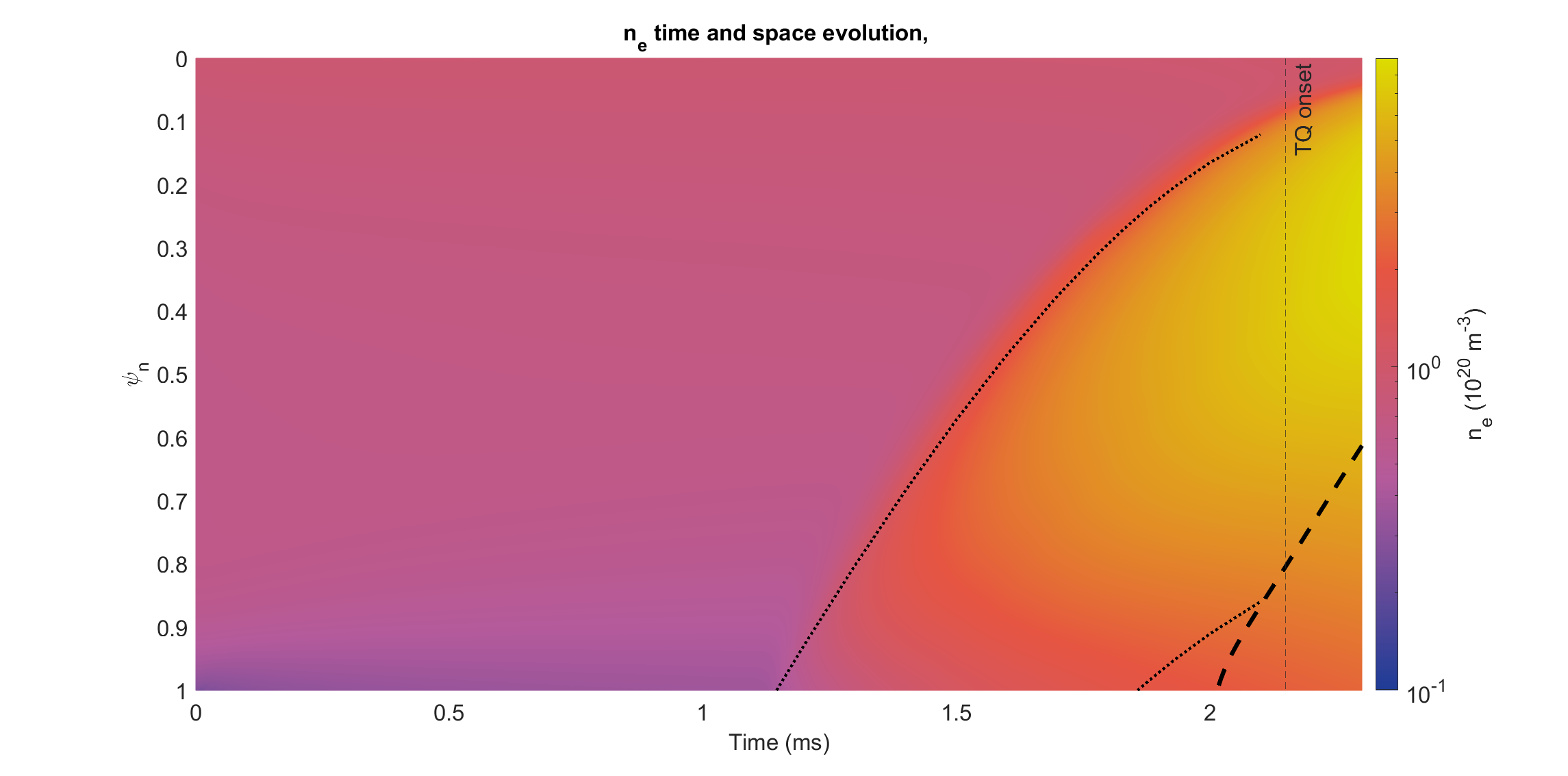}
         \caption{Electron density space and time evolution}
         \label{fig:examp_2D_ne}
     \end{subfigure}     
     \caption{Density and temperature profile evolution in space and time. Black dotted lines mark the trajectories of the fastest and slowest fragments. Black dashed line marks the movement of the cold front. Vertical black dotted line marks the crossing of the cold front beyond the $q=2$ surface hereby satisfying the TQ onset condition.}
     \label{fig:examp_2D_profs}
\end{figure}

With the fragments entering at 0.8 ms, the pre-TQ duration for this example simulation is 1.35 ms. This method of estimating the pre-TQ duration is used later in this thesis for comparing trends of simulated pre-TQ duration(s) with experimental pre-TQ measurements. The simulation stops at $\sim$ 2.6 ms due to difficulties in plasma equilibrium convergence.  \\\\
As discussed above, a common feature of the mixed deuterium/neon injections is that the increase in ablation and radiation follows the movement of the fastest fragments however, the cold front lags behind in space. While this can be seen in the plasma profiles in \autoref{fig:exampRemainingsubplots} (A,D), this distinction can be made more clear by studying the electron temperature and density profiles in space and time as shown in \autoref{fig:examp_2D_profs}. One of the black dotted lines represent the trajectory of one of the fastest fragments in the distribution which appears at the plasma edge at $\sim$ 1.1 ms and the other black dotted line appearing at $\sim$ 1.8 ms shows the trajectory of the fragment whose speed was at the 25\textsuperscript{th} percentile within the distribution indirectly indicating the position of the slowest few fragments. While the plasma density starts increasing immediately after the entry of the fastest fragments, plasma cooling happens on a longer timescale. The edge and intermediate plasma regions have a temperature on the order of 100 eV after the first fragments and eventually the formation of a cold front occurs at $\sim$ 2 ms as marked by the black dashed line. When the cold front reaches the $q=2$ surface which corresponds to $\psi_N \approx 0.86$, the onset of the TQ is expected. 

\subsection{Ablation dynamics for pure deuterium injections}
\label{ssec:backavgTS}
Pure deuterium pellet injections experience a plasmoid drift effect when injected from the magnetic LFS \cite{matsuyama_enhanced_2022}. In this section, I will briefly describe the plasmoid drift, which was first observed with fueling pellets \cite{muller_high_2002}. The injected fragments start to ablate when they enter the plasma which leads to the formation of a cigar shaped plasmoid along the field lines. The plasmoid expansion rate is slower than the rate of energy input into the fragments leading to over-pressurised plasmoids. The plasmoids become vertically polarised due to charge separation in a non-uniform magnetic field and experience a $\Vec{E} \times \Vec{B}$ drift \cite{rozhansky_evolution_1995} in the direction of the major radius. This drift can reduce the assimilation (ionized material within the plasma volume) of the injected material, even limiting the assimilation to the edge of the plasma \cite{matsuyama_enhanced_2022}. This effect is undesirable as increasing core density is important to avoid runaway electron formation. This drift is not observed for mixed D/Ne injections. Modelling \cite{matsuyama_neutral_2022} results indicate that the line radiation from neon for a mixed neon pellet has a significant impact on limiting the temperature and pressure of the ionized plasmoid. As a result, the ionized plasmoid is expected to homogenize along the magnetic field lines where the ablated material is released, without causing substantial cross-field drift motion.\\\\
In INDEX, the drift estimation is not based on first principles but instead on a back-averaging model \cite{jardin_fast_2000}. A user-defined fraction of the ablated atoms at radius $\rho^\prime = r/a$ is deposited uniformly over the exterior plasma volume where $\rho > \rho^\prime$. The back-averaged density increase caused by single shards is then calculated as 
\begin{equation}
    \Delta n_\text{dep} = \Delta N / \left[(1+\beta) V_p (1) - V_p(\rho^\prime) \right]
\label{eq:backavg}
\end{equation}
where $\Delta N$ is the number of ablated atoms and $V_p (\rho^\prime)$ is the interior plasma volume with $\rho < \rho^\prime$. The extent of the radial drift displacement is determined by the factor $\beta$. Physically, $\beta$ corresponds to the ratio of the volume outside the plasma where the particles can be deposited to the volume inside the plasma. Hence, it controls the spatial extent of the particle deposition. Due to the lack of a self-consistent drift model in INDEX, an interpretative approach is required to determine the value of $\beta$. In the present setup, the same $\beta$ value is used for all the fragments. A user-defined fraction of shifted atoms can be set to any value between 0 and 1 where 0 corresponds to all the deposited atoms being shifted from the ablation location (the extent of the shift depending on $
\beta$) and 1 corresponds to the atoms ablating at the fragment location. For pure deuterium injections presented later in this report, the fraction of locally deposited particles was set to 0. \\\\

\begin{figure}[H]
    \centering
    \includegraphics[width = 0.9\linewidth]{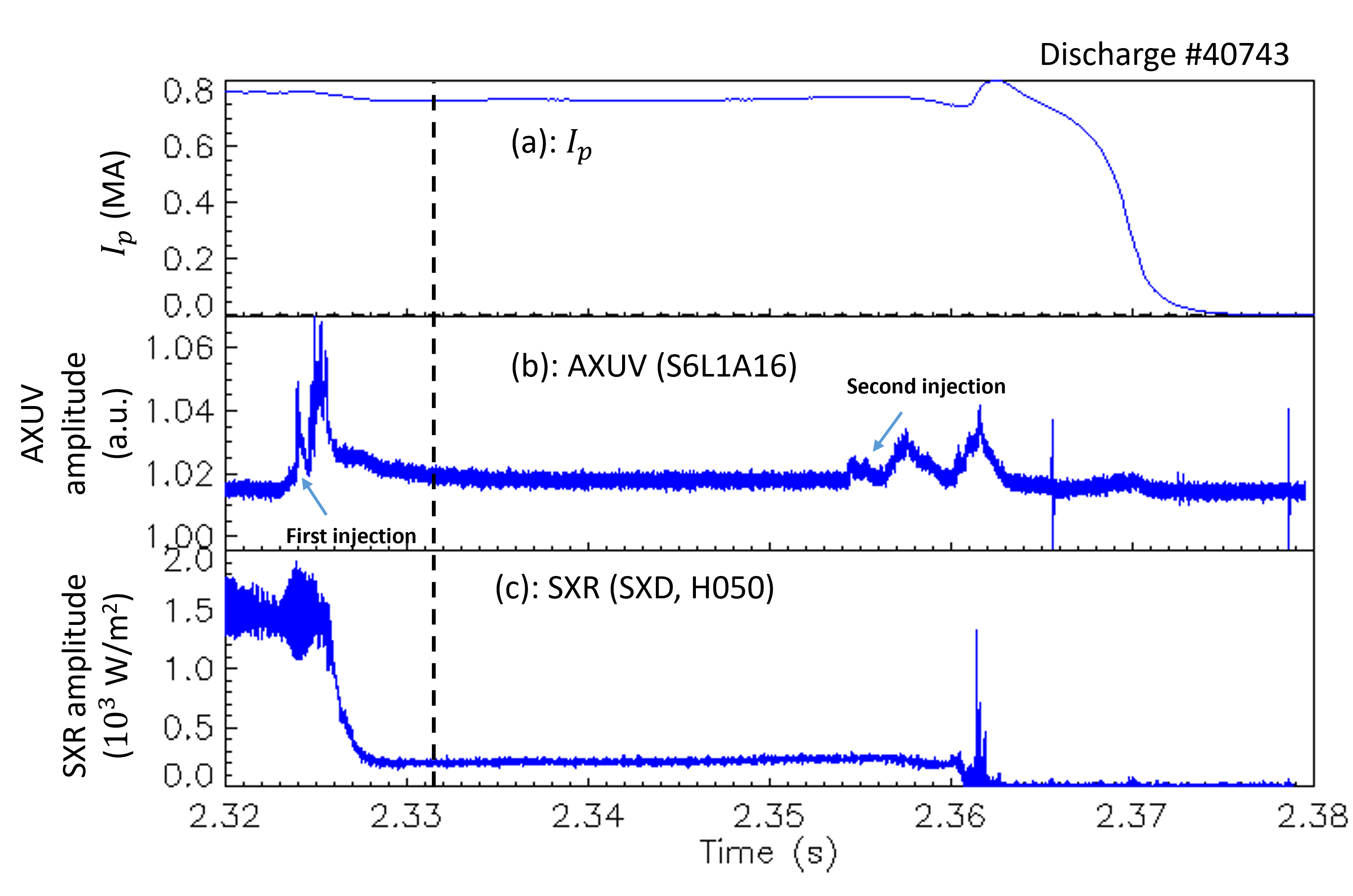}
    \caption{Time traces of plasma and diagnostic signals for \#40743: (a) Plasma current in MA, (b) AXUV bolometer signal whose LOS passes very close to the injection location, (c) Soft X-Ray (SXR) amplitude from a channel passing through the core of the plasma. Black dashed line marks the time stamp of Thomson scattering measurement.}
    \label{fig:40743_discharge_time_trace}
\end{figure}
\begin{figure}[H]
    \centering
    \includegraphics[width=0.8\linewidth]{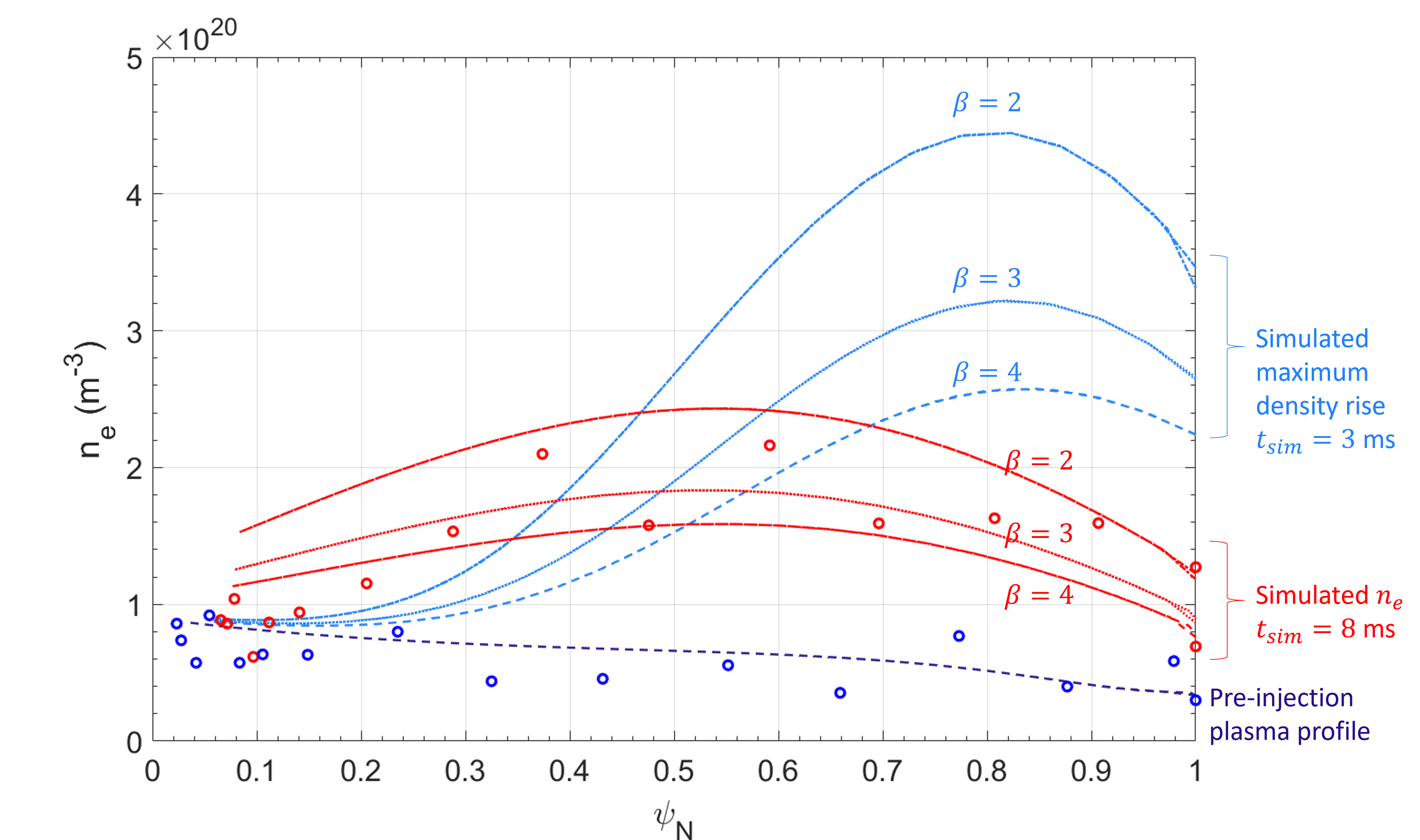}
    \caption{Comparison of experimental (\#40743) and synthetic Thomson scattering measurements. Blue and red circular scatter points show the experimental plasma density measurements before and after SPI respectively. Plotted lines show the simulated plasma density profiles at different times.}
    \label{fig:TS_SimExpCompar}
\end{figure}
\begin{figure}[H]
   \centering
     \includegraphics[width=0.8\linewidth]{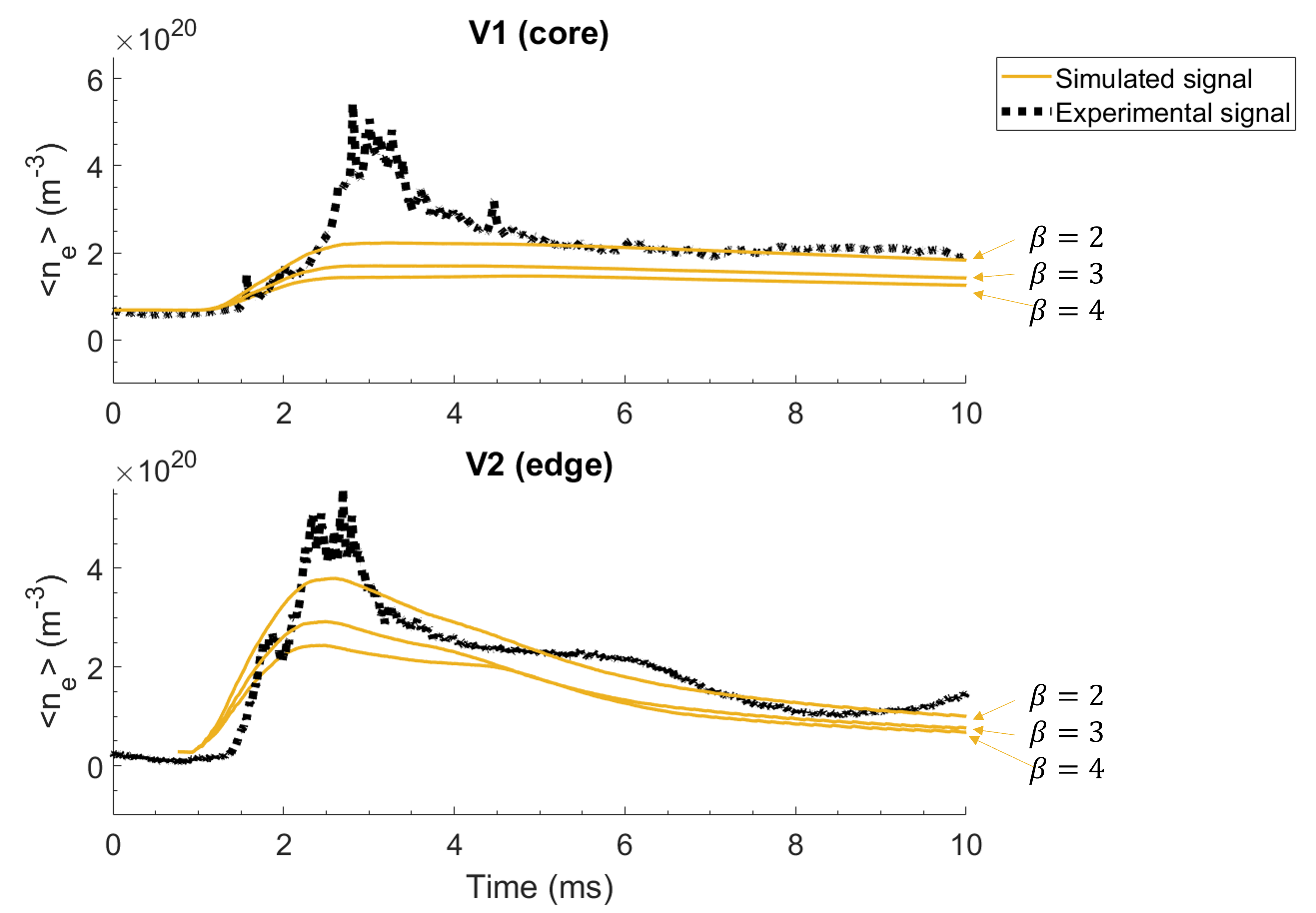}
     \caption{Synthetic interferometer signals (green solid) for different $\beta$ values compared with experimental signals (black dotted) from the core channel (top) and edge channel (bottom) of \#40743.}
     \label{fig:TS_interf_SimExpCompar}
\end{figure}

I carried out a set of simulations for this thesis with values of $\beta = 2,3,4$  (refer to \autoref{eq:backavg}) to match to the Thomson scattering measurements of \#40743. Plasma and diagnostic parmeters of \#40743 are shown in \autoref{fig:40743_discharge_time_trace}. Relevant input parameters for the simulation are shown in \autoref{table:AUGparams_SOL_scan}. The fragments first enter the plasma in the simulation at $\sim$ 0.6 ms ($t_\text{sim} = 0$ is the shattering time). The corresponding comparison of synthetic and experimental Thomson scattering profiles is plotted in \autoref{fig:TS_SimExpCompar}. For reference, the pre-injection plasma profile is also plotted and compared with the experimental pre-injection profiles shown by dark blue points and dark blue line profiles. The plasma profiles after most of the material assimilation has finished are shown with light blue colors which happens around $\sim 3$ ms, 2.4 ms after the fragments enter the plasma.  It can be noted that a lower $\beta = 2$ value leads to higher assimilation in the plasma compared to $\beta = 4$. However, the initial assimilation is limited to the edge in all 3 cases. Following the maximum material assimilation towards the edge, the density profile evolution is mainly linked to the particle diffusion. The density profiles at $t_{sim} = 8$ ms are also shown with red lines in the same figure and the corresponding experimental Thomson scattering profile is plotted with red scatter points. At $t_\text{sim} = 8$ ms, the figure shows that the core assimilation is best matched for $\beta = 4$ with the edge density being slightly underestimated. However, a precise match of the simulated density will also be affected by the value of the particle diffusion coefficient and the recycling from outside the last closed flux surface. This would require additional inputs from experiments or full MHD simulations. \\\\ 
Comparison of the assimilation measurements for the core and edge interferometer channels in the experiments and the synthetic diagnostic are shown in \autoref{fig:TS_interf_SimExpCompar}. First, in the experimental interferometer signals, a peak density rise can be observed for the core and edge interferometer channels. This density rise is presumed to be associated with the characteristic of the line of sight measurements during SPI. As the fragments ablate, they form plasmoids and the interferometer line of sight passing through a dense plasmoid can lead to overestimation of the line integrated measurement. Eventually, as the plasmoid completely assimilates in the background plasma, the interferometer signals decay to an intermediate value and the final assimilated material can be obtained. This behaviour will be confirmed with 3D JOREK simulations \cite{w_tang_non-linear_2023}. The INDEX synthetic diagnostic signals are closer to the final density assimilation measurements rather than the peak density measurements. The implications of this observation for comparison of experimental and simulated assimilation measurements are discussed in \autoref{ssec:sim_exmp_comp_assim}. 
\\\\
For the optimal determination of the $\beta$ parameter, the synthetic Thomson scattering diagnostic was preferred to the matching of the interferometer signal due to its localized measurement characteristic. Hence, the optimal value of $\beta$ was fixed to be $\beta \approx 4$. The same $\beta$ value was also used to match Thomson scattering profiles after SPI in DIII-D experiments \cite{lvovskiy_evolution_2022}. A different value of $\beta$ might be required for different fragment sizes and fragment speeds, however, due to the lack of Thomson scattering measurements to corroborate, a fixed value is utilized. 

\begin{table}[h]
\caption{Input parameters for the $\beta$ scan.}
\rule[0.2ex]{10 cm}{1.5pt}
\centering
\begin{tabular}{p{5.2cm} || p{3cm}}
Mean number of fragments & 400\\
Mean fragment size & 0.6781 mm\\
Mean fragment velocity & 270 m/s\\
Pellet composition(\% Ne/D) & 0/100 \\
Velocity dispersion ($\Delta v/ v$) & 40\% \\
Local deposition fraction & 0 \\
Pellet length & 4.67 mm\\
Pellet diameter & 8 mm\\
\end{tabular}
\label{table:AUGparams_SOL_scan}
\end{table}

\subsection{Example: pure deuterium injection}
\label{ssec:example_pure_deuterium}
\begin{figure}[H]
    \centering
    \includegraphics[width = 0.6\linewidth]{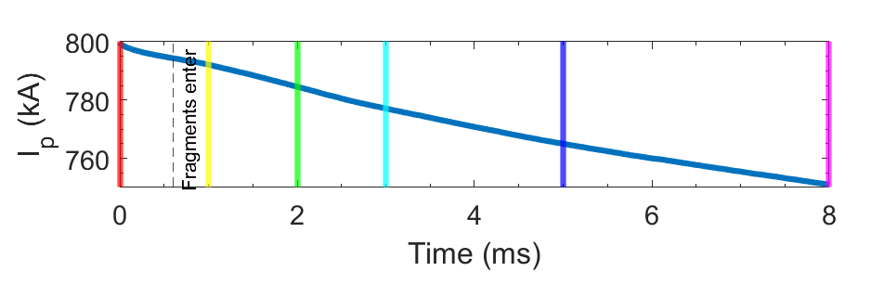}
    \caption{Time trace of plasma current evolution for pure deuterium injection.}
    \label{fig:pure_D_examp_Ip}
\end{figure}

\begin{figure}[H]
    \centering
    \includegraphics[width = 0.6\linewidth]{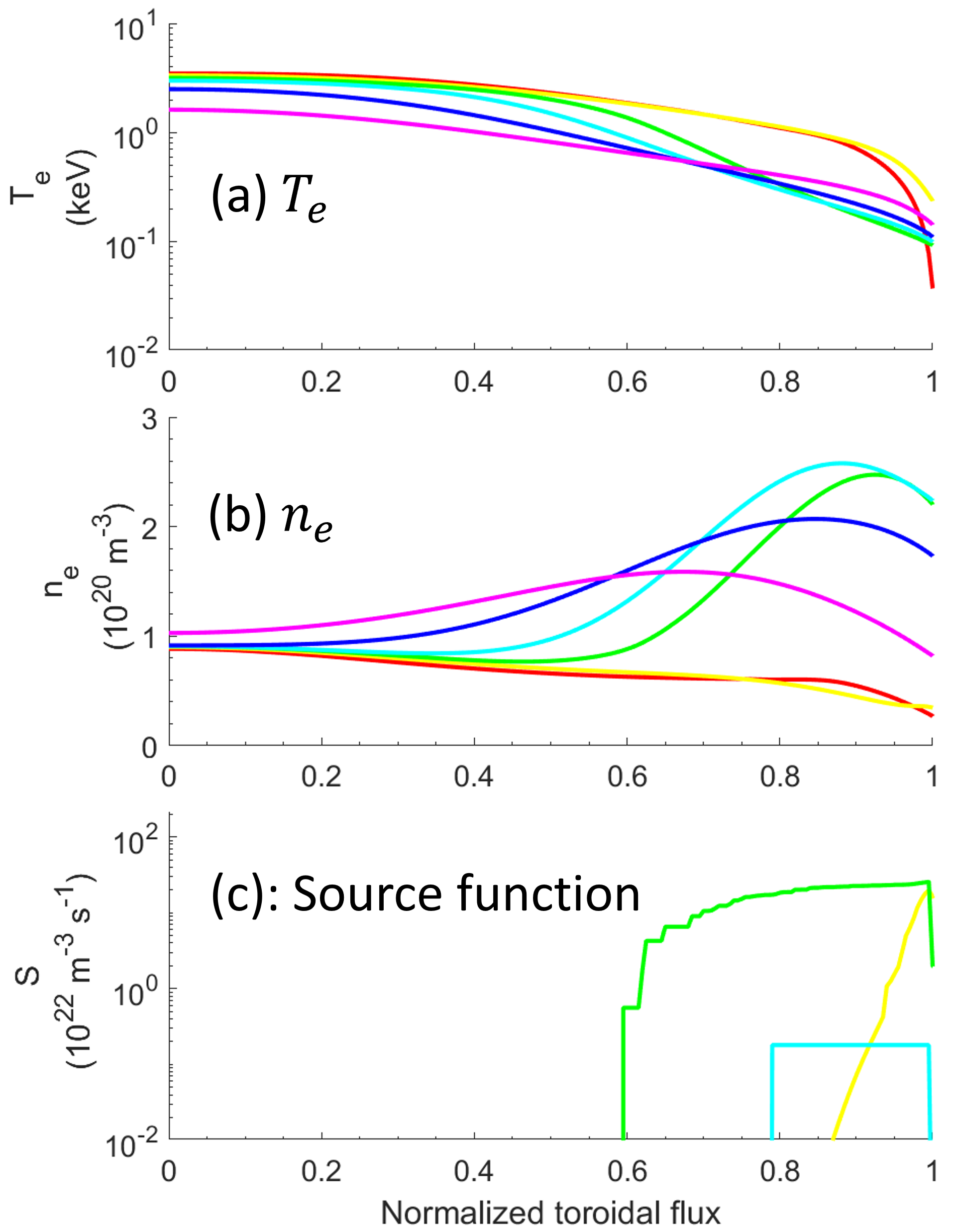}
    \caption{Profiles for various SPI relevant parameters against normalised poloidal flux for a pure deuterium injection.}
    \label{fig:pure_D_examp_remaining_subplots}
\end{figure}

With the back-averaging parameter, pure deuterium injection simulations can be carried out. Since the plasma response to pure deuterium injections is significantly different to mixed deuterium-neon injections, an example simulation is described in this section. The same input SPI parameters were used as in \autoref{ssec:backavgTS} with $\beta = 4$.  Particle recycling from outside the last closed flux surface (LCFS) is turned off for all pure deuterium simulations. The justification for this choice is that the back-averaging model artificially deposits the ablated particles outside the LCFS and in the simulations, they can diffuse inside after deposition leading to a constant fuelling of particles and increase in plasma density. While experimentally, no significant rise in the plasma density is observed after the ablation of the fragments has finished within 3-5 ms \cite{ansh_patel_internship_2023}. The time trace of the simulated plasma current is shown in \autoref{fig:pure_D_examp_Ip}. Profiles at the time stamps marked in \autoref{fig:pure_D_examp_Ip} with different colour vertical lines are shown in \autoref{fig:pure_D_examp_remaining_subplots}. Again, the physical events are described below in chronological order.   

\begin{enumerate}
    \item The fragments enter the plasma at $\sim$ 0.6 ms as marked in \autoref{fig:pure_D_examp_Ip}. As the fragments enter from the plasma edge, a rise in the neutral source function can be observed in \autoref{fig:pure_D_examp_remaining_subplots} (c) in the yellow curve. 
    \item At a later time step at 2 ms, the neutral source function penetrates to a deeper location as marked by the green curve in \autoref{fig:pure_D_examp_remaining_subplots} (c). Since the back-averaging model displaces the ablated material radially outwards, the penetration of the source function is limited. Green curves in \autoref{fig:pure_D_examp_remaining_subplots} (a), (b) show a rise in plasma density and a drop in plasma temperature at the edge of the plasma as a result of dilution cooling. 
    \item Further along in the simulation at 3 ms, the neutral particle source function in turquoise is strongly reduced indicating that the majority of the ablation has finished. The ablation is limited to the edge plasma resulting from the last arriving fragments. Subsequently, all the injected material is deposited however only a limited fraction is deposited inside the plasma (determined by $\beta$). Afterwards, the changes in the plasma density and temperature profiles are mainly caused by diffusion. 
    \item Finally, at 8 ms (corresponding to magenta colour), no source function can be observed in \autoref{fig:pure_D_examp_remaining_subplots}(c). As the density profile evolves, the electron density diffuses inwards and the electron temperature drops towards the core. Due to the nature of dilution cooling, an inversion in the profiles can be observed i.e. as the material diffuses inwards leading to a density rise and temperature drop in the core, the edge density drops and edge temperature increases slightly.  
\end{enumerate}

Based on the observations above, the main difference in the plasma response for pure deuterium injections, compared to mixed deuterium/neon injections, is the absence of a radiative cold front as no impurities are injected. With negligible radiation, the current profile also remains intact. Furthermore, due to the back-averaging model, the material assimilation is limited to the edge and core density increases by diffusion. An excluded element in the simulations that might affect the onset of the TQ in pure deuterium simulations is the inclusion of intrinsic impurities. 

\newpage
\section{Simulation results}
Having established the transport and SPI models of INDEX and various AUG inputs, the main goal of this thesis, the effect of SPI parameters on penetration and assimilation can be assessed. The results shown in this thesis mainly focus on the effect of SPI on the pre-TQ phase. The following subsections address the effect of fragment sizes, speeds and pellet composition on penetration and assimilation in the plasma. The effects of different fragment sizes and speeds are categorised by the pellet composition and studied separately for mixed neon/deuterium and pure deuterium pellets due to the difference in plasma response.

\subsection{Effect of fragment sizes}
\label{sec:frag_size_results}

\subsubsection{Mixed Ne/D pellets}
\label{ssec:frag_size_results_mixed_large_neon}
\begin{figure}[H]
    \centering        
    \includegraphics[width=\linewidth]{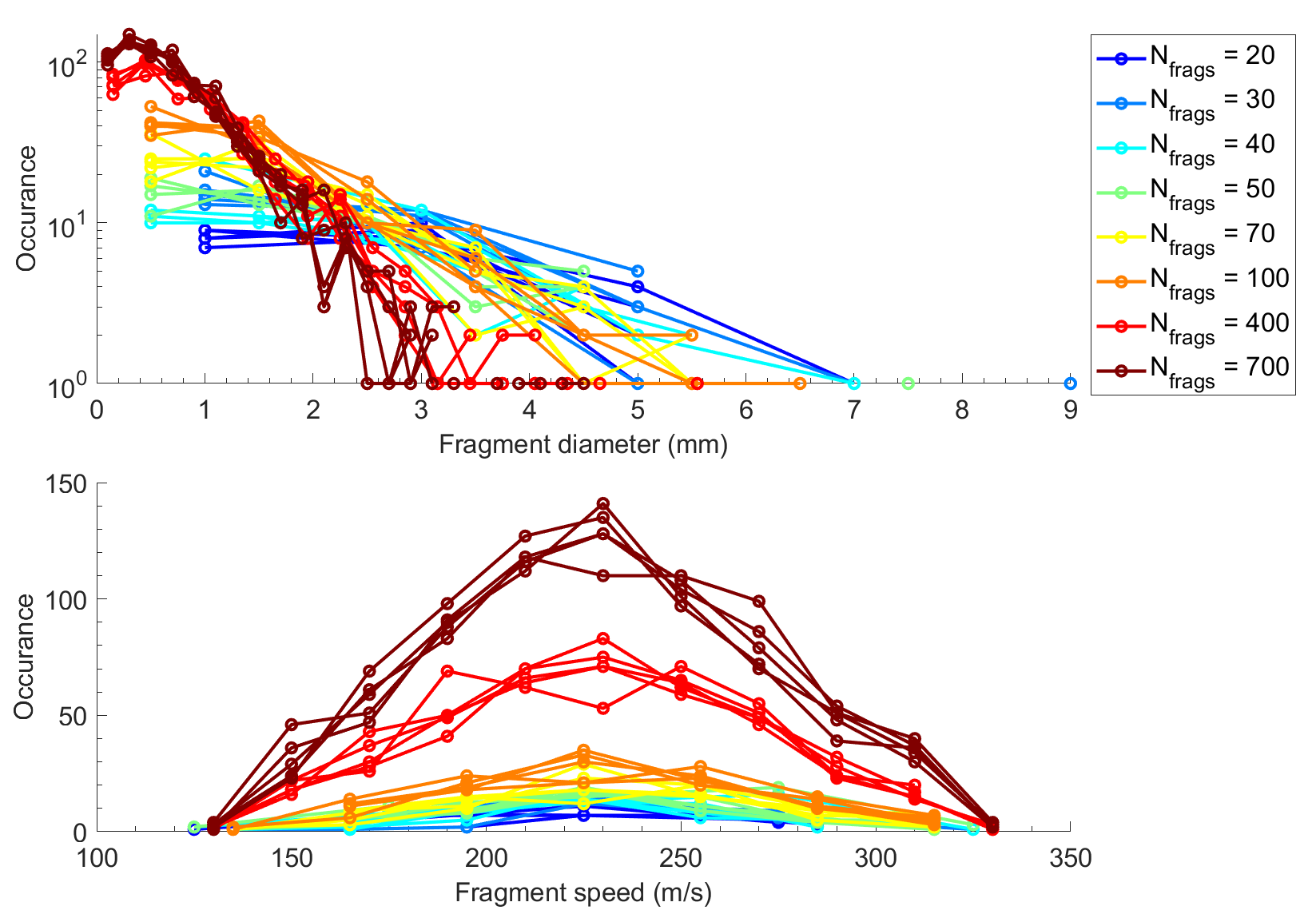}
    \caption{Different fragment size distributions (top) and fragment speed distributions (bottom) for varying number of sampled fragments. Different distributions with the same color indicate different realizations of the probability distribution for a fixed number of sampled fragments.}
    \label{fig:size_var_frag_size_speed_distributions}
\end{figure}

\begin{figure}[H]
    \centering
    \includegraphics[width=0.5\linewidth]{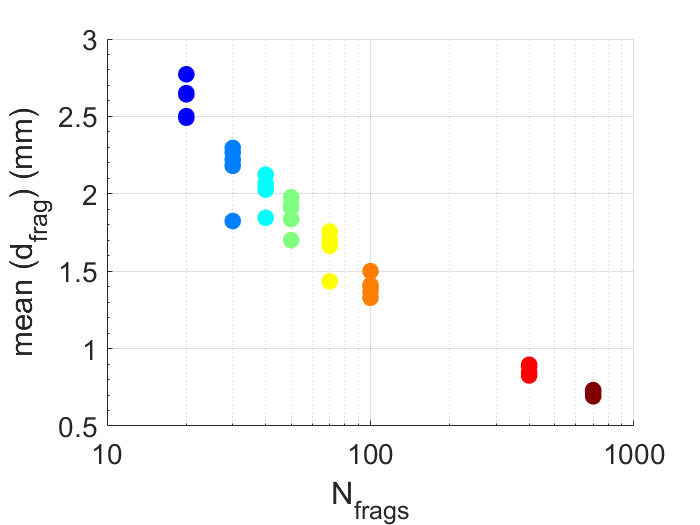}
    \caption{Relation between number of fragments and the mean fragment diameter of the sampled fragments.}
    \label{fig:size_var_n_frags_mean_diameter_relation}
\end{figure}

For mixed Ne/D injection, the effect of fragment sizes is first reported in this section. A set of simulations with injections of 10\% Ne and 90\% D was carried out. The fragment sizes were varied indirectly by sampling a different number of fragments. For constant pellet volume, sampling more fragments leads to a smaller average fragment size. Binned distributions of fragment sizes and speeds are shown in \autoref{fig:size_var_frag_size_speed_distributions}. 5 different realizations of fragment size and speed distributions were generated for all the cases. The mean fragment velocity was set to be 230 m/s for all the cases. The relation of the size distributions for varying number of fragments and the mean fragment diameter is shown in \autoref{fig:size_var_n_frags_mean_diameter_relation}. The relevant parameters for this set of simulations are summarised in \autoref{table:frag_size_statics}. 

\begin{table}[H]
\caption{Parameters for mixed deuterium/neon injections fragment size scan.}
\rule[0.2ex]{10 cm}{1.5pt}
\centering
\begin{tabular}{p{5.2cm} || p{3cm}}
Mean fragment velocity & 230 m/s\\
Pellet composition(\% Ne/D) & 10\%/90\% \\
Velocity dispersion ($\Delta v/ v$) & 40\% \\
Local deposition fraction & 1 \\
Pellet length & 10 mm\\
Pellet diameter & 8 mm\\
\end{tabular}
\label{table:frag_size_statics}
\end{table}

\begin{figure}[H]
    \centering
    \includegraphics[width=\linewidth]{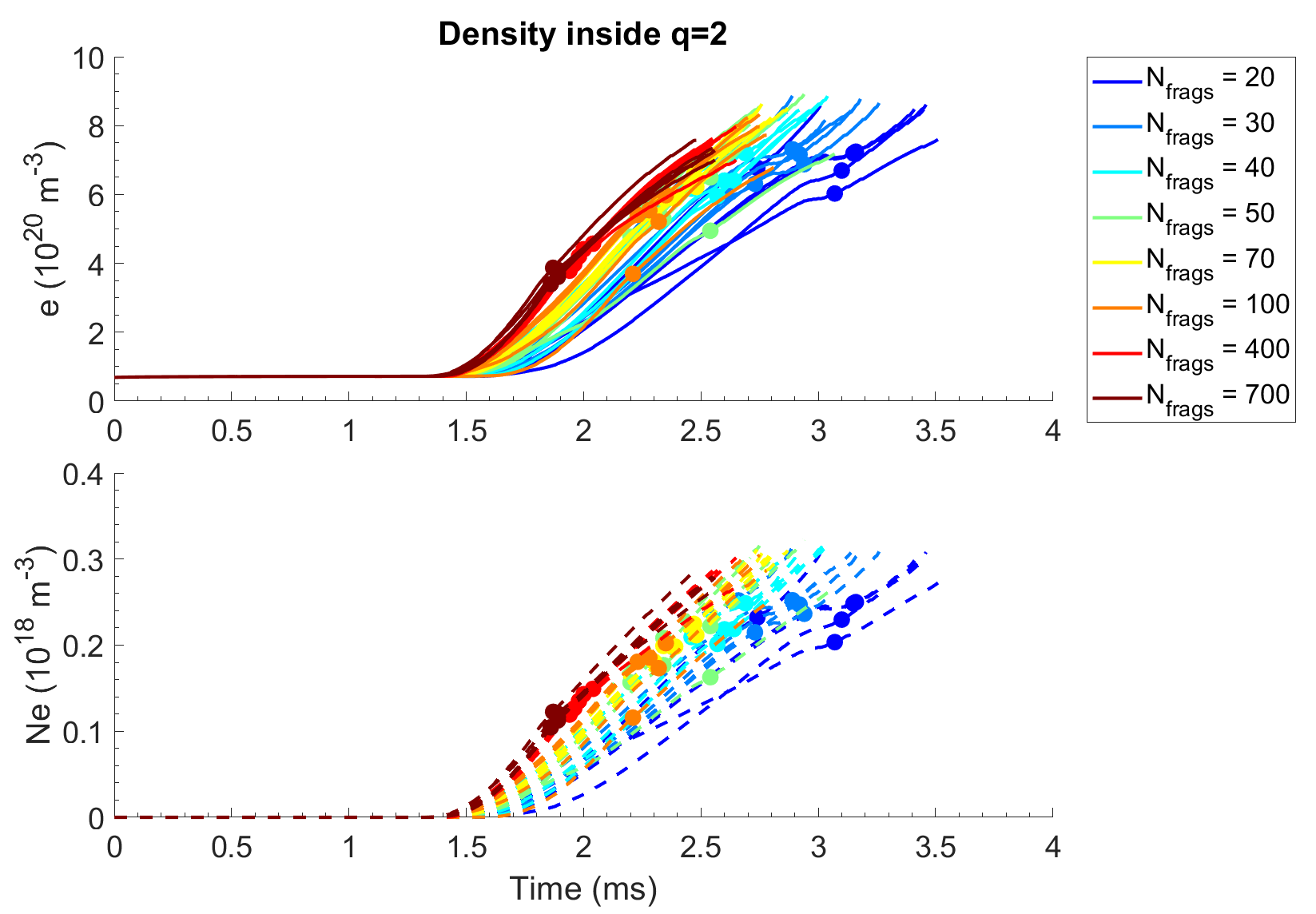}
    \caption{Average electron density (a) and neon density (b) inside the q=2 surface. Faster fragments start to assimilate quicker than larger fragments however maximum assimilation before the TQ onset (circular markers) is lower for smaller fragments.}
    \label{fig:frag_size_var_density_inside_q2}
\end{figure}

\begin{figure}[H]
   \centering
     \begin{subfigure}[b]{0.7\textwidth}
         \centering
         \includegraphics[width=\textwidth]{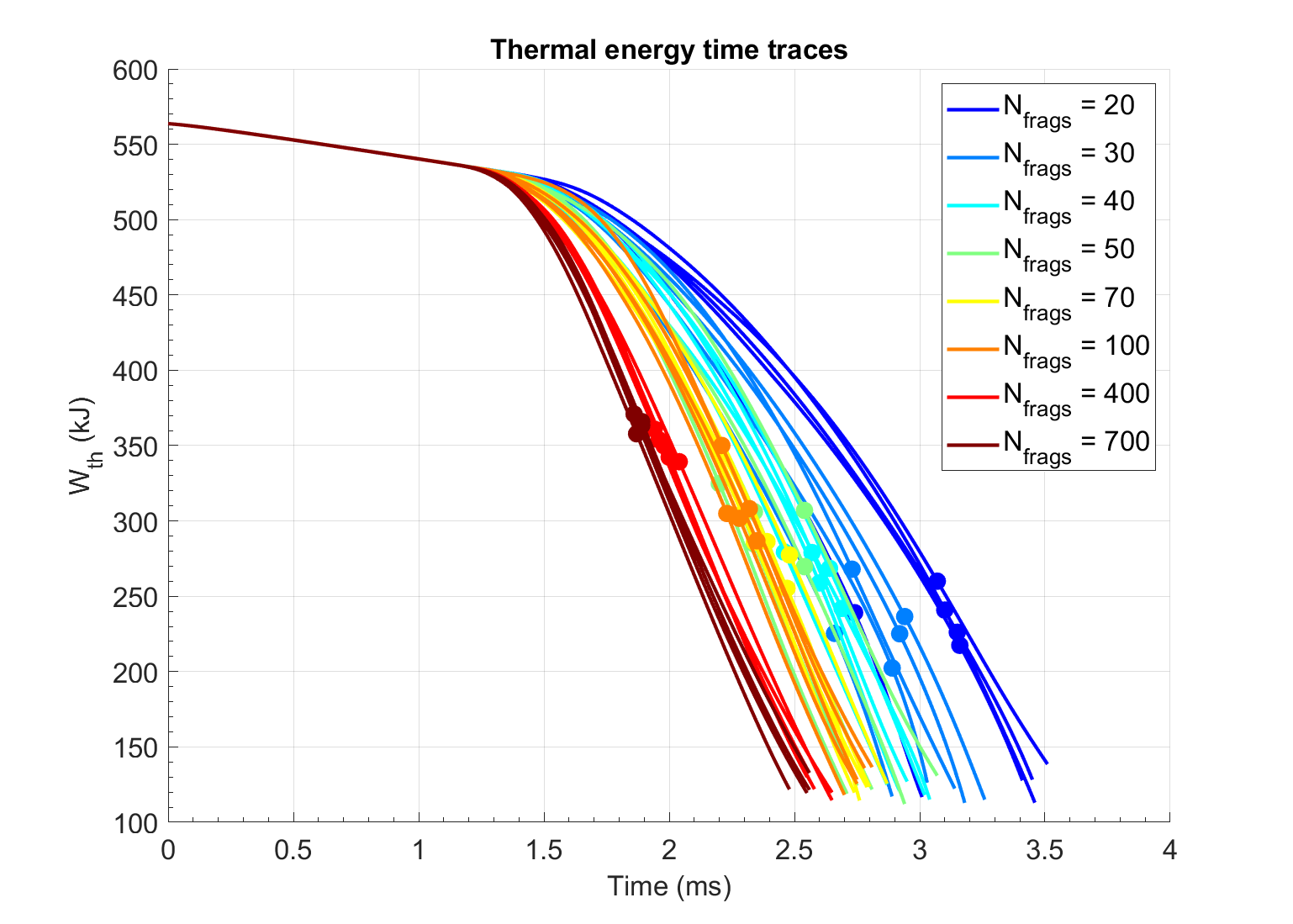}
         \caption{}
         \label{fig:frag_size_var_thermal_time_trace}         
     \end{subfigure}
     \begin{subfigure}[b]{0.7\textwidth}
         \centering
         \includegraphics[width=\textwidth]{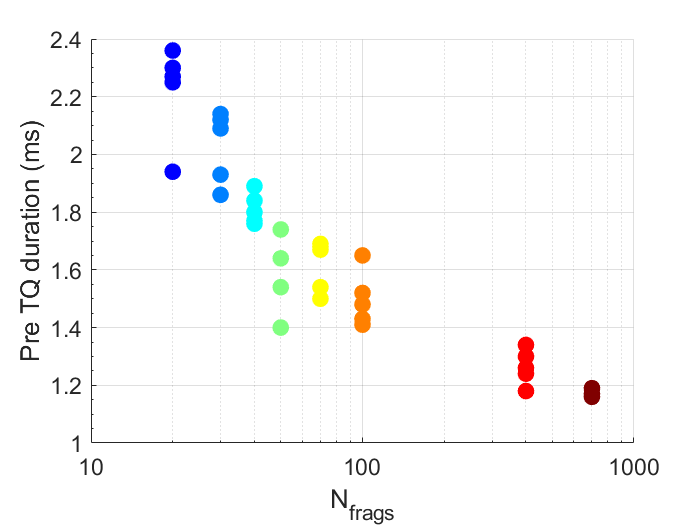}
         \caption{}
         \label{fig:frag_size_var_preTQ_duration}
     \end{subfigure}
\caption{Time traces of plasma thermal energy (top) and pre-TQ duration for different fragment sizes (bottom). Smaller fragments lead to a shorter pre-TQ duration.}
\label{fig:frag_size_var_thermal_energy_and_pre_TQ_duration}
\end{figure}

As mentioned before, the role of the DMS before disruption is to radiate away the plasma energy and increase the plasma density to avoid runaway electron formation. To study the assimilation of material for different fragment sizes in the core of the plasma, the time traces of volume averaged electron and neon density (sum of all charge states) are shown in \autoref{fig:frag_size_var_density_inside_q2}. It can be observed that smaller fragments start to assimilate faster in the plasma as can be observed with the earlier increase in free electron and neon density inside the $q=2$ surface. This behaviour can be attributed to smaller fragments having a relatively higher surface area that leads to faster ablation (also refer to \autoref{eq:ablation_recession}). As a result, the thermal energy of the plasma also starts to drop faster for smaller fragments as shown in \autoref{fig:frag_size_var_thermal_time_trace}. It should be noted that the electron and neon density time traces in \autoref{fig:frag_size_var_density_inside_q2}, for the case of large fragments (N=20,30,40,50,70), can cross over each other at a given point in simulation time. Due to larger statistical variations associated with sampling a small number of fragments, such behaviour can be expected depending on the distribution of fragment sizes and their corresponding speeds. Regardless, a general trend of faster assimilation for smaller fragments still holds when the fragment sizes vary drastically. The circular markers in \autoref{fig:frag_size_var_density_inside_q2} indicate the time when the TQ onset condition (defined in \autoref{sec:methodology}) is satisfied. The pre-TQ duration(s) are separately plotted against the number of fragments in \autoref{fig:frag_size_var_preTQ_duration}. 

\begin{figure}[H]
    \centering
    \includegraphics[width = 0.8\linewidth]{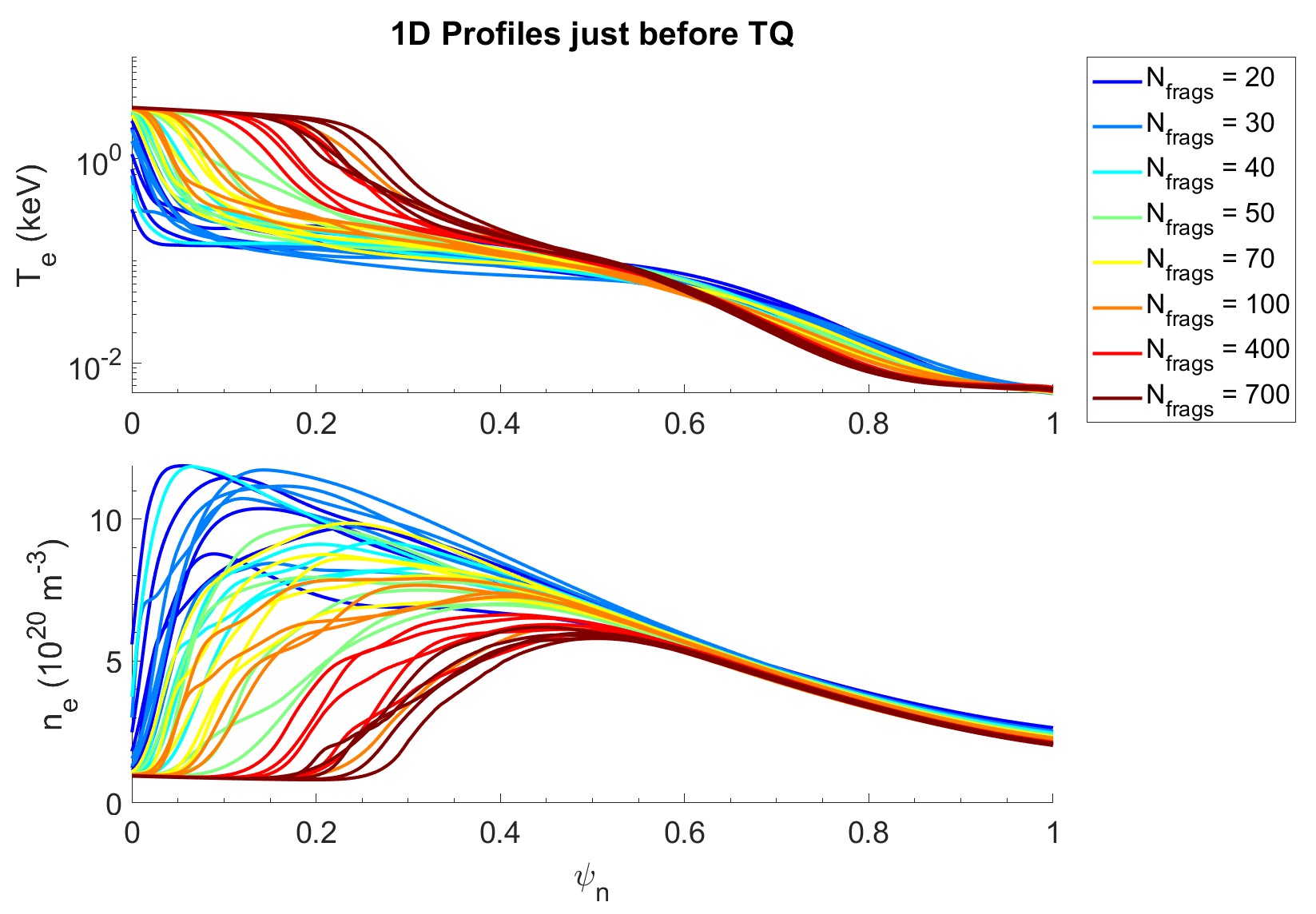}
    \caption{Electron temperature profiles (top) and density profiles (bottom) at TQ onset condition. Larger fragments can penetrate deeper and radiate away energy from deeper in the plasma leading to lower core temperatures and higher core density at TQ onset. q=2 surface is at $\psi_N \approx 0.87$.}    \label{fig:frag_size_var_Te_ne_profiles_at_TQ}
\end{figure}

In \autoref{fig:frag_size_var_preTQ_duration}, smaller fragments lead to an earlier TQ onset. This behaviour can be understood by the faster deposition of neon leading to quicker radiative cooling of the edge plasma. As smaller fragments cool down the plasma edge faster than larger fragments, the penetration of material before the TQ onset is decreased leading to limited core density rise for smaller fragments. The radial profiles of temperature and density at the onset of the TQ condition are plotted against the normalised poloidal flux in \autoref{fig:frag_size_var_Te_ne_profiles_at_TQ}. Larger fragments are better for deeper penetration and higher assimilation before the onset of the TQ. 

\subsubsection{Pure deuterium pellets}
\label{sssec:frag_size_results_pure_D}
\begin{figure}[H]
    \centering        
    \includegraphics[width=0.8\linewidth]{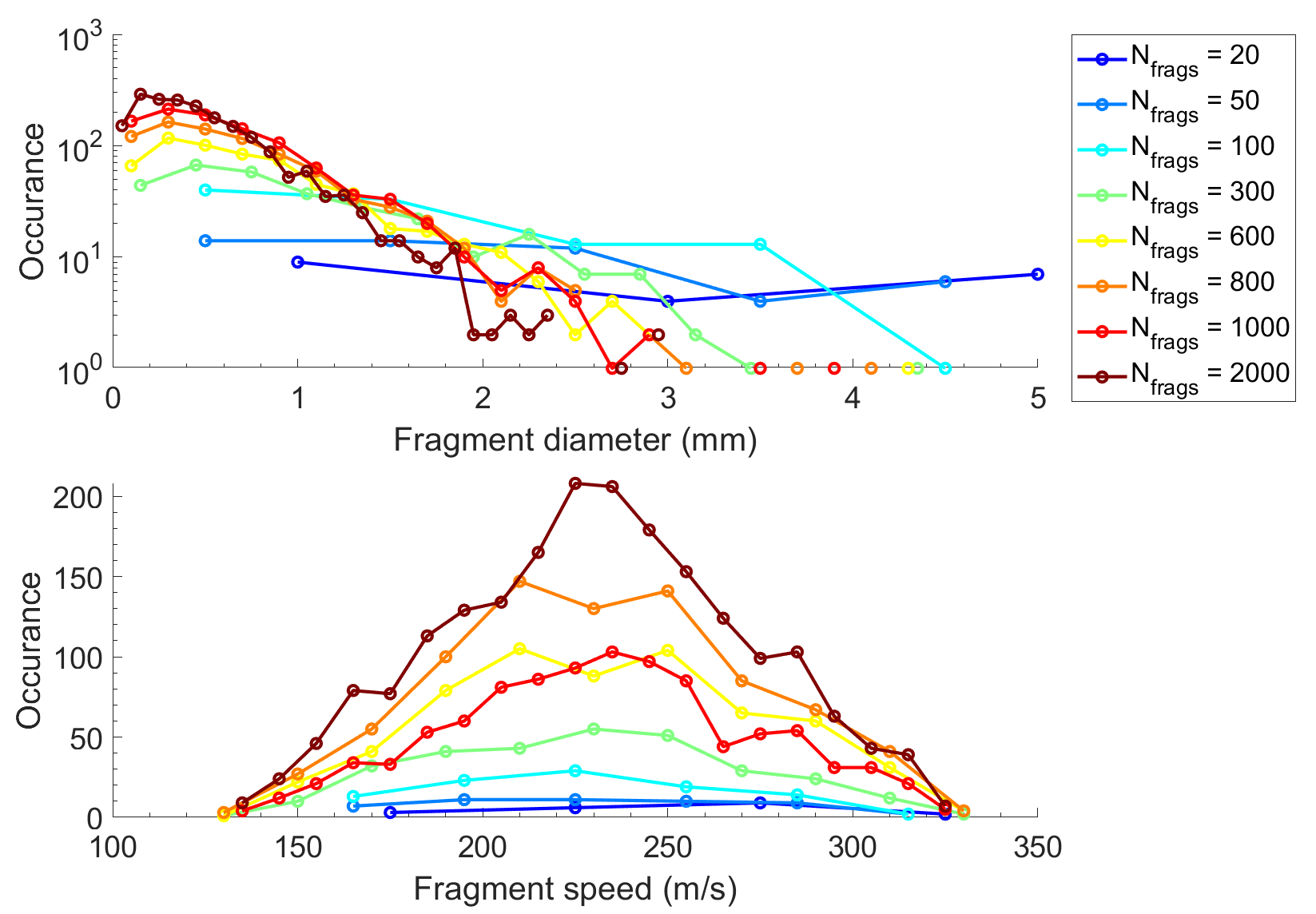}    
    \caption{Different fragment size distributions (top) and fragment speed distributions (bottom) for varying number of sampled fragments.}
    \label{fig:deuterium_injections_frag_size_speed_distributions}
\end{figure}

Pure deuterium injections would be utilised in the staggered injection scheme as discussed in \autoref{sec:DMS_ITER}. Since pure deuterium fragments experience a plasmoid drift, their simulation was carried out with a back-averaging model which is discussed in \autoref{ssec:backavgTS}. Similar to the mixed D/Ne injections discussed in the \autoref{ssec:frag_size_results_mixed_large_neon}, a scan of different fragment sizes was carried out by sampling a different number of fragments for a fixed pellet volume. The mean fragment speed was also set to 230 m/s. The main input parameters of the simulation are summarized in \autoref{table:AUGparams_deuterium_size_var} and the fragment size and speed distributions are shown in \autoref{fig:deuterium_injections_frag_size_speed_distributions}.  Since these simulations do not involve injecting neon, no radiative cooling front is developed at the plasma edge. As a result, the TQ onset condition discussed in \autoref{ssec:exampleSim} is not satisfied in any of the simulations. Experimentally, it was found that pure deuterium injections with large 8 mm diameter pellets mostly lead to a disruption. A possible reason for disruptions in pure deuterium injections could be due to density limits \cite{greenwald_new_1988}, however, this has not been concluded yet. Hence, for the comparisons of the material assimilation, it should be kept in mind that experimental discharges might disrupt before the maximum simulated material assimilation is achieved. Another element in the simulations that might affect the onset of the TQ in pure deuterium simulations is the presence of intrinsic impurities which is not considered in the present simulations. 

\begin{table}[H]
\caption{Static parameters for fragment size scan for pure deuterium injections.}
\rule[0.2ex]{14 cm}{1.5pt}
\centering
\begin{tabular}{p{5.2cm} || p{8cm}}
Number of fragments & [20, 50, 100, 300, 600, 800, 1000, 2000]\\
Mean fragment diameters & [2.77, 1.94, 1.50, 1.00, 0.77, 0.69, 0.64, 0.51] mm\\
Pellet composition(\% Ne/D) & 10\%/90\% \\
Mean fragment velocity & 230 m/s\\
Velocity dispersion ($\Delta v/ v$) & 40\% \\
Local deposition fraction & 0 \\
Back averaging parameter $\beta$ & 4 \\
Pellet length & 10 mm\\
Pellet diameter & 8 mm\\
\end{tabular}
\label{table:AUGparams_deuterium_size_var}
\end{table}

\begin{figure}[H]
    \centering
    \includegraphics[width=0.8\linewidth]{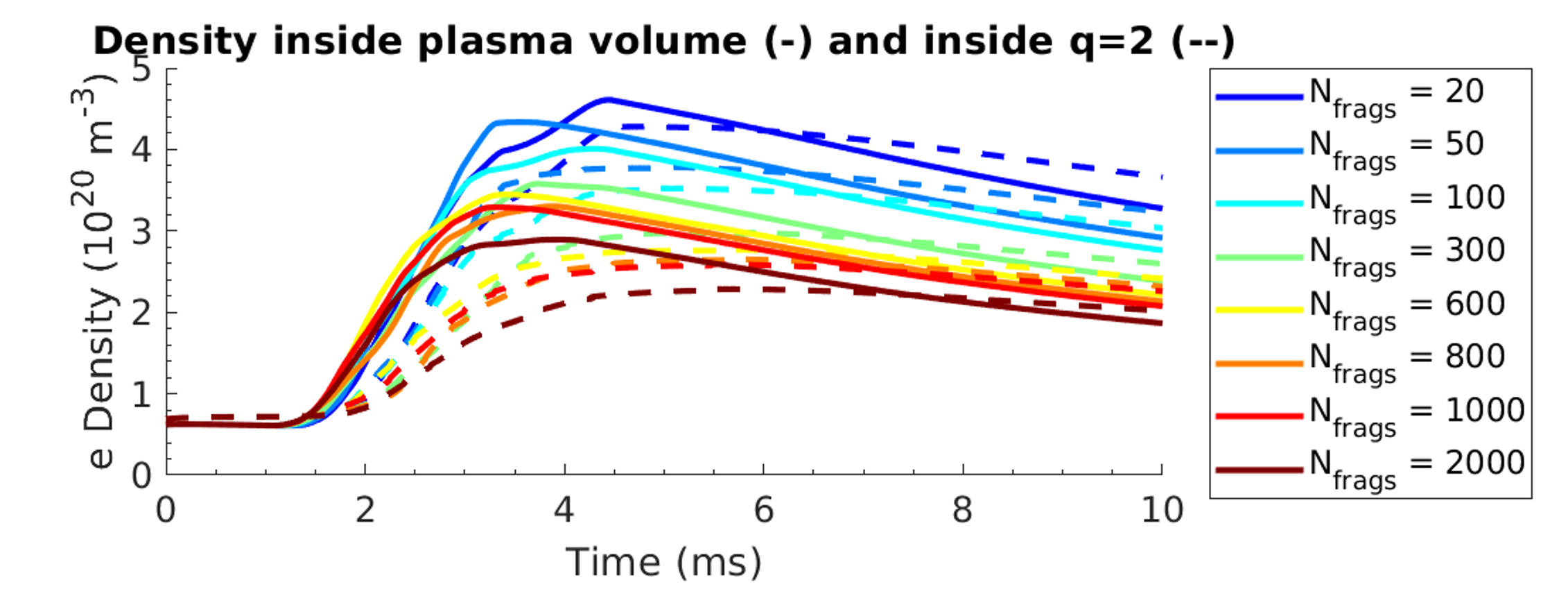}
    \caption{Time traces of electron density inside the plasma volume (solid lines) and inside q=2 surface (dashed lines). Larger fragments lead to a higher maximum density rise.}
    \label{fig:deuterium_size_var_density_inside_vol_and_q_2}
\end{figure}

The average electron density inside the q=2 surface and inside the plasma volume is shown in \autoref{fig:deuterium_size_var_density_inside_vol_and_q_2}. A trend of higher assimilation for larger fragments (corresponding to fewer number of fragments) can be observed. Additionally, as discussed in \autoref{ssec:example_pure_deuterium}, material deposition in these simulations is limited to the plasma edge and core density only rises later due to diffusion. This phenomenon is indicated in \autoref{fig:deuterium_size_var_density_inside_vol_and_q_2} as the volume averaged density is initially higher than the density inside q=2 surface for a given simulation. The maximum volume averaged density is reached when the the majority of ablation has taken place. After this stage, the electron density diffuses inwards and at one point the density inside q=2 surface overcomes the volume averaged density. The crossover happens earlier for larger fragments. 

\begin{figure}
    \centering
    \includegraphics[width = 0.8\linewidth]{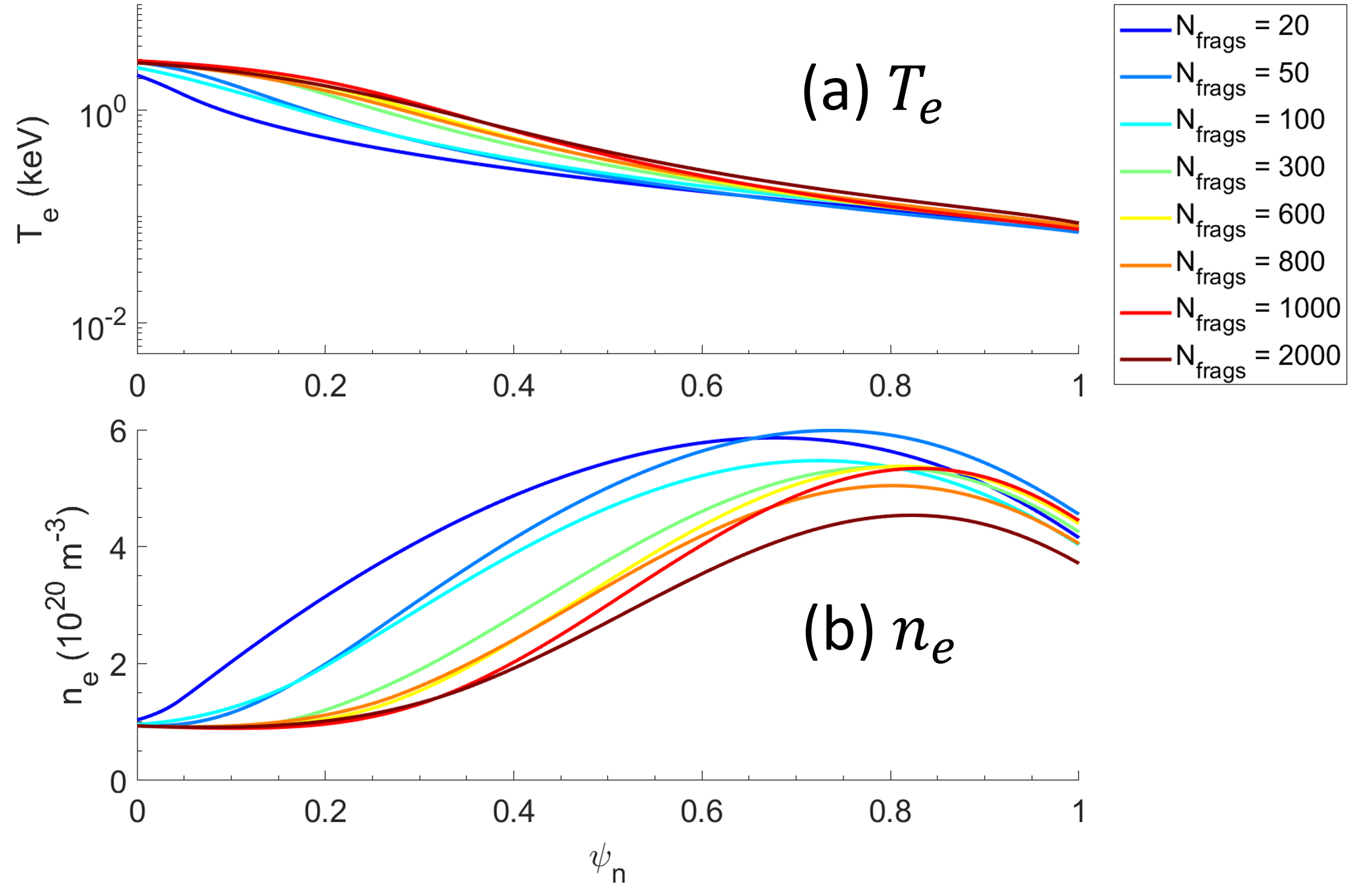}
    \caption{(a) Electron temperature profiles, (b) electron density profiles at the time of maximum volume averaged density for each case of different fragment sizes. Larger fragments can penetrate deeper and lead a higher core density rise.}
    \label{fig:deuterium_size_var_profiles_at_max_vol_density}
\end{figure}

While the smaller fragments reach the maximum assimilated density prior to larger fragments, the magnitude of maximum assimilated material is lower compared to larger fragments. This is because smaller fragments ablate away quicker than faster fragments and penetration is shallower compared to larger fragments. To show this, the electron temperature and density profiles at time of maximum volume averaged density for each simulation is shown in \autoref{fig:deuterium_size_var_profiles_at_max_vol_density}. It can be noticed that the smaller fragments lead to only edge assimilation while larger fragments can lead to comparatively deeper penetration when majority of the ablation has finished. Consequently, the plasma temperature is slightly lower towards the core for larger fragments as a result of dilution cooling of the plasma. After the maximum volume averaged density is reached, the plasma density profile is governed mainly by diffusion. 

\subsection{Effect of fragment speeds}
\label{section:frag_speed_var_mix_pellets}

\subsubsection{Mixed D/Ne pellets}
\begin{figure}[H]
   \centering
     \begin{subfigure}[t]{0.8\textwidth}
       \centering
        \includegraphics[trim={0 9cm 0 0},clip,width=\linewidth]{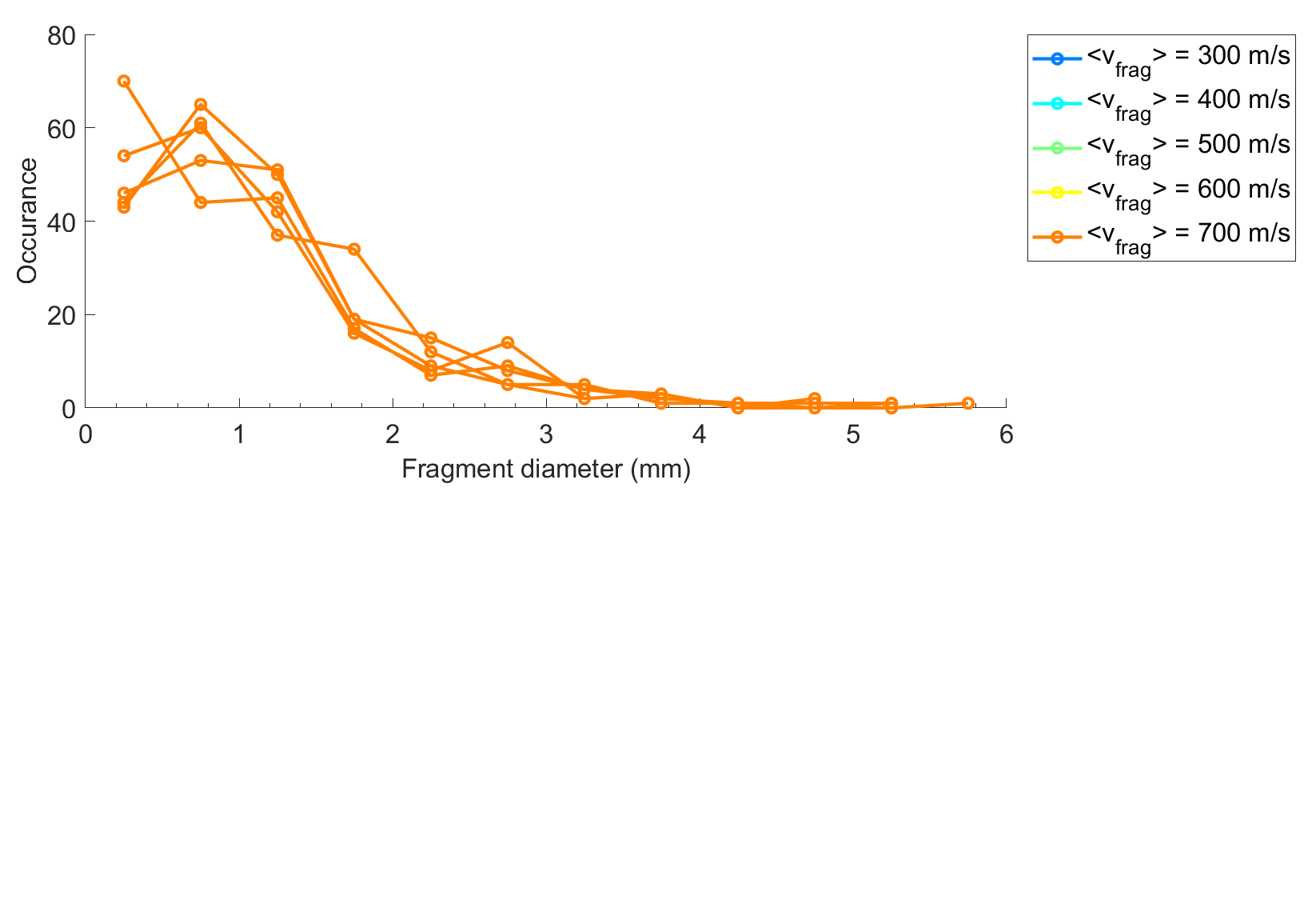}
        \caption{}
        \label{fig:frag_speed_var_size_distribution}             
     \end{subfigure}
     \begin{subfigure}[t]{0.8\textwidth}
        \centering
        \includegraphics[trim={0 0 0 9cm},clip,width=\linewidth]{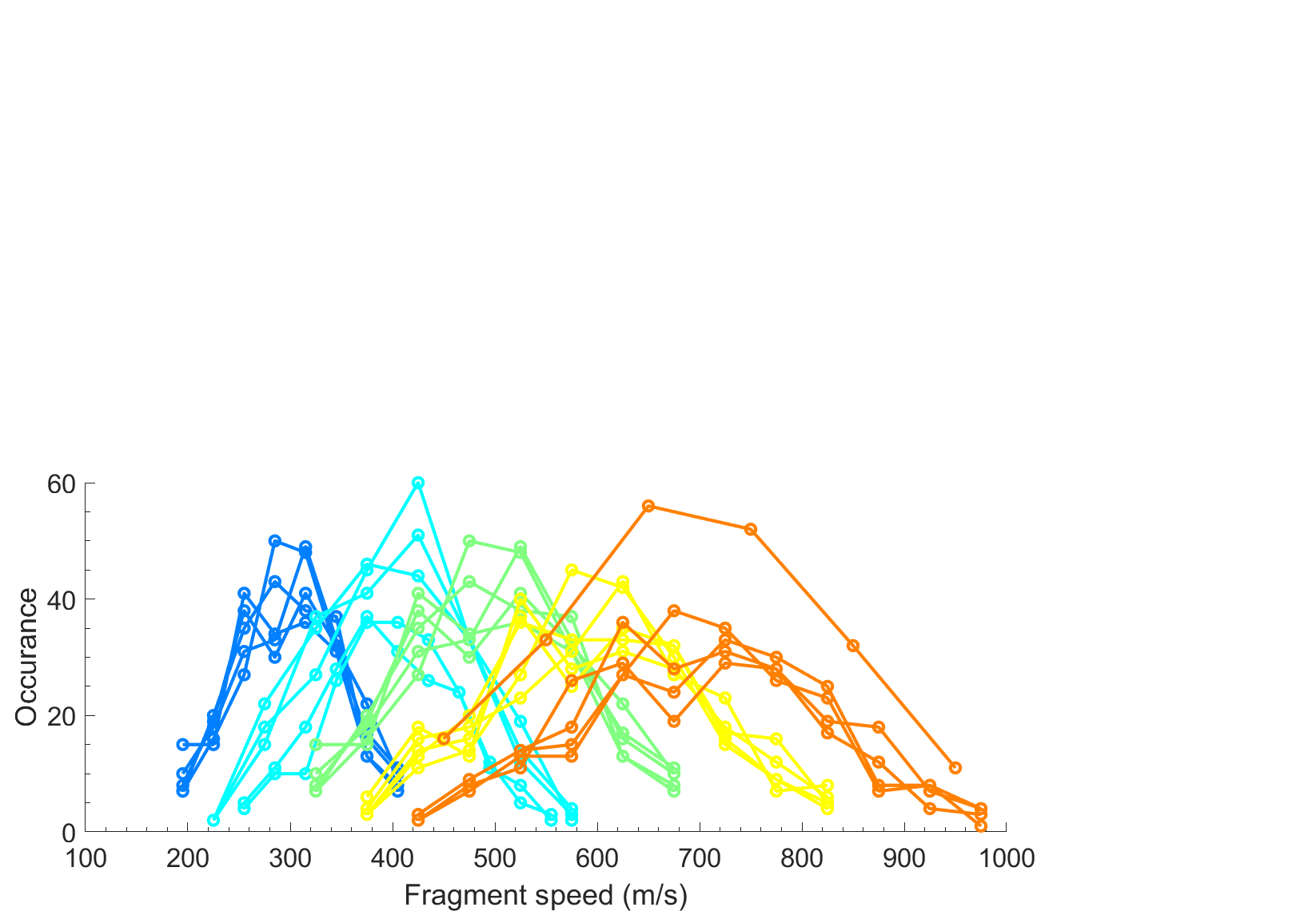}
        \caption{}
        \label{fig:frag_speed_var_speed_distribution}                 
     \end{subfigure}
     \caption{Different fragment size distributions (top) and fragment speed distributions (bottom) for varying number of sampled fragment speeds. Different distributions with the same color indicate different realizations of the probability distribution for a fixed number of sampled fragments.}
     \label{fig:frag_speed_var_size_speed_distributions}
\end{figure}

Apart from fragment sizes, I also studied the effect of fragment speeds by carrying out a set of simulations with varying mean fragment speeds but the same fragment sizes for 10\% neon and 90\% deuterium composition. 200 fragments were sampled for each simulation corresponding to a mean fragment diameter of 1.137 mm. The resulting fragment size and speed distributions are shown in \autoref{fig:frag_speed_var_size_speed_distributions}. A fixed velocity dispersion of $\Delta v/v = 40\%$ is used for all the fragment speed distributions in \autoref{fig:frag_speed_var_speed_distribution} so larger mean fragment speeds also have a larger standard deviation. This can be observed by comparing the speed distributions of $\langle v_\text{{frag}} \rangle = 300,700$ m/s in blue and orange. The main input parameters of the simulation are summarized in \autoref{table:frag_speed_var}. 

\begin{table}[H]
\caption{Parameters for fragment speed scan for mixed deuterium/neon injections.}
\rule[0.2ex]{12 cm}{1.5pt}
\centering
\begin{tabular}{p{5.2cm} || p{5cm}}
Number of fragments & 200\\
Mean fragment diameters & $\sim 1.11$ mm\\
Mean fragment velocity & [300, 400, 500, 600, 700] m/s\\
Pellet composition(\% Ne/D) & 10\%/90\% \\
Velocity dispersion ($\Delta v/ v$) & 40\% \\
Local deposition fraction & 1\\
Pellet length & 10 mm\\
Pellet diameter & 8 mm\\
\end{tabular}
\label{table:frag_speed_var}
\end{table}

\begin{figure}[H]
   \centering
     \begin{subfigure}[b]{0.7\textwidth}
         \centering
         \includegraphics[trim={0 9cm 0 0},clip,width=\linewidth]{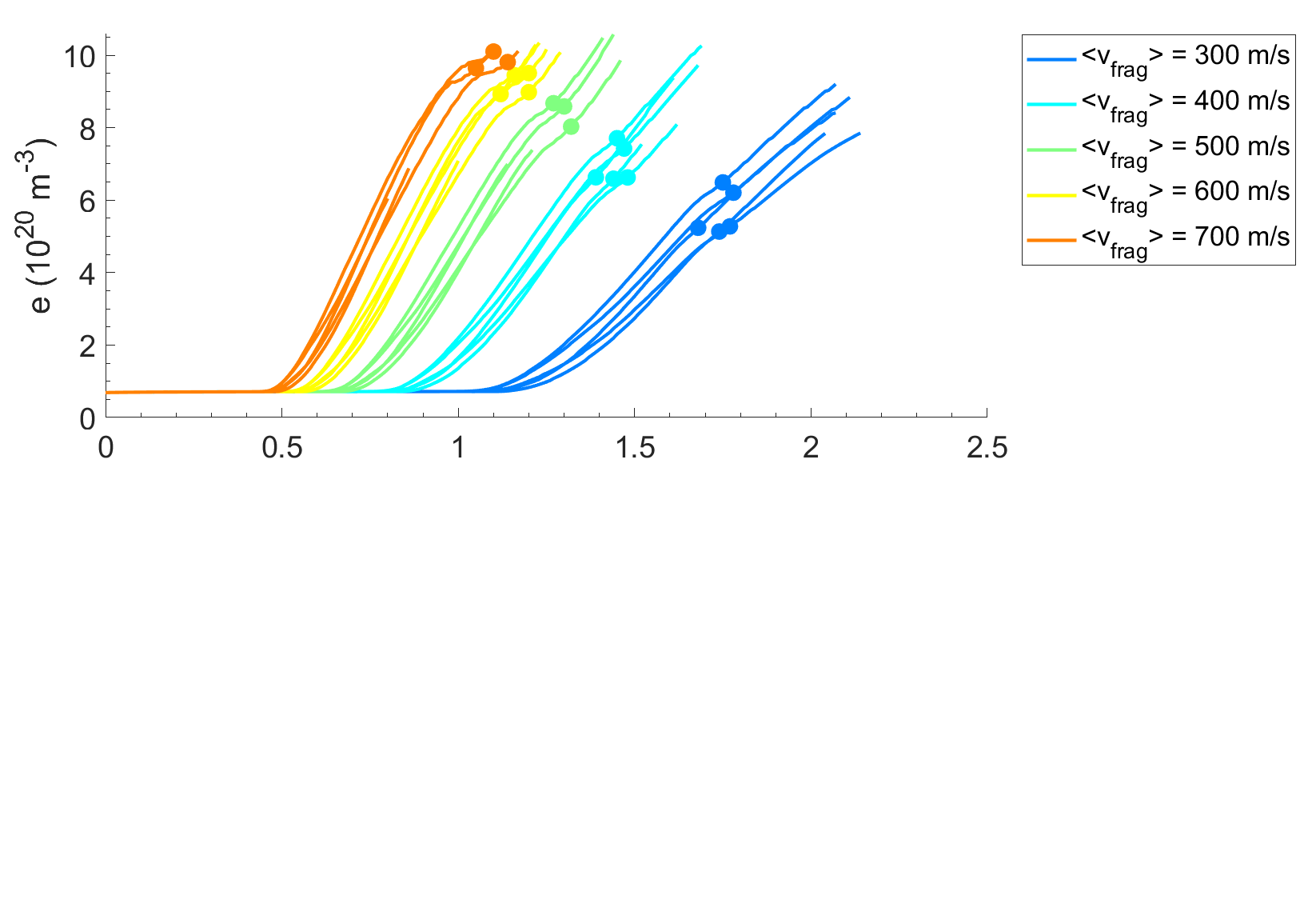}
         \phantomcaption
         \label{fig:frag_speed_var_e_density_inside_q_2}         
     \end{subfigure}
     \begin{subfigure}[b]{0.7\textwidth}
         \centering
         \includegraphics[trim={0 0 0 9cm},clip,width=\linewidth]{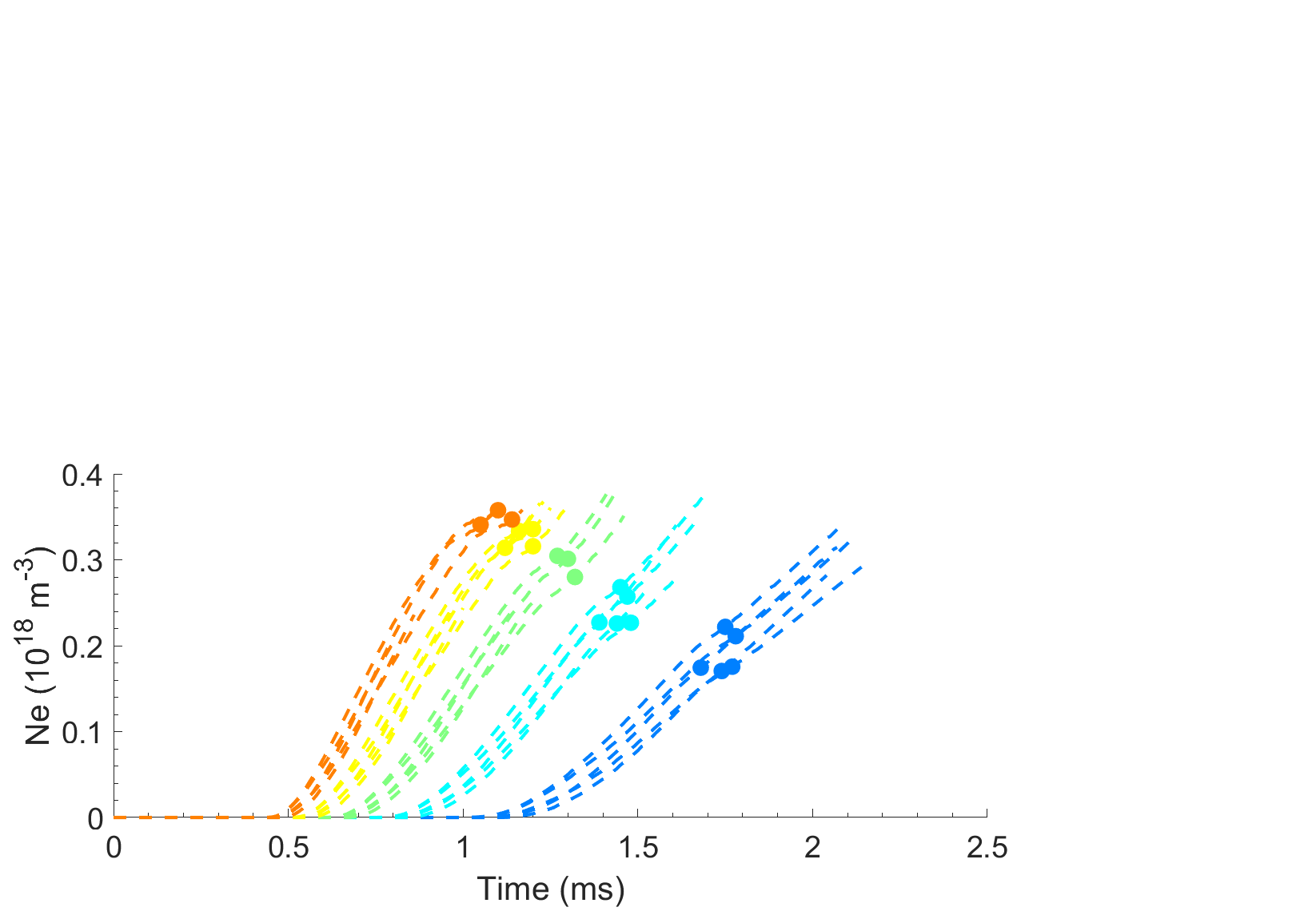}
         \phantomcaption
         \label{fig:frag_speed_var_ne_density_inside_q_2}         
     \end{subfigure}
\caption{Average electron density (top) and neon density (bottom) inside the q=2 surface. Faster fragments start to assimilate quicker and lead to a higher material assimilation at TQ onset (circular markers).}
\label{fig:frag_speed_var_density_inside_q_2}
\end{figure}

\begin{figure}[H]
    \centering
    \includegraphics[width = 0.5\linewidth]{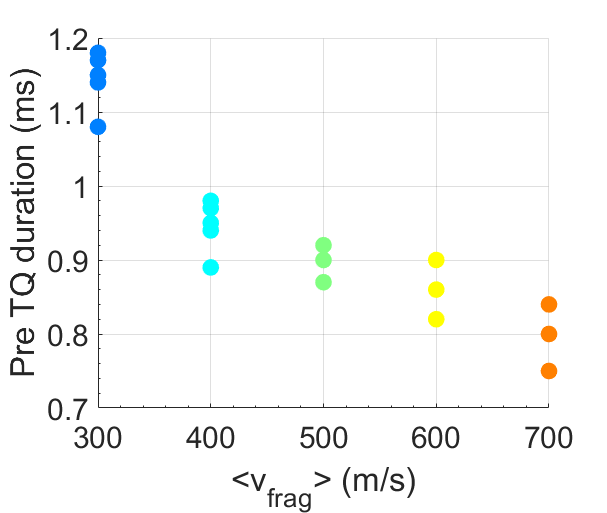}
    \caption{Pre-TQ duration as a function of mean fragment velocity. Faster fragments lead to a shorter pre-TQ duration.}
    \label{fig:frag_speed_var_pre_TQ_duration}
\end{figure}

To study the assimilation, the electron and neon density is plotted inside the q=2 surface in \autoref{fig:frag_speed_var_density_inside_q_2}. Observing the time traces for different mean fragment speeds, it can be noticed that faster fragments lead to higher core assimilation and the assimilation happens quicker as well. This trend can be understood by the movement of faster fragments penetrating the plasma quicker and being able to penetrate deeper before the deposited neon cools down the plasma hereby reducing ablation. Similar to \autoref{fig:frag_size_var_density_inside_q2}, the circular markers indicate the time points where the TQ condition is satisfied. The dependence of pre-TQ duration on the mean fragment velocity is shown in \autoref{fig:frag_speed_var_pre_TQ_duration}. A decrease in the pre-TQ duration can be observed for faster fragments. 

\begin{figure}[H]
   \centering
     \begin{subfigure}[b]{0.6\textwidth}
         \centering
         \includegraphics[width=\textwidth]{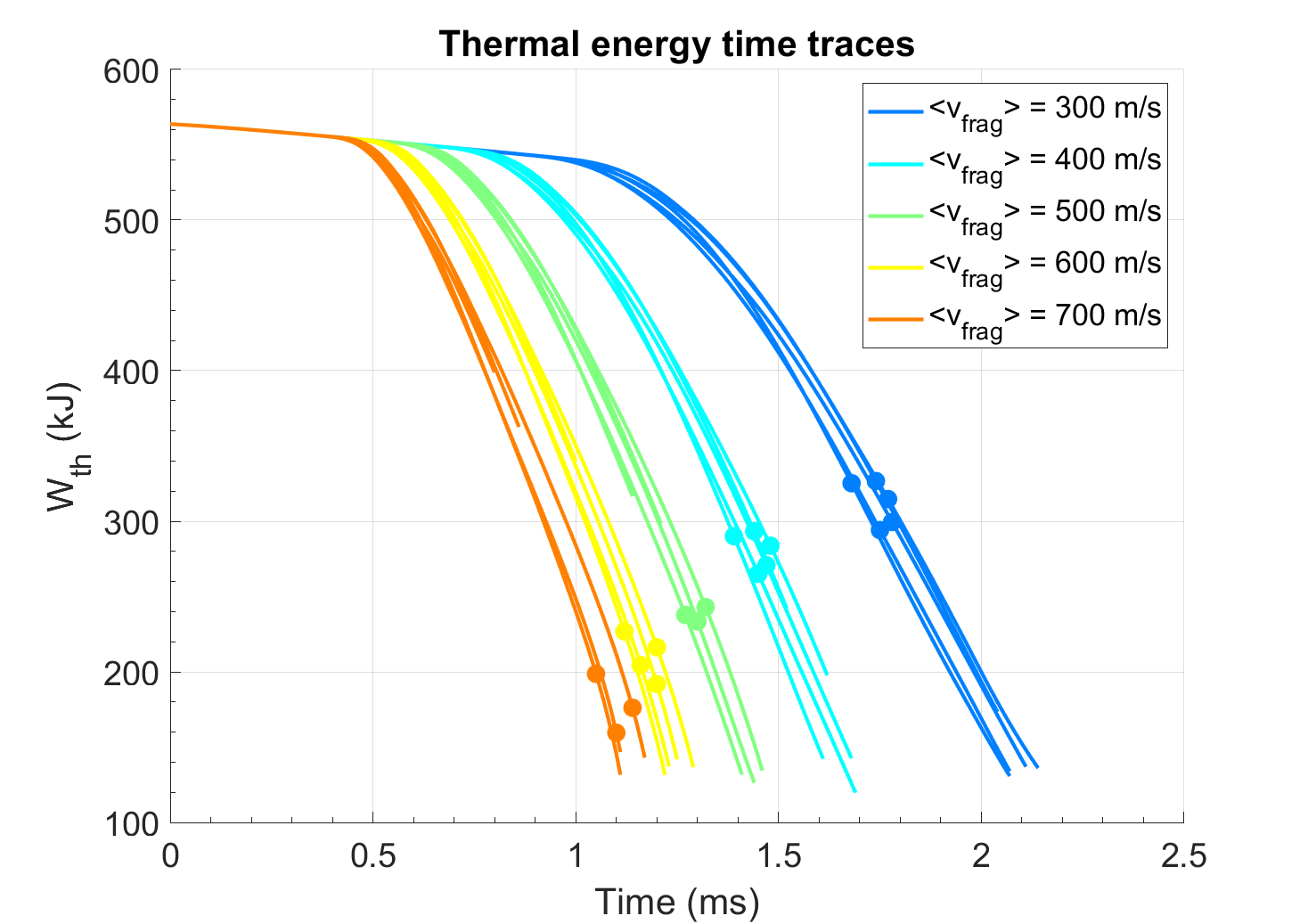}
         \caption{Plasma thermal energy time traces for different mean fragment velocities.}
         \label{fig:frag_speed_var_thermal_time_trace}         
     \end{subfigure}
     \begin{subfigure}[b]{0.6\textwidth}
         \centering
         \includegraphics[width=\textwidth]{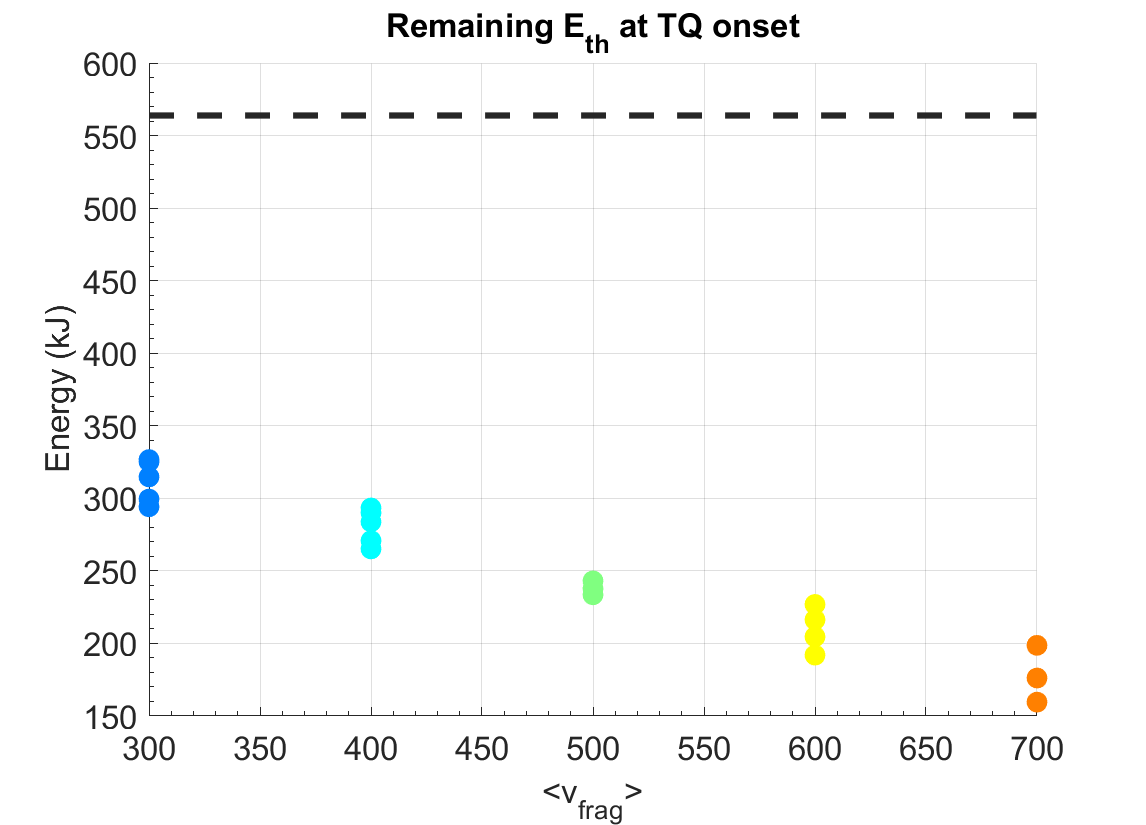}
         \caption{Remaining thermal energy at TQ onset for different mean fragment speeds. Black dashed line shows the initial thermal energy. Faster fragments radiate away more energy before TQ onset.}
         \label{fig:frag_speed_var_remaining_E_th_at_TQ}        
     \end{subfigure}
\caption{}
\label{fig:frag_speed_var_thermal_energy_time_trace_and_remaining_W_th_at_TQ}
\end{figure}

\begin{figure}[H]
    \centering
    \includegraphics[width = 0.8\linewidth]{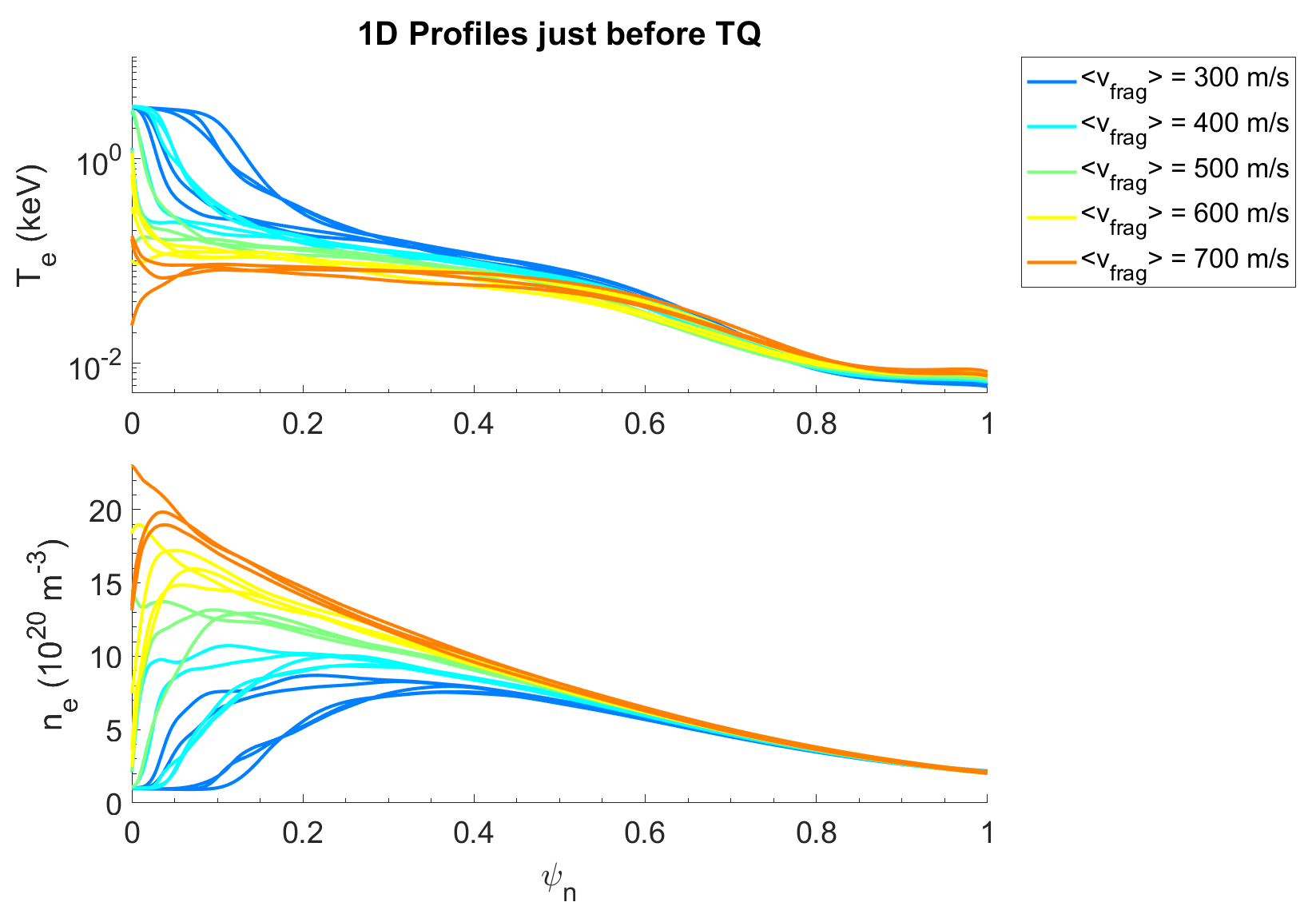}
    \caption{Electron temperature profiles (top) and density profiles (bottom) at TQ onset condition. Faster fragments can penetrate deeper and radiate away energy from deeper in the plasma leading to lower core temperatures and higher core density at TQ onset. q=2 surface is at $\psi_N \approx 0.87$.}
    \label{fig:frag_speed_var_Te_ne_profiles_at_TQ}
\end{figure}

The dependence of radiated thermal energy on the mean fragment velocity is noted in \autoref{fig:frag_speed_var_thermal_energy_time_trace_and_remaining_W_th_at_TQ}. Faster fragments lead to a quicker drop in the plasma thermal energy due to the quicker neon accumulation, which can also be noticed in \autoref{fig:frag_speed_var_ne_density_inside_q_2} and hence also starts to decrease faster due to radiative cooling. The dropping rate of thermal energy is slightly faster for faster mean fragments. Similar to previous analysis, the circular (scatter) points indicate the TQ onset markers. By plotting the remaining thermal energy at TQ onset against the mean fragment speed in \autoref{fig:frag_speed_var_remaining_E_th_at_TQ}, faster fragments are noted to radiate away more energy before TQ onset. The higher fraction of radiated energy before the TQ for faster fragments is understood by the radiation of energy from the core because of the deeper core neon deposition by faster fragments. This is evident in the plasma profiles at the TQ onset condition, shown in \autoref{fig:frag_speed_var_Te_ne_profiles_at_TQ} where faster fragments lead to a larger reduction in the plasma temperature towards the core compared to slower fragments. Hence, even though faster fragments lead to an earlier TQ onset, more material can be assimilated and more of the plasma thermal energy can be radiated before the TQ onset as compared to slower fragments. 

\subsubsection{Pure deuterium pellets}
\label{sssec:frag_speed_results_pure_D}
\begin{figure}[H]
   \centering
     \begin{subfigure}[b]{0.7\textwidth}
         \centering\includegraphics[width=\linewidth]{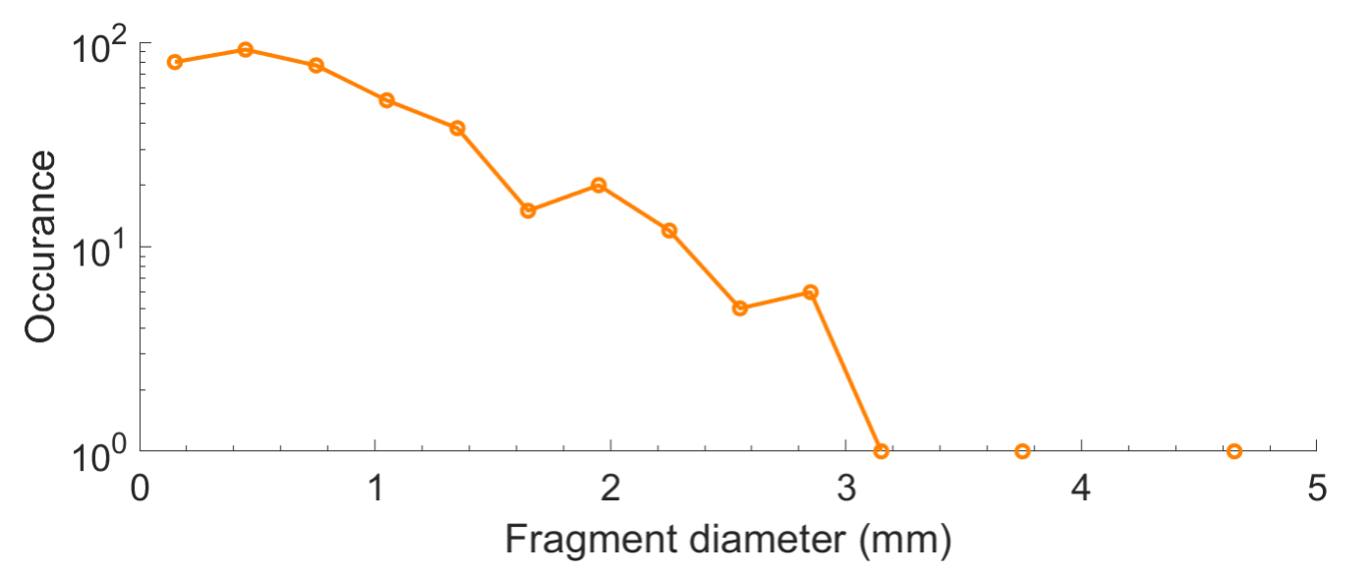}
         \phantomcaption
         \label{fig:deuterium_speed_var_frag_size_distributions}         
     \end{subfigure}
     \begin{subfigure}[b]{0.7\textwidth}
         \centering\includegraphics[width=\linewidth]{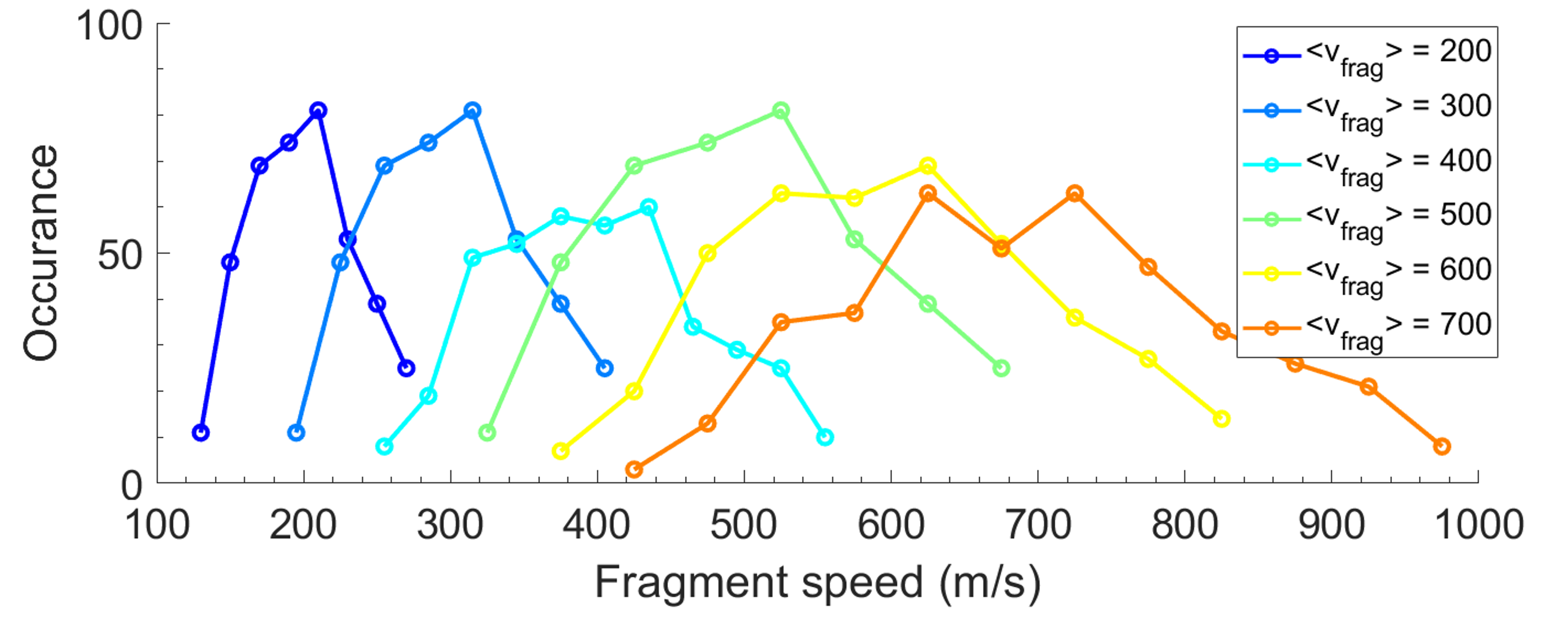}
         \phantomcaption
         \label{fig:deuterium_speed_var_frag_speed_distributions} 
     \end{subfigure}
\caption{Fragment size (top) and speed (bottom) distributions for varying mean fragment velocity.}
\label{fig:deuterium_speed_var_frag_size_speeds}
\end{figure}

Again using the back-averaging model, pure deuterium injection simulations were carried out by sampling different fragment speed distributions with the same fragment size distribution. The input fragment size and speed distributions for this set of simulations is shown in \autoref{fig:deuterium_speed_var_frag_size_speeds} . Other relevant parameters of the simulation are mentioned in \autoref{table:frag_speed_statics}. Similar to the deuterium injections in \autoref{sssec:frag_size_results_pure_D}, there is no formation of a cold front due to the lack of injected and intrinsic impurities.

\begin{table}[H]
\caption{Simulation parameters for fragment speed scan of pure deuterium injections.}
\rule[0.2ex]{14 cm}{1.5pt}
\centering
\begin{tabular}{p{5.2cm} || p{5cm}}
Number of fragments & 400\\
Mean fragment size & 0.874 mm\\
Mean fragment velocity & [200, 300, 400, 500, 600, 700] m/s\\
Pellet composition(\% Ne/D) & 0\%/100\% \\
Velocity dispersion ($\Delta v/ v$) & 40\% \\
Local deposition fraction & 0 \\
$\beta$ & 4 \\
Pellet length & 10 mm\\
Pellet diameter & 8 mm\\
\end{tabular}
\label{table:frag_speed_statics}
\end{table}

\begin{figure}[H]
    \centering
    \includegraphics[trim={0 9cm 0 0},clip,width=0.8\linewidth]{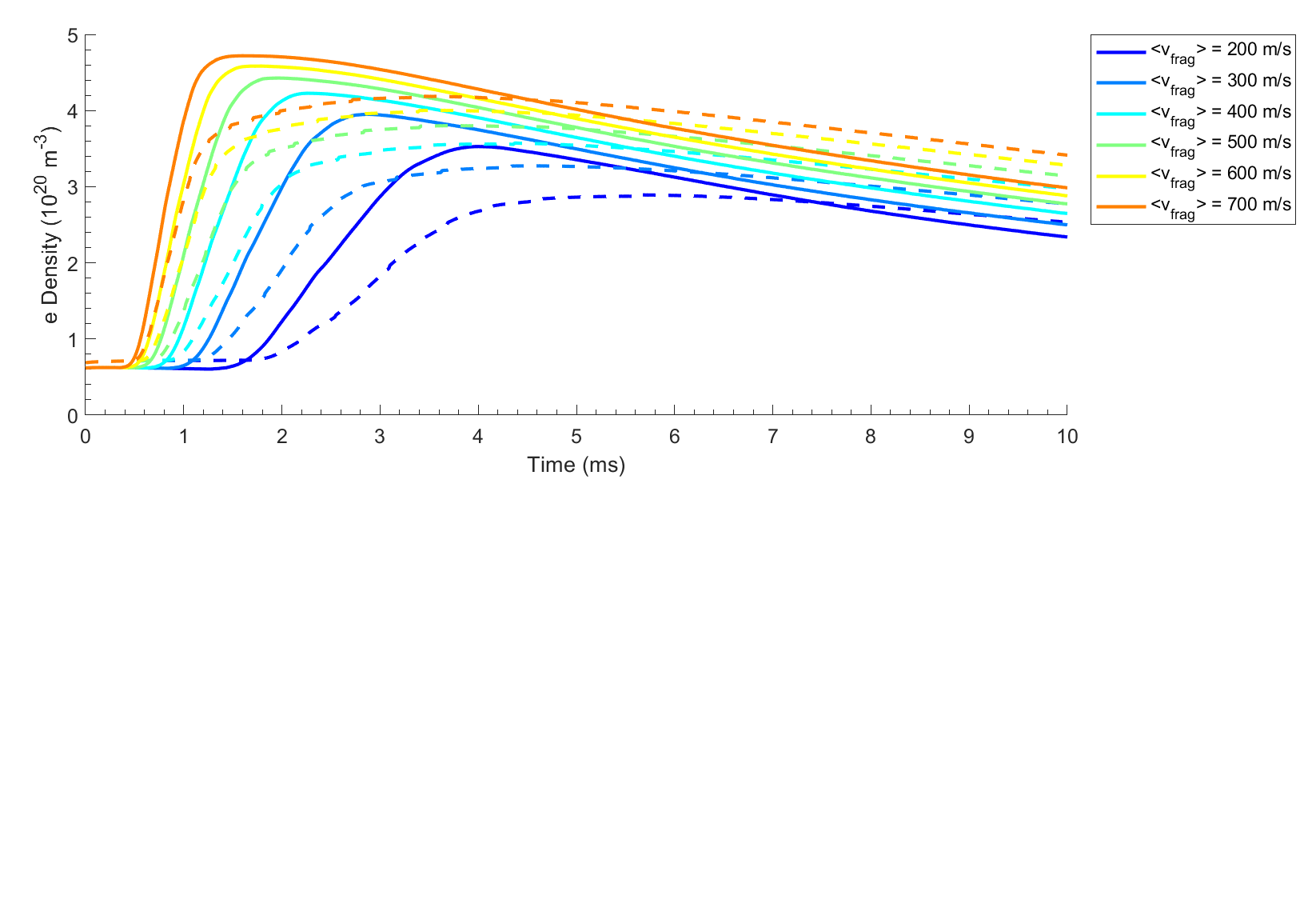}
    \caption{Time traces of electron density inside the plasma volume (solid lines) and inside q=2 surface (dashed lines). Faster fragments lead to higher maximum assimilation compared to slower fragments.}
    \label{fig:deuterium_speed_var_density_inside_vol_and_q2}
\end{figure}

To study the dependence of assimilation of the injected deuterium, the time traces of the average electron density inside the plasma volume and inside the q=2 surface for simulations with different mean fragment velocities are plotted in \autoref{fig:deuterium_speed_var_density_inside_vol_and_q2}. Faster fragments start to assimilate faster as a result of entering and penetrating the plasma quicker than slower fragments. The maximum material assimilation inside the plasma volume is also higher for faster fragments as compared to slower ones. In line with previous deuterium simulations discussed in \autoref{sssec:frag_size_results_pure_D}, initial assimilation of the injected material is limited to the plasma edge which can be indirectly observed in \autoref{fig:deuterium_speed_var_density_inside_vol_and_q2} with the volume-averaged density being higher than the average density inside the q=2 surface. In conformity with the simulations discussed in \autoref{sssec:frag_size_results_pure_D}, the maximum volume averaged density is reached when majority of the ablation has finished after which the plasma density evolves only due to diffusion. As the assimilated material diffuses inwards, the average density inside the q=2 surface overcomes the volume averaged density, this crossover happening quicker for faster fragments. 

\begin{figure}[H]
   \centering
     \begin{subfigure}[t]{0.8\textwidth}
       \centering
        \includegraphics[trim={0 9cm 0 0},clip,width=\linewidth]{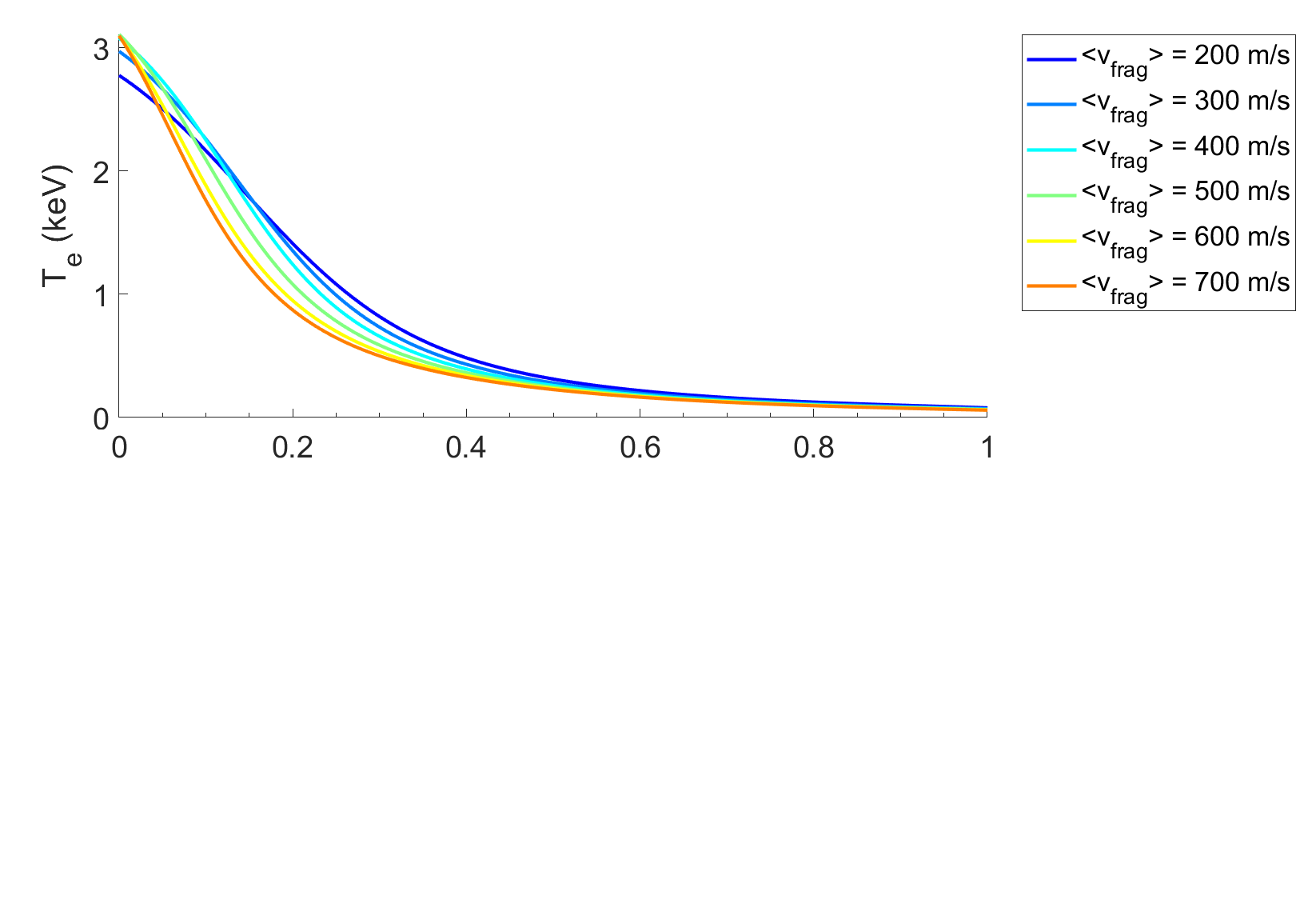}
        \phantomcaption
        \label{fig:deuterium_speed_var_Te_profiles_at_TQ}   
     \end{subfigure}
     \begin{subfigure}[t]{0.8\textwidth}
        \centering
        \includegraphics[trim={0 0 0 9cm},clip,width=\linewidth]{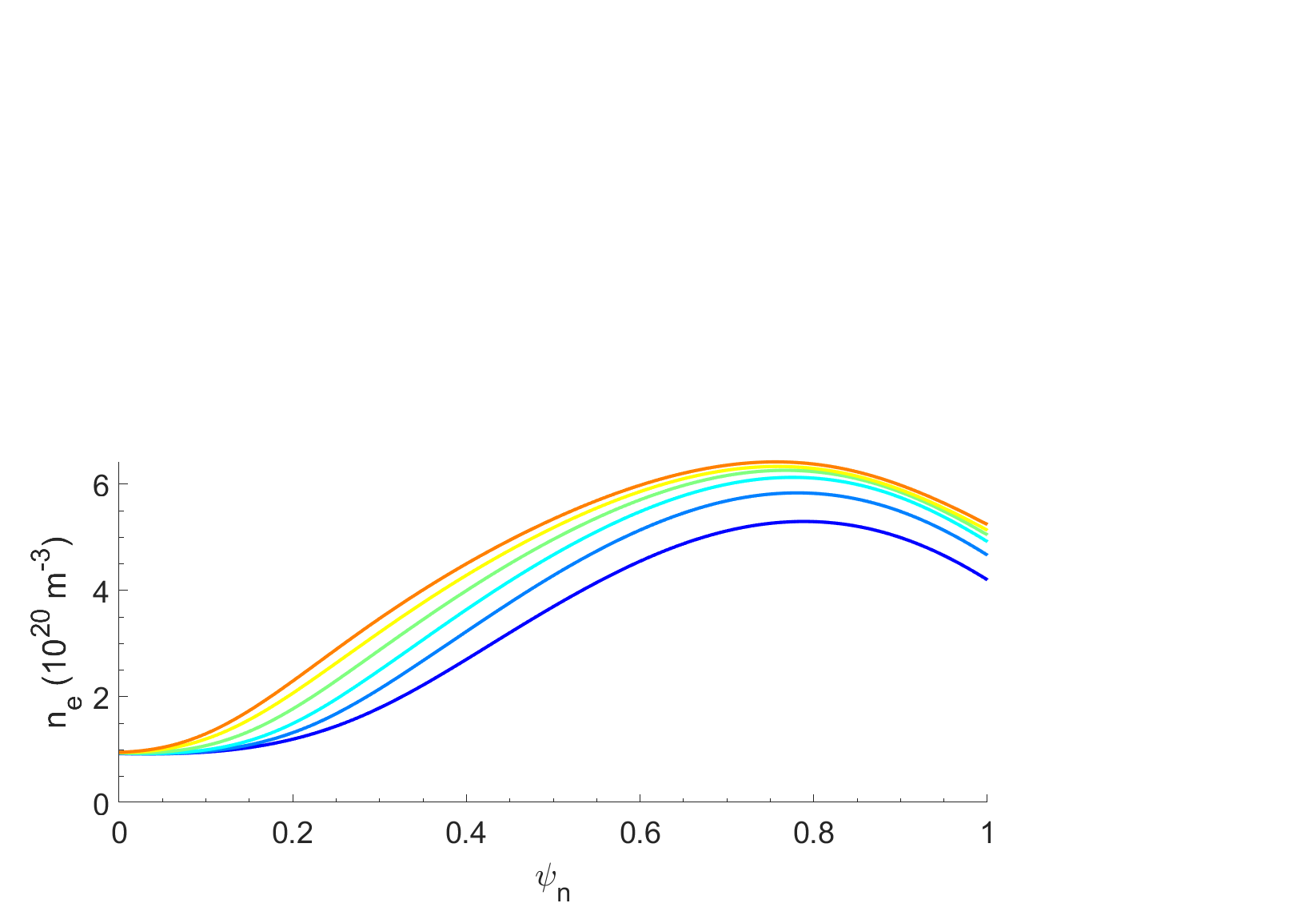}
        \phantomcaption
        \label{fig:deuterium_speed_var_ne_profiles_at_TQ}   
     \end{subfigure}
     \caption{Electron temperature profiles (top) and electron density profile (bottom) at TQ onset for mean fragment velocities.}
     \label{fig:deuterium_speed_var_ne_Te_profiles_at_TQ}
\end{figure}

The role of fragment speed is very straightforward; faster fragments can penetrate deeper (and also quicker) in the plasma before completely ablating or drifting out. Because of the higher plasma temperature towards the core, more material is ablated and assimilated compared to slower fragments. This effect can be directly observed by comparing the plasma density profiles at the point of maximum volume averaged density for each simulation as shown in \autoref{fig:deuterium_speed_var_ne_Te_profiles_at_TQ}. A higher electron density can be observed for faster fragments as compared to slower fragments consistently across the plasma. 

\subsection{Effect of pellet composition}
\label{sec:pellet_composition_results}
Finally, I carried out a scan of the the pellet composition by varying the neon fraction in the pellet from 0.085\% to 100\% with multiple simulations with trace amounts of neon ($\le 1\% $). The lower limit in the neon composition was decided based on the pellet composition in the 2022 AUG experimental campaign. The main parameters of the simulation are shown in \autoref{table:AUGparams_neoncomp_var}. Five different realisations of fragment size and speed distributions were tested for each case. It should be kept in mind that while the addition of small amounts of neon has been shown to suppress the plasmoid drift, it isn't clear yet what fraction of neon would be sufficient to suppress the drift completely and research is ongoing. In the present simulations, it has been assumed that mixed D/Ne fragments with even trace neon fractions will have no plasmoid drift. The effect of this assumption on the simulation results is discussed later in this section. 

\begin{table}[H]
\caption{Parameters for pellet composition scan.}
\rule[0.2ex]{15 cm}{1.5pt}
\centering
\begin{tabular}{p{5.2cm} || p{8cm}}
Number of fragments & 200\\
Mean fragment diameter & $\sim$ 1.11 mm\\
Mean fragment velocity & 230 m/s\\
Velocity dispersion ($\Delta v/ v$) & 40\% \\
Local deposition fraction & 1 \\
Pellet composition(\% Ne) & [0.085, 0.1, 0.25, 0.5, 1, 5, 10, 25, 50] \% \\
\end{tabular}
\label{table:AUGparams_neoncomp_var}
\end{table}

\begin{figure}[H]
    \centering
    \includegraphics[width = 0.8\linewidth]{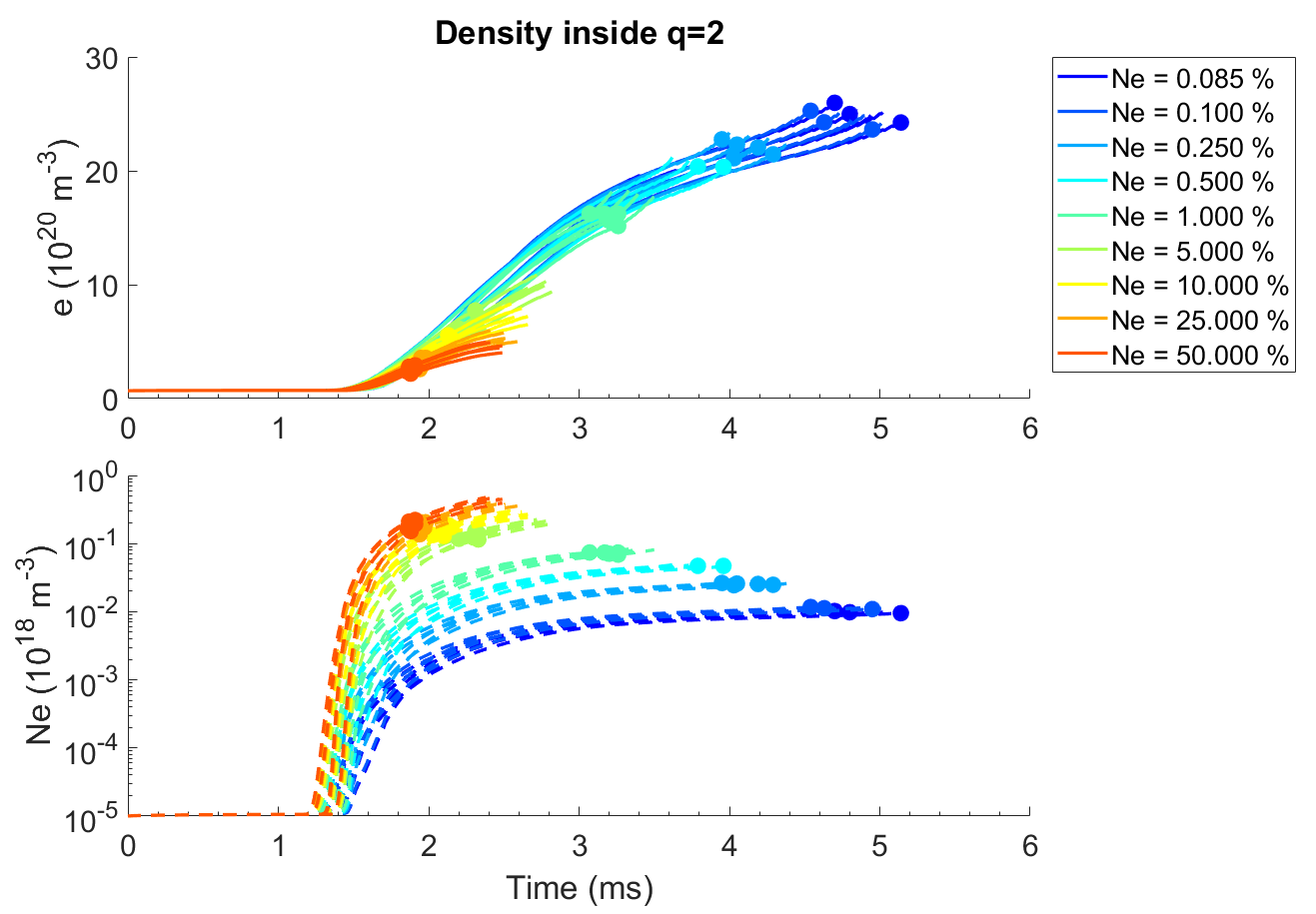}
    \caption{Average electron density (top) and neon density (bottom) inside the q=2 surface. Electron assimilation is higher for injections with less neon while neon assimilation is higher for injections with more neon.}
    \label{fig:xmol_var_density_inside_q2}
\end{figure}

\begin{figure}[H]
    \centering
    \includegraphics[width = 0.5\linewidth]{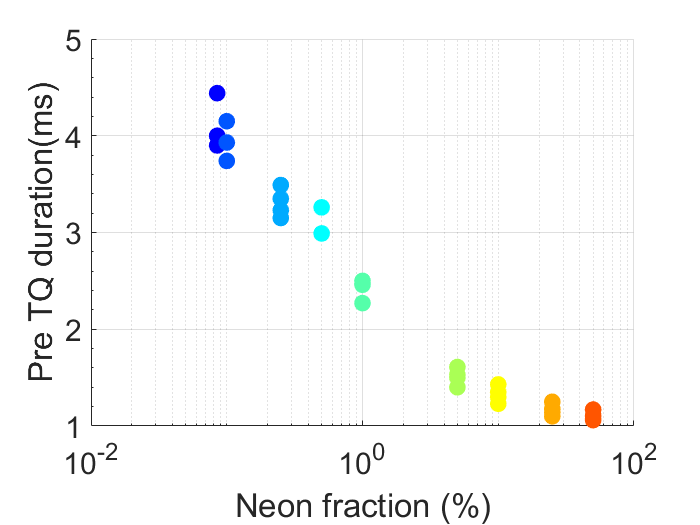}
    \caption{Pre-TQ duration as a function of neon fraction. Higher neon fraction injections lead to a shorter pre-TQ duration.}
    \label{fig:xmol_var_preTQ_duration}
\end{figure}

\begin{figure}[H]
   \centering
     \begin{subfigure}[t]{0.45\textwidth}
       \centering
        \includegraphics[width=0.8\linewidth]{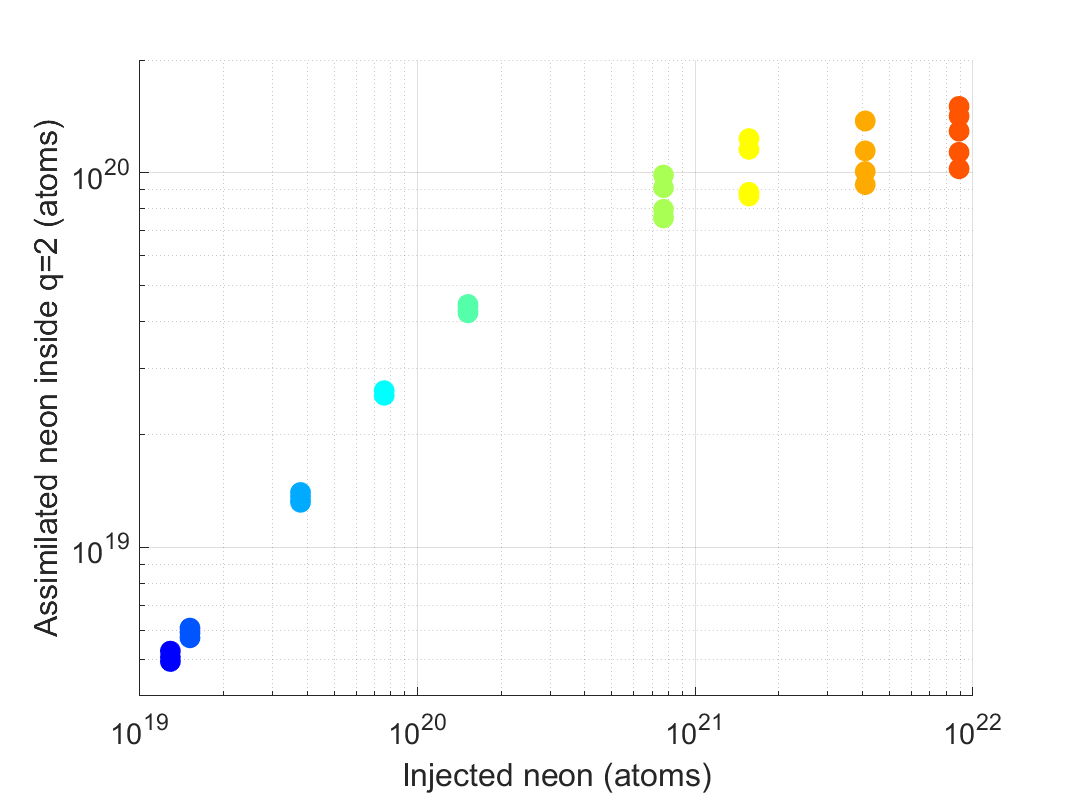}
        \caption{}
        \label{fig:xmol_var_absolute_neon_scatter}             
     \end{subfigure}
     \begin{subfigure}[t]{0.45\textwidth}
        \centering
        \includegraphics[width=1\linewidth]{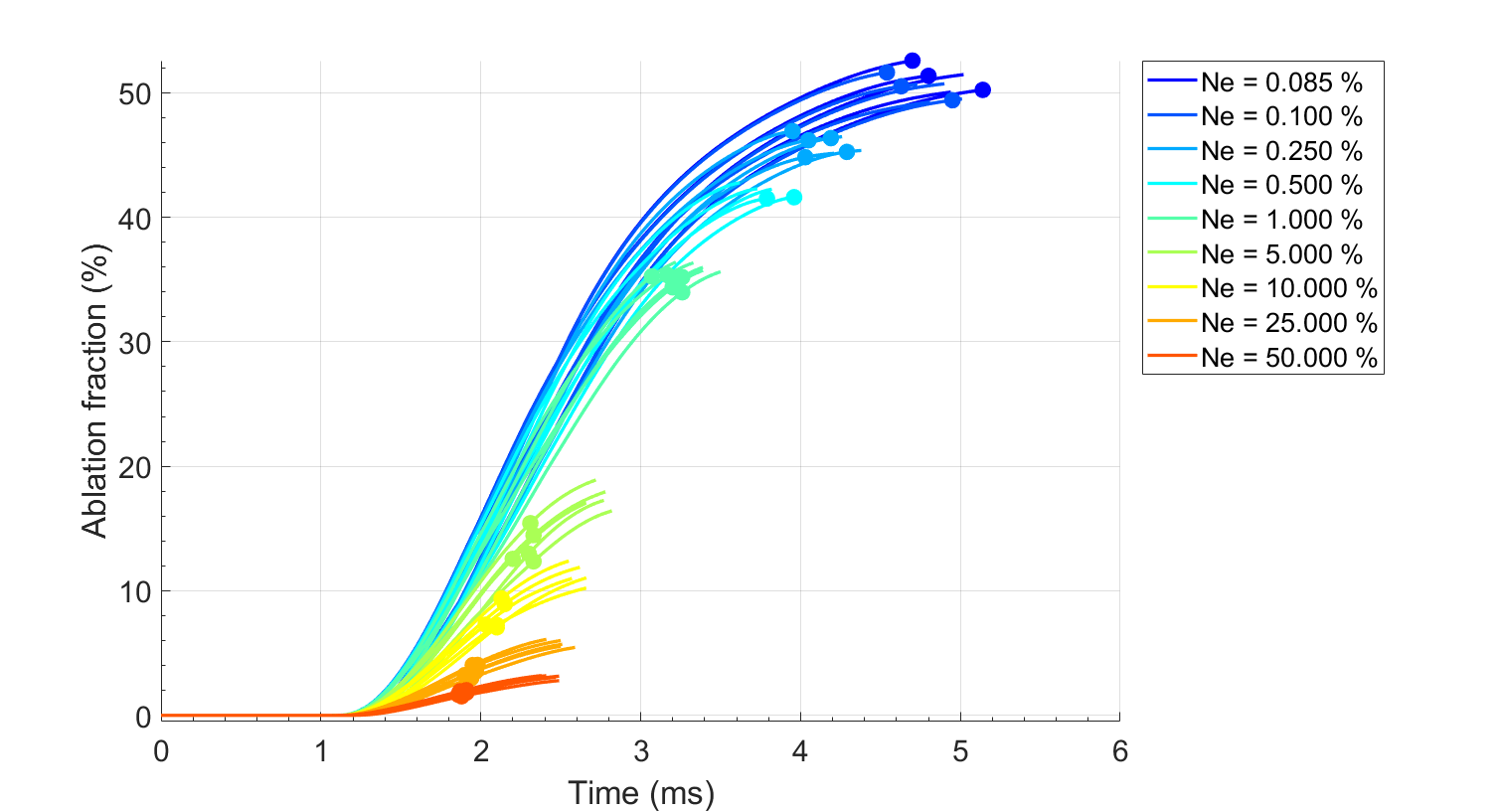}
        \caption{}  
        \label{fig:xmol_var_ablation_fraction}        
     \end{subfigure}
     \caption{(a) Neon atoms inside the q=2 surface at the TQ onset against neon fraction in the injected pellet. The assimilated neon increases linearly with injection neon amount but saturates for higher injected neon quantities; (b) Ablation fraction time traces for varying neon fractions.}
\end{figure}

\begin{figure}[H]
    \centering
    \includegraphics[width=0.8\linewidth]{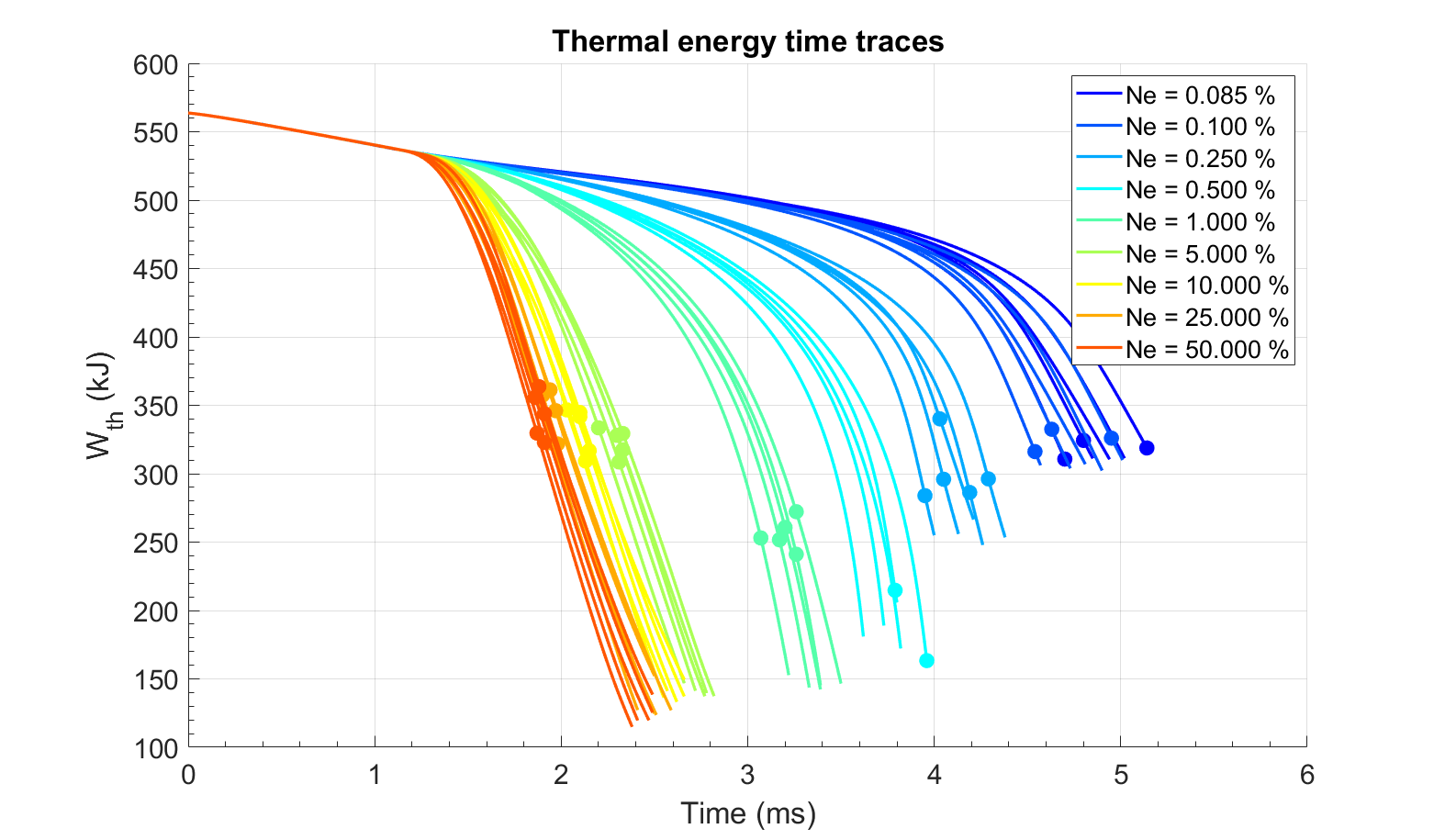}
    \caption{Time traces of plasma thermal energy evolution for varying neon fraction.}
    \label{fig:xmol_var_thermal_energy_time_traces}
\end{figure}

In line with the previous comparisons, first, the time traces of electron density and neon density inside the q=2 surface for injections with different neon fractions are shown in \autoref{fig:xmol_var_density_inside_q2}. Similar to \autoref{fig:frag_size_var_density_inside_q2}, the circular markers indicate the time point when the TQ onset condition is satisfied. The pre-TQ duration is plotted separately against the neon fraction in \autoref{fig:xmol_var_preTQ_duration}. It can be observed that the electron assimilation is higher for injected fragments with less neon and the neon assimilation is higher for injected fragments with more neon. To study the assimilation of neon in the core of the plasma, \autoref{fig:xmol_var_absolute_neon_scatter} shows the the assimilated neon atoms in the plasma inside the q=2 surface as a function of the injected neon atoms. For trace neon fractions, the assimilated neon atoms increases linearly with increasing neon fraction. However, at larger neon fractions, the assimilated neon quantity increases at a much slower rate and is almost saturated. This trend can be explained by swift cooling of the plasma for high amount of injected neon. Beyond a certain amount of neon, 5\% 8 mm pellets or $>10^{21}$ atoms from \autoref{fig:xmol_var_absolute_neon_scatter}, the plasma cools down very drastically (also refer to \autoref{fig:xmol_var_thermal_energy_time_traces}) and hence only a limited amount of material can be ablated leading to a self-regulated saturation. To study the dependence of the ablation fraction i.e. ratio of ablated neutral atoms to the number of injected atoms on the injected neon, time traces of ablation fraction for varying neon fractions are plotted in \autoref{fig:xmol_var_ablation_fraction}. The reduced ablation fraction for larger neon fractions is very apparent. Correspondingly, decreasing pre-TQ duration for larger neon fraction can also be observed in \autoref{fig:xmol_var_preTQ_duration}. 

\begin{figure}[H]
    \centering
    \includegraphics[width=0.8\linewidth]{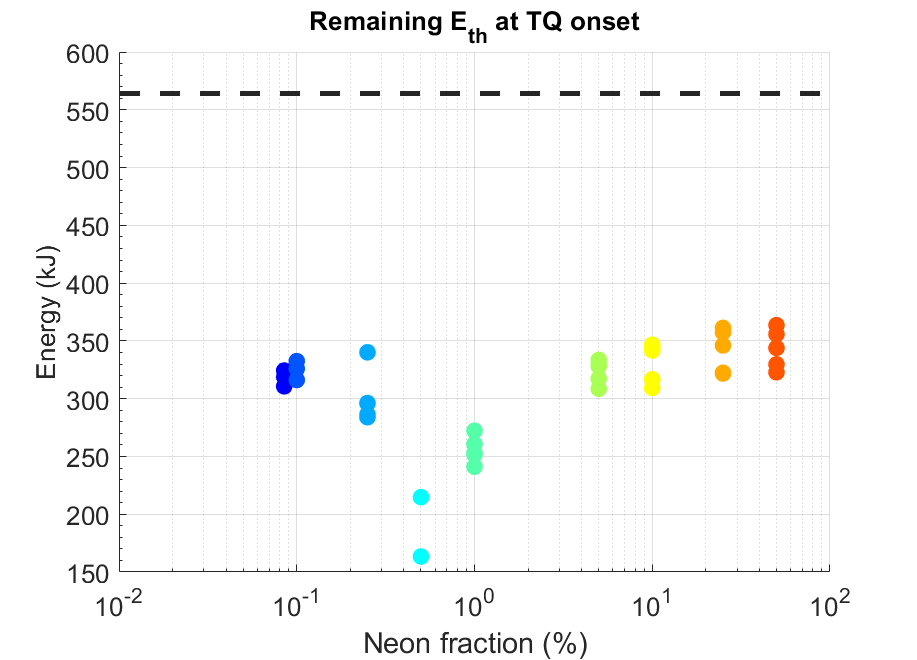}
    \caption{Remaining thermal plasma energy at TQ onset condition. Black solid line shows initial thermal energy. A non-monotonic behaviour can be observed with minimum values for trace neon injections.}
    \label{fig:xmol_var_thermal_energy_remaining_TQ}
\end{figure}

Since large fraction mixed neon pellets will have to radiate away plasma energy before the TQ onset, the thermal energy time evolution was studied for different neon fractions and is shown in \autoref{fig:xmol_var_thermal_energy_time_traces}. While the initial fragments enter the plasma at $\sim$ 0.7 - 0.8 ms, the decay of the thermal energy due to radiative cooling starts at $\sim$ 1.1 ms. A faster rate of thermal energy loss due to radiative cooling can be observed for increasing neon fractions.  A behaviour similar to the assimilation can be seen i.e. for trace neon quantities ($<1\%$), large differences are observed in the rate at which thermal energy decreases. However, for larger neon fractions, the thermal energy is radiated at very similar rates although the trend of higher decay rate for higher neon fraction persists. Looking at the TQ onset markers in \autoref{fig:xmol_var_thermal_energy_time_traces}, a non-monotonic trend of remaining thermal energy can be observed when going from small to large neon fractions which has a minima at $0.5\%$ neon cases. This can be more clearly seen by relating the remaining thermal energy at TQ onset to the neon fraction as plotted in \autoref{fig:xmol_var_thermal_energy_remaining_TQ}. 

\begin{figure}[H]
   \centering
     \begin{subfigure}[t]{0.7\textwidth}
       \centering
        \includegraphics[trim={0 9cm 0 0},clip,width=\linewidth]{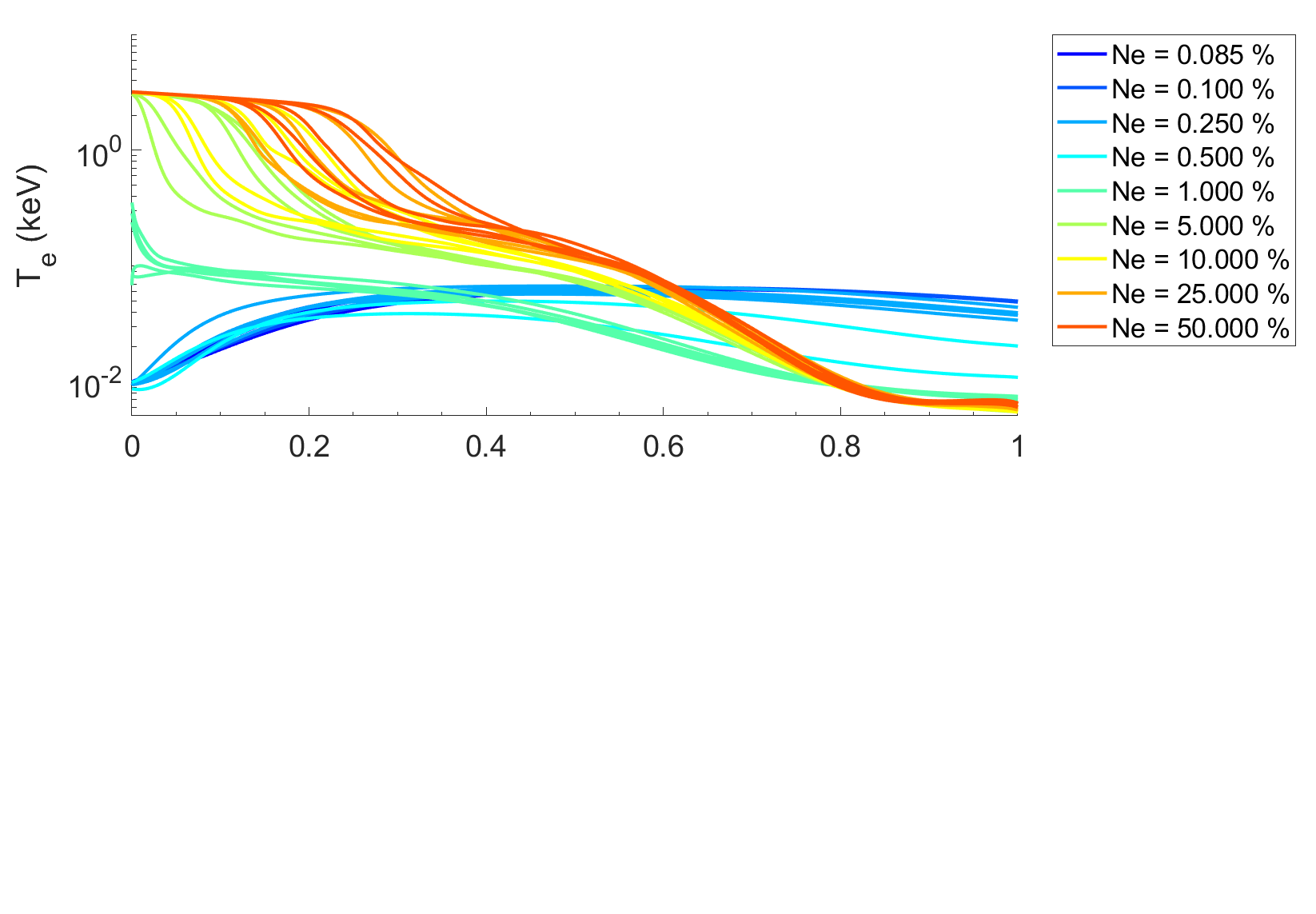}
        \phantomcaption
        \label{fig:xmol_var_Te_profile_at_TQ}   
     \end{subfigure}
     \begin{subfigure}[t]{0.7\textwidth}
        \centering
        \includegraphics[trim={0 0 0 9cm},clip,width=\linewidth]{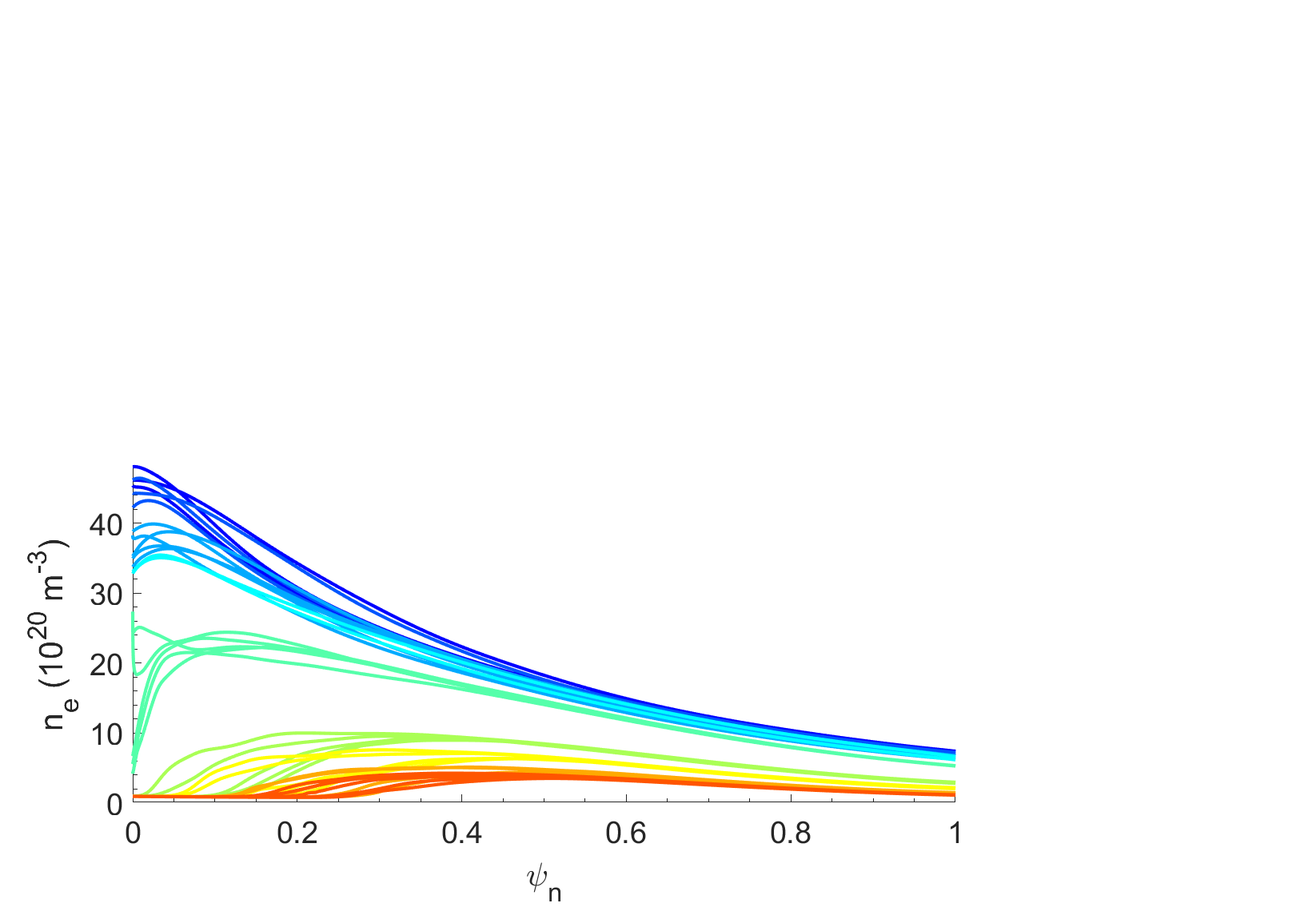}
        \phantomcaption
        \label{fig:xmol_var_ne_profile_at_TQ}   
     \end{subfigure}
     \caption{Electron temperature profiles (top) and electron density profile (bottom) at TQ onset for different neon fractions. An inside-out temperature collapse can be observed for trace neon fraction injections in contrast with higher neon fraction injections where the cold front moves in from the edge.}
     \label{fig:xmol_var_profiles_at_TQ}
\end{figure}

The cause of the non-monotonic behaviour can be understood by looking at the plasma temperature and radial profiles at the TQ onset which are plotted in \autoref{fig:xmol_var_profiles_at_TQ}. For the trace neon injections (in blue), \autoref{fig:xmol_var_Te_profile_at_TQ} shows an inside out temperature collapse where the core temperature drops lower than 10 eV. While for large neon fractions, an outside-in temperature collapse occurs as a result of the radiative cold front moving in. The inside-out temperature collapse for trace neon injections can be understood as follows. As the fragments penetrate the plasma edge, they start to ablate, however, the ablation rate remains low (refer to \autoref{eq:ablation_recession}), leading to a slow deposition of neon atoms for trace neon fragments. Further, as the fragments travel deeper in the core, the ablation rate increases due to the hot plasma leading to a higher core deposition of neon atoms as compared to the edge. As a result, the neon starts radiating strongly from the core while the edge still remains relatively hot compared to the higher neon fraction injections. As a result of the inside-out temperature collapse, a larger fraction of the thermal energy can be radiated before the TQ onset condition is satisfied. As mentioned before, the plasmoid drift in the simulations has been enabled for trace neon injections. This assumption can affect the trend of the stronger core deposition for trace neon injections as discussed above. The trends that would likely be affected would be the non-monotonic trends in remaining thermal energy at TQ onset (\autoref{fig:xmol_var_thermal_energy_remaining_TQ}) and the inside-out temperature collapse for trace neon injection (\autoref{fig:xmol_var_profiles_at_TQ}).

\newpage
\section{Experimental comparisons}
\label{ssec:sim_exp_comp}
A preliminary comparison with experimental data is reported in this section. I carried out some of the experimental analysis regarding penetration and assimilation and the findings are available in my ITER internship report \cite{ansh_patel_internship_2023}. However, I also rely on other experimental results carried out by other members of the AUG and ITER teams. Comparisons between the simulations and experiments are carried out for dependence of fragment sizes, speeds and composition on penetration, material assimilation. Additionally, the simulated pre-TQ duration is also compared with experimental pre-TQ duration. 

\subsection{Penetration}
\begin{figure}[H]
\centering
\begin{subfigure}[t]{0.70\textwidth}
    \centering
    \includegraphics[width=\textwidth]{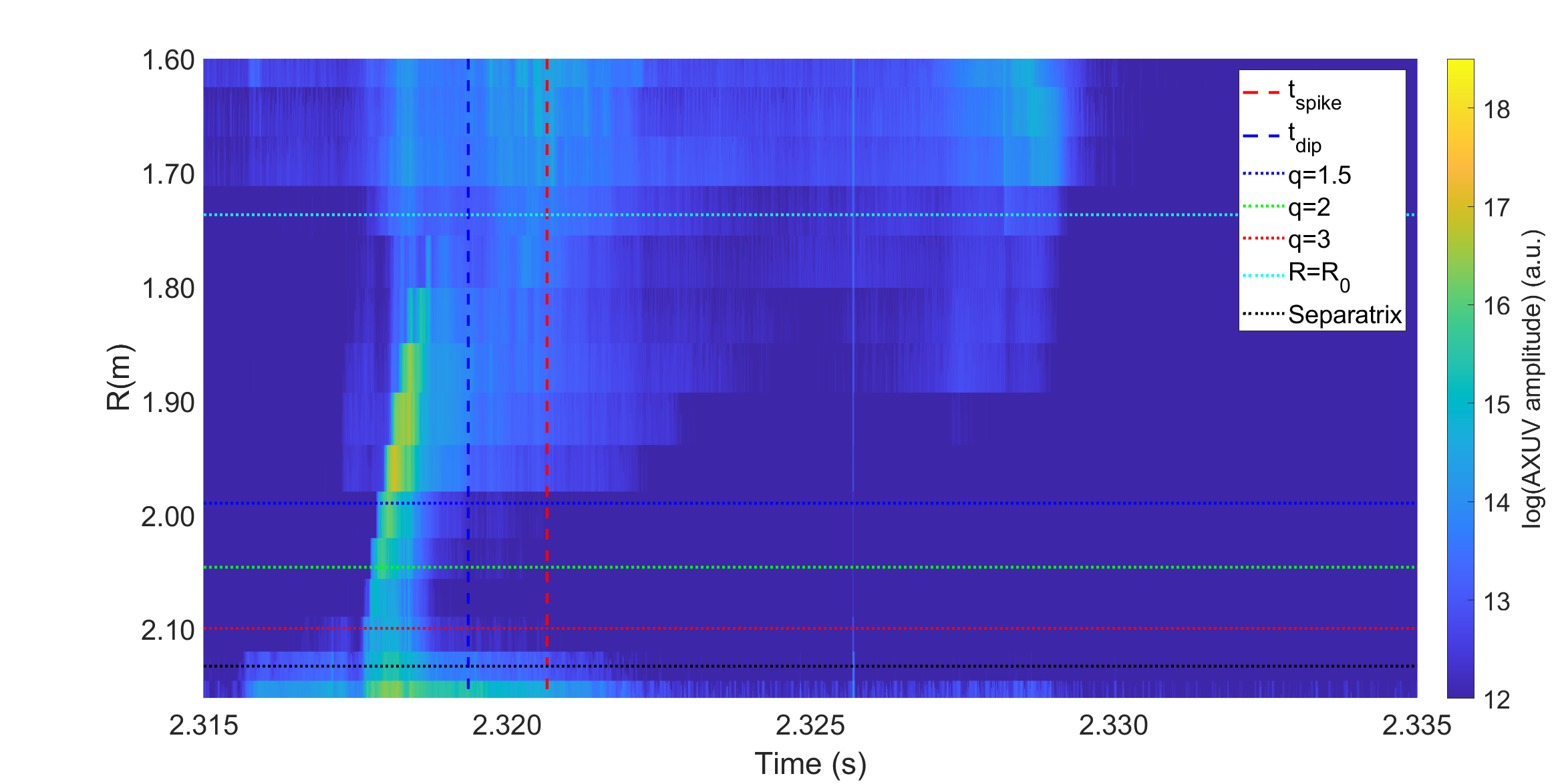}
    \caption{\#40655: 2D AXUV mesh plot, larger fragments ($D_2$ SPI, $12.5^\circ$ rect., 442.9 m/s)}
    \label{fig:40655_cam}
    \end{subfigure}

    \begin{subfigure}[b]{0.70\textwidth}
    \centering
    \includegraphics[width=\textwidth]{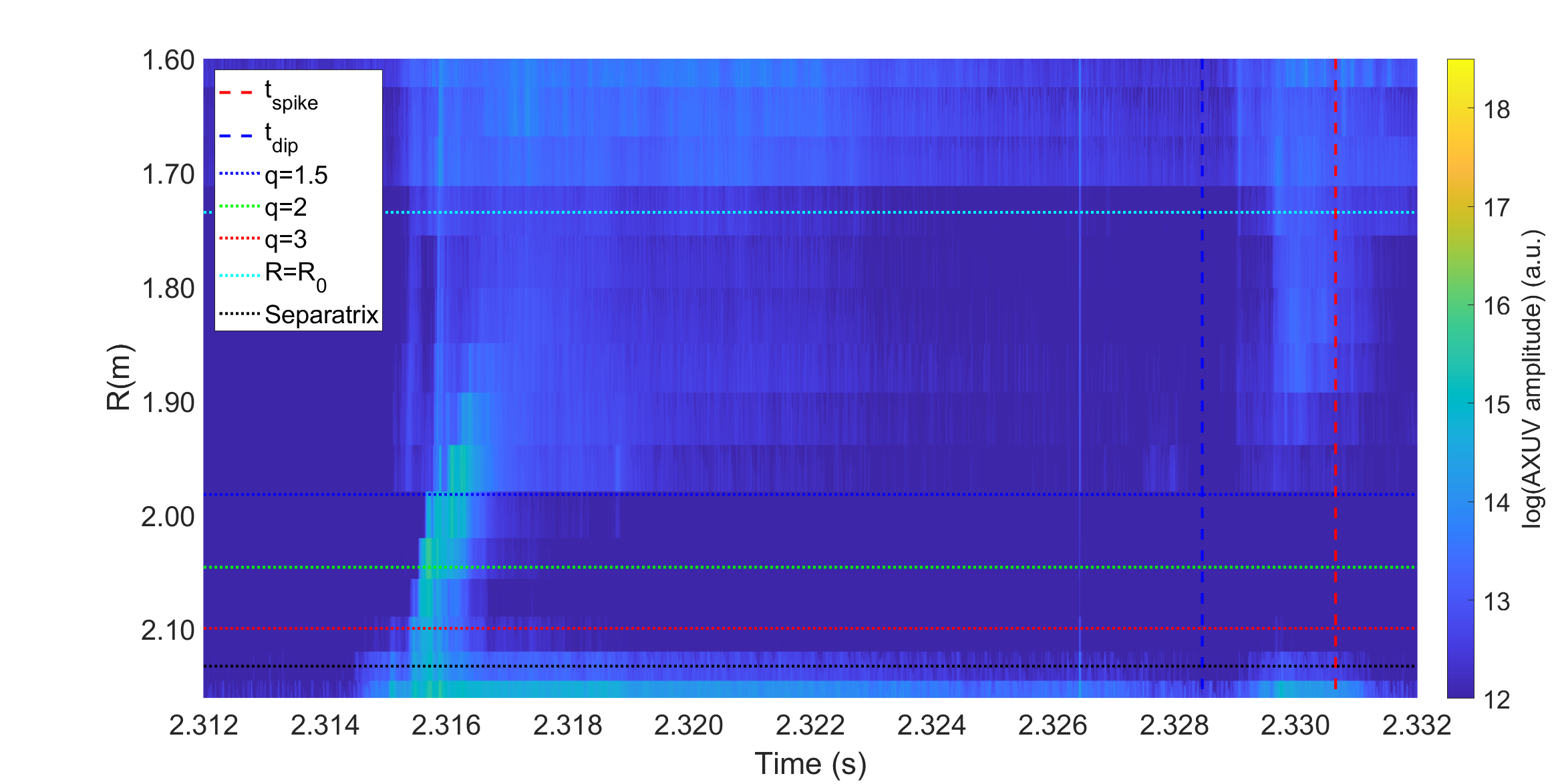}
    \caption{\#40656: 2D AXUV mesh plot, smaller fragments ($D_2$ SPI, $25^\circ$ rect., 471 m/s)}
    \label{fig:40656_cam}
    \end{subfigure}
\caption{2D AXUV plots for \#40655 [top] and for \#40656 [bottom]. Larger fragments (\#40655) lead to deeper penetration as compared to smaller fragments (\#40656).
In both plots, vertical blue dashed and red dashed lines mark the $I_p$ dip and spike respectively. Horizontal lines mark different radial locations of interest; Black dotted: Separatrix, Red dotted: q=3, yellow dotted: q=2, blue dotted: q=1.5 and cyan dotted: Magnetic axis $R_0$.}
\label{fig:40655_40656_axuv}
\end{figure}

\begin{figure}[H]
    \centering
    \begin{subfigure}[b]{0.7\textwidth}
    \centering
    \includegraphics[width=\textwidth]{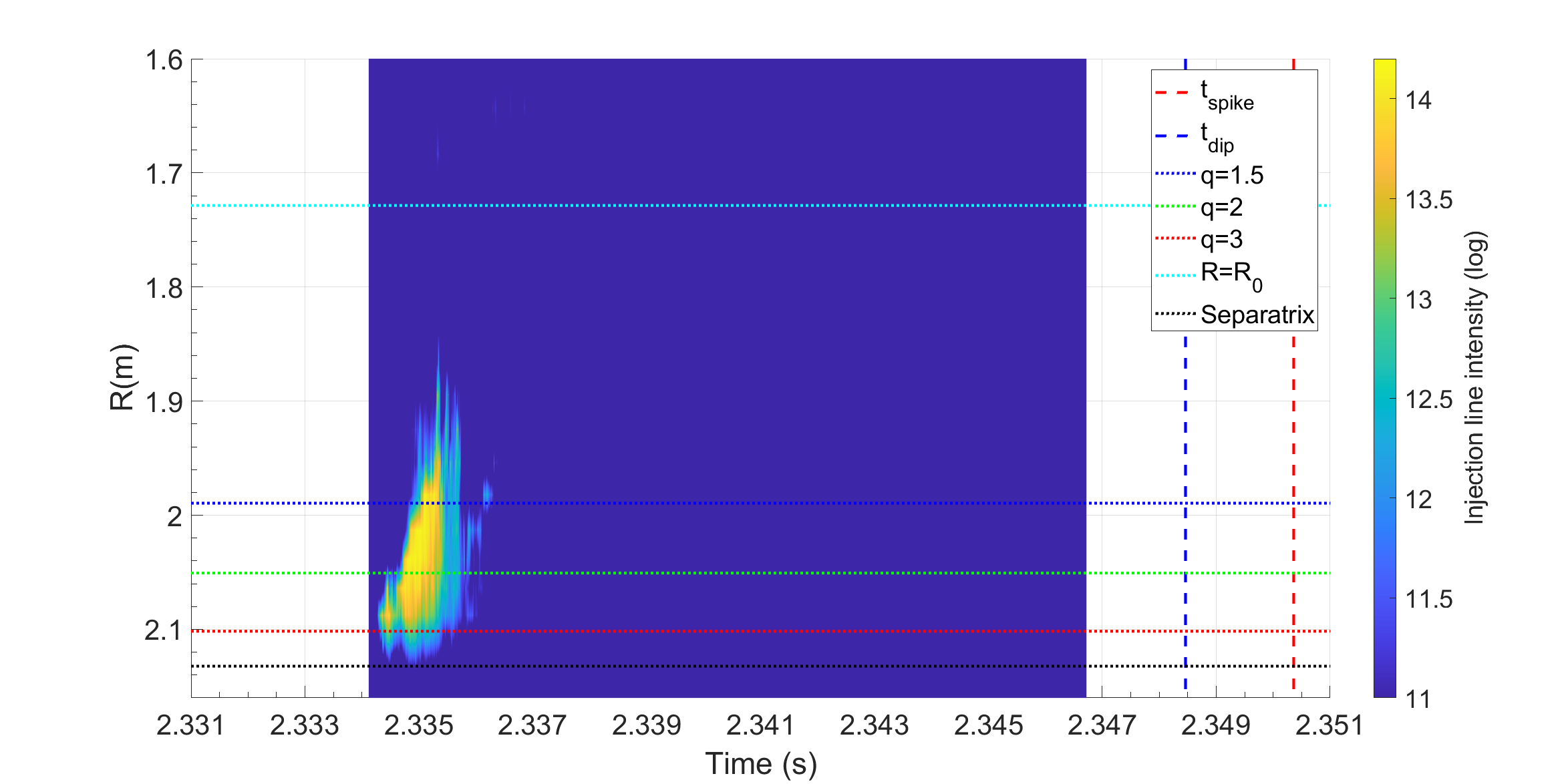}
    \caption{41012 2D camera plot, slower fragments ($D_2$ SPI, $25^\circ$  rect., 342 m/s, with $D_\alpha$ filter)}
    \label{fig:41012_cam}
    \end{subfigure}
    
    \begin{subfigure}[b]{0.7\textwidth}
    \centering
    \includegraphics[width=\textwidth]{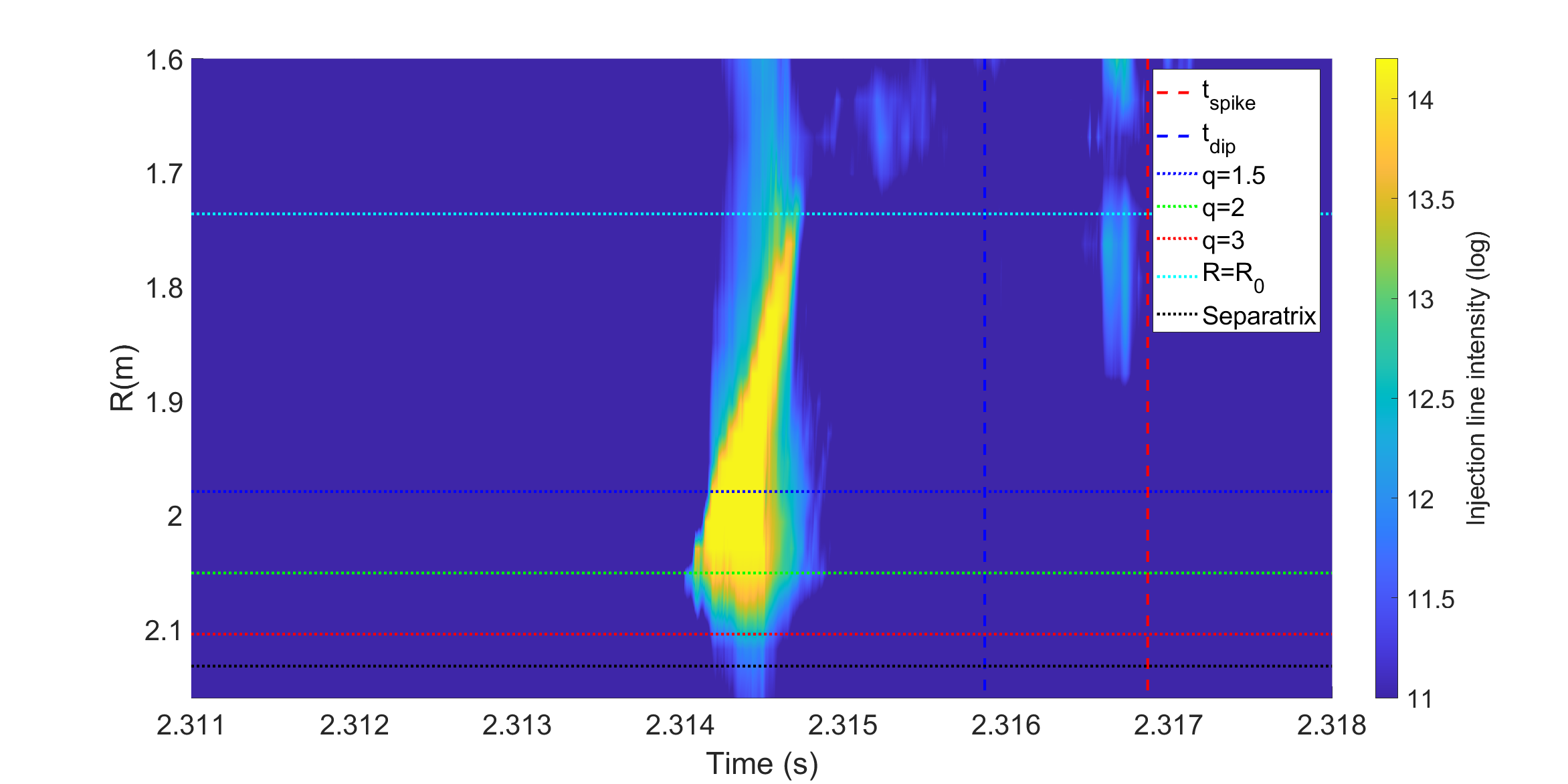}
    \caption{40772 2D camera plot, faster fragments ($D_2$ SPI, $12.5^\circ$  rect., 700 m/s, without $D_\alpha$ filter)}
    \label{fig:40772_cam}
    \end{subfigure}
    \caption{2D camera plots for \#41012 [top] and for \#40772 [bottom]. Faster fragments (\#40772) lead to deeper penetration as compared to slower fragments (\#41012). In both plots, vertical blue dashed and red dashed lines mark the $I_p$ dip and spike respectively. Horizontal lines mark different radial locations of interest; Black dotted: Separatrix, Red dotted: q=3, yellow dotted: q=2, blue dotted: q=1.5 and cyan dotted: Magnetic axis $R_0$. Note the the two plots have different time duration.}
    \label{fig:41012_40772_cam}
\end{figure}

Starting with penetration, fast visible cameras \cite{papp_asdex_2020, dibon_design_2023, ansh_patel_internship_2023} measured $D_\alpha$ radiation in the experimental campaign and were used to gauge the penetration of the fragments for pure deuterium injections. I found that larger fragments (with similar mean fragment velocities) showed stronger ablation towards the core for pure deuterium injections. In comparing fragment speeds, higher fragment speeds penetrated much deeper and also showed stronger ablation towards the core as compared to slower fragments. Representative experimental penetration plots are showed in \autoref{fig:40655_40656_axuv} and \autoref{fig:41012_40772_cam}. For both these figures, the penetration was studied by mapping the fast camera and AXUV bolometer line of sights to a radial location (y-axis in  \autoref{fig:40655_40656_axuv}, \autoref{fig:41012_40772_cam}) based on their intersection with the injection vector. More details of the mapping can be found in my ITER internship report \cite{ansh_patel_internship_2023}. \autoref{fig:40655_40656_axuv} compares the penetration for different fragments sizes. Higher radiation was observed from the intermediate plasma region for larger fragments as compared to smaller fragments. Similarly, faster fragments penetrated much deeper in the plasma compared to slower fragments as shown in \autoref{fig:41012_40772_cam}. 

\begin{figure}[H]
   \centering
     \begin{subfigure}[t]{0.45\textwidth}
         \centering
         \includegraphics[width=\textwidth]{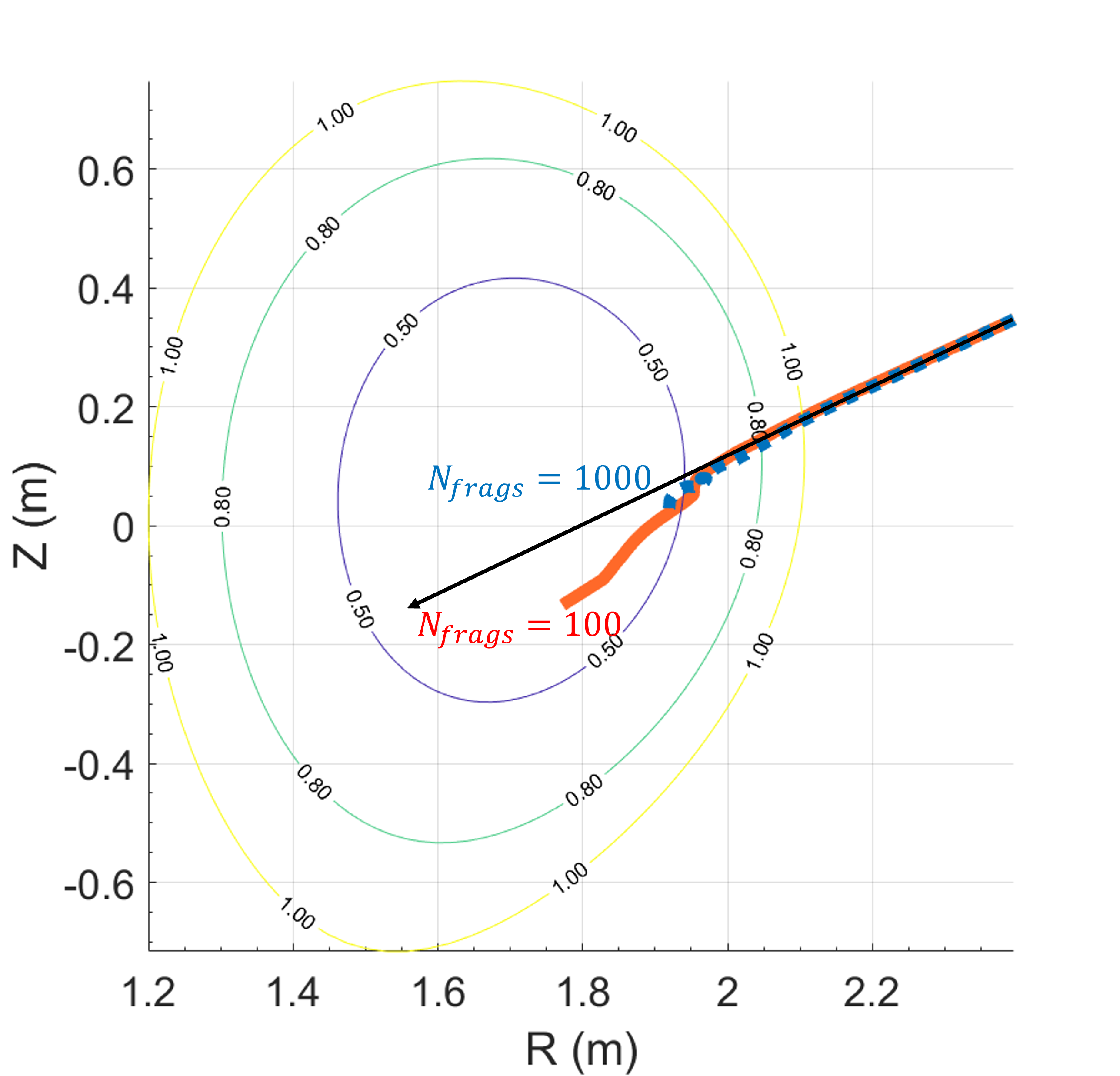}
         \caption{Variations in fragment sizes (small sizes for a higher number of fragments).}         
         \label{fig:COM_size_var}         
     \end{subfigure}
     \begin{subfigure}[t]{0.45\textwidth}
         \centering
         \includegraphics[width=\textwidth]{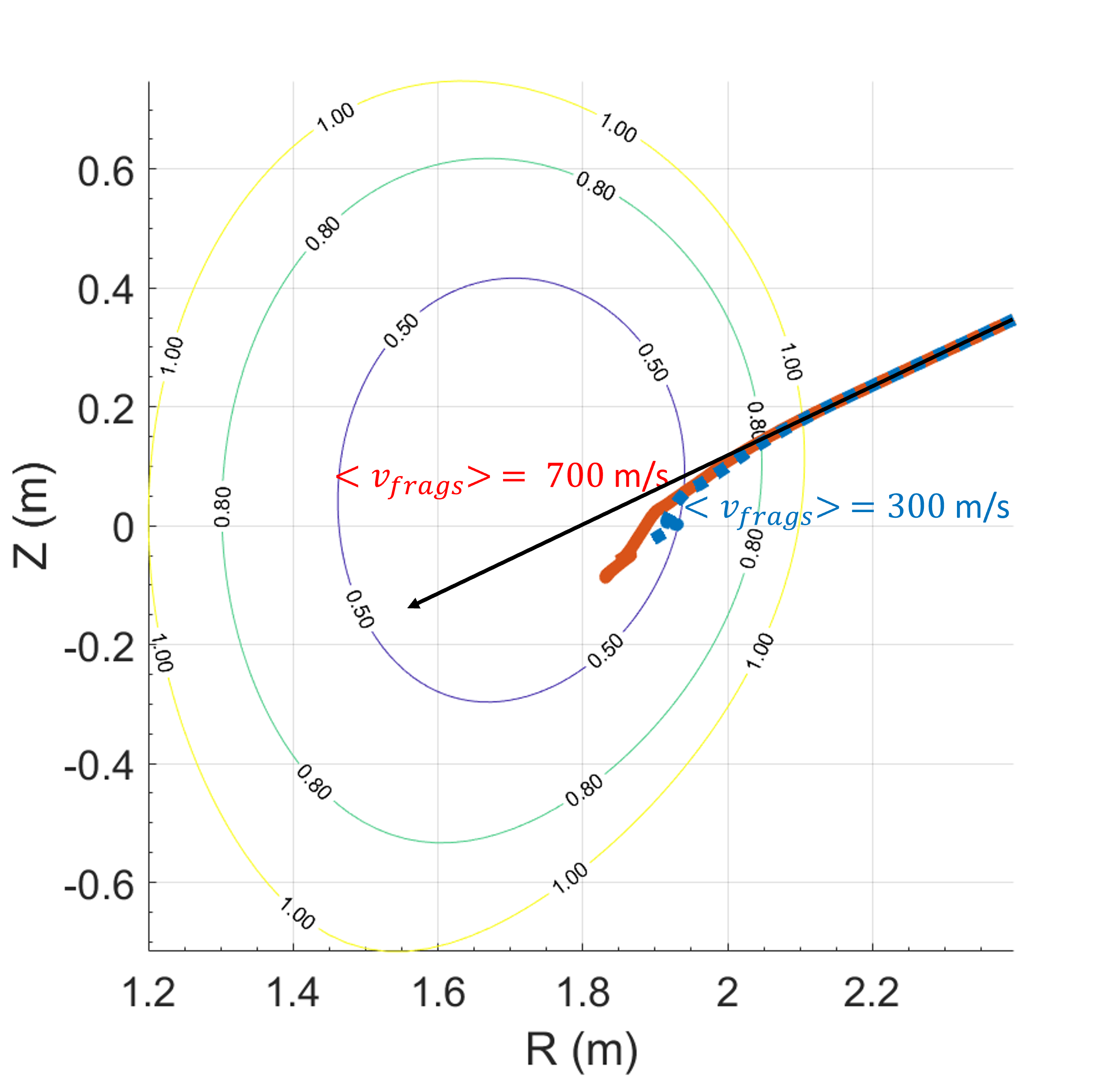}
         \caption{Variation in fragment speeds.}
         \label{fig:COM_speed_var}        
     \end{subfigure}
\caption{Simulated centre of mass of the injected fragments for variations in fragment sizes and speeds plotted in the poloidal cross section. Different contour levels are surfaces of constant toroidal flux. Black arrow shows the injection vector. Deeper penetration of the injected material can be observed for larger and faster fragments.}
\label{fig:COM_frag_sizes_speeds}
\end{figure}

Comparing the pure deuterium simulation results in  \autoref{fig:deuterium_size_var_profiles_at_max_vol_density} and \autoref{fig:deuterium_speed_var_ne_profiles_at_TQ}, larger and faster fragments penetrate and deposit material deeper in the plasma at the time when most of the ablation has finished. A more direct indication of fragment penetration can be obtained from the centre of mass of the fragments shown in \autoref{fig:COM_frag_sizes_speeds}. Again, larger and faster fragments lead to higher penetration as can be seen by the final location of the centre of mass along the injection vector. Quantitative matches of material penetration are challenging to perform due to limitations in experimental measurements of penetration of individual fragments and limited viewing angles. 

\subsection{Assimilation}
\label{ssec:sim_exmp_comp_assim}
\begin{figure}[H]
    \centering
    \includegraphics[width = 0.6\linewidth]{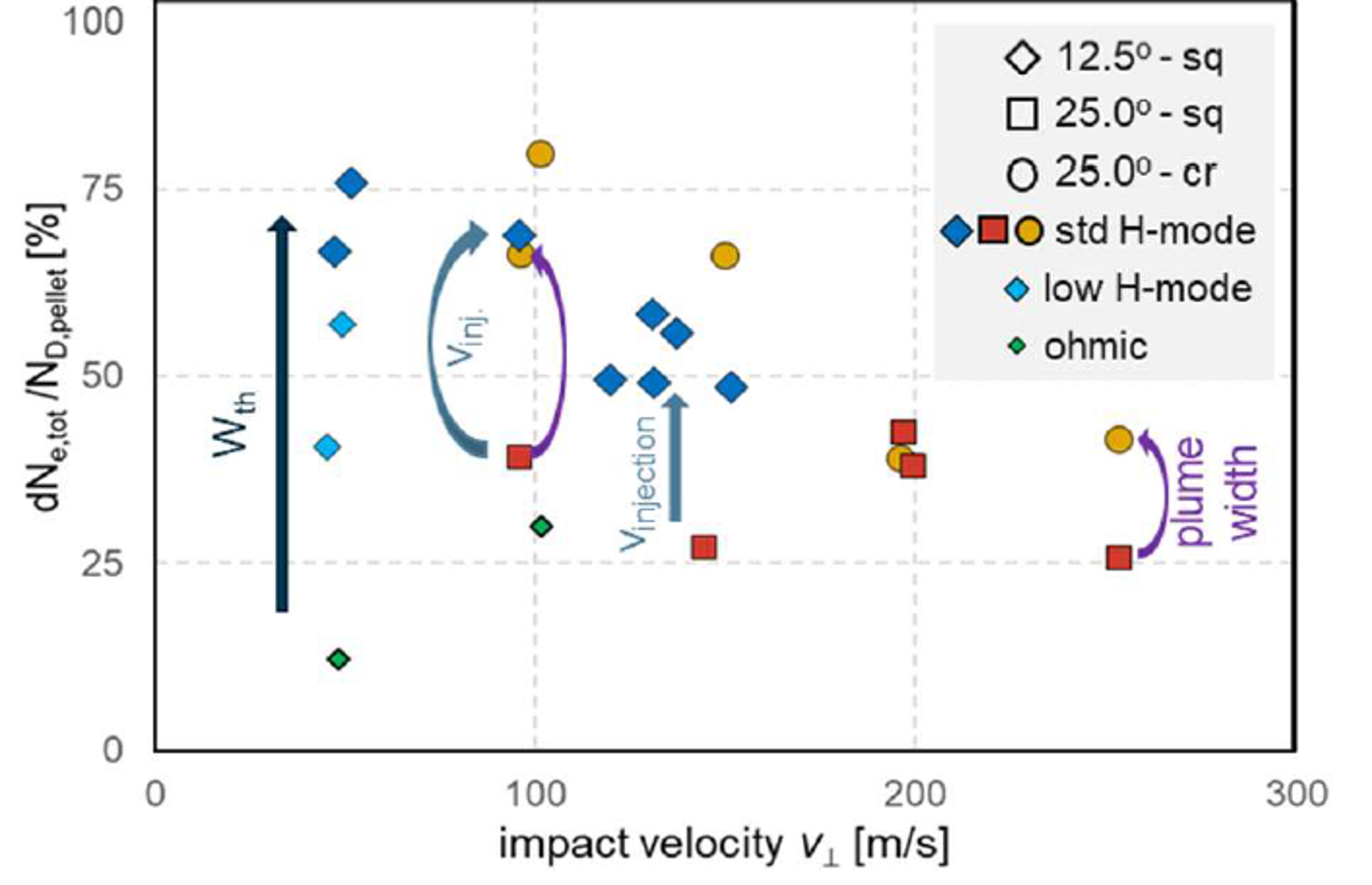}
    \caption{Peak rise in electron density from core interferometer channel (refer to \autoref{fig:AUGDiagnosticLOS}) normalized to injected deuterium atoms against impact velocity. Figure taken from S. Jachmich et. al. \cite{s_jachmich_shattered_2023}. With increasing impact velocity ($\sim$ decreasing fragment sizes) the absolute assimilation decreases. Comparing data points with higher injection velocities, marked by '$V_\text{inj}$', higher assimilation is found for faster injections.}
    \label{fig:Stefan_assimilation}
\end{figure}

Assimilation measurements in the AUG experimental campaign for the pure deuterium injections were made using the COO core and edge interferometer channels shown in \autoref{fig:AUGDiagnosticLOS}. Trends in absolute assimilation for different fragment sizes and speeds in pure deuterium injections were studied in Patel, 2023 and Jachmich et. al., 2023 \cite{s_jachmich_shattered_2023, ansh_patel_internship_2023} and are shown in \autoref{fig:Stefan_assimilation}. It should be noted that the assimilation estimates are provided with the peak density rise from the interferometer (refer to \autoref{fig:TS_interf_SimExpCompar}). With increasing impact velocity, i.e. decreasing mean fragments size, an overall decrease in assimilation can be observed. At the same perpendicular velocity, the $12.5^\circ$ sq. shatter tube indicated by diamond data points would have a higher mean fragment velocity indicated by a higher injection speed. 

\begin{figure}[H]
   \centering
     \begin{subfigure}[t]{0.45\textwidth}
         \centering
         \includegraphics[width=\textwidth]{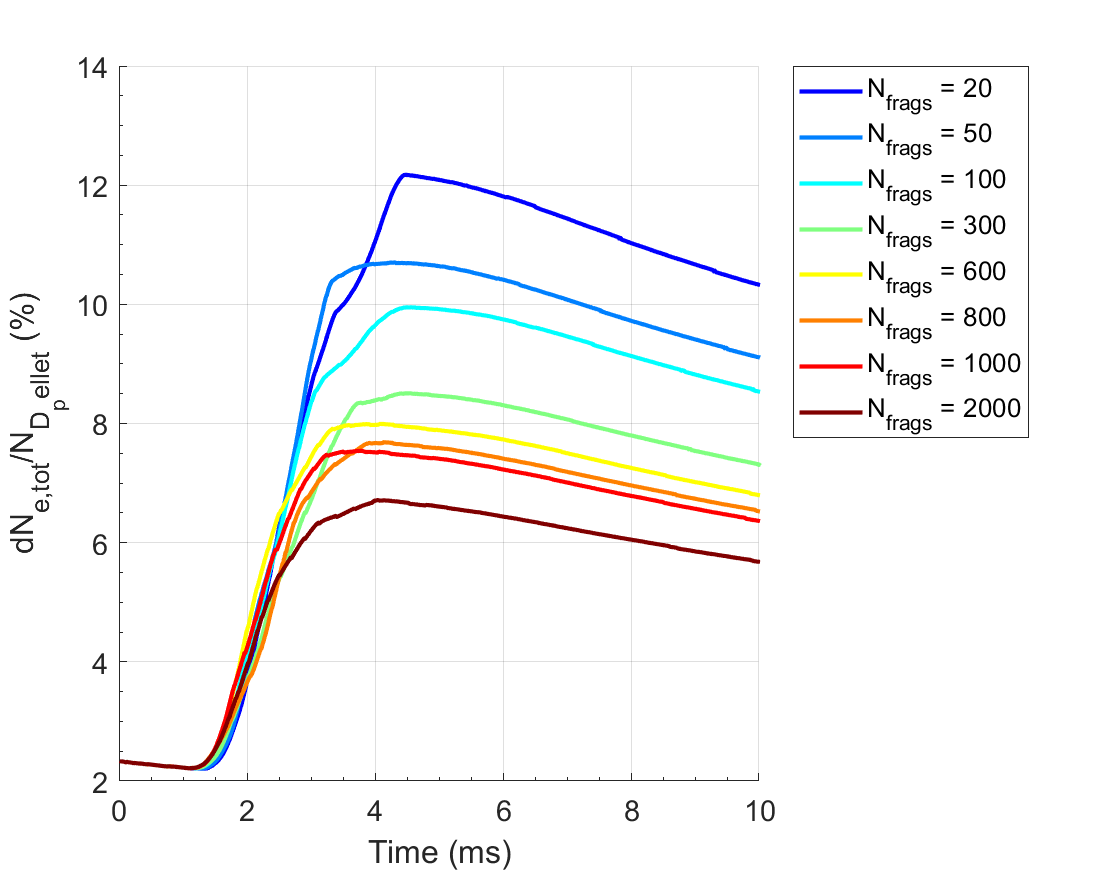}
         \caption{Fragment sizes.}                 \label{fig:deuterium_size_var_absolute assim}         
     \end{subfigure}
     \begin{subfigure}[t]{0.45\textwidth}
         \centering
         \includegraphics[width=\textwidth]{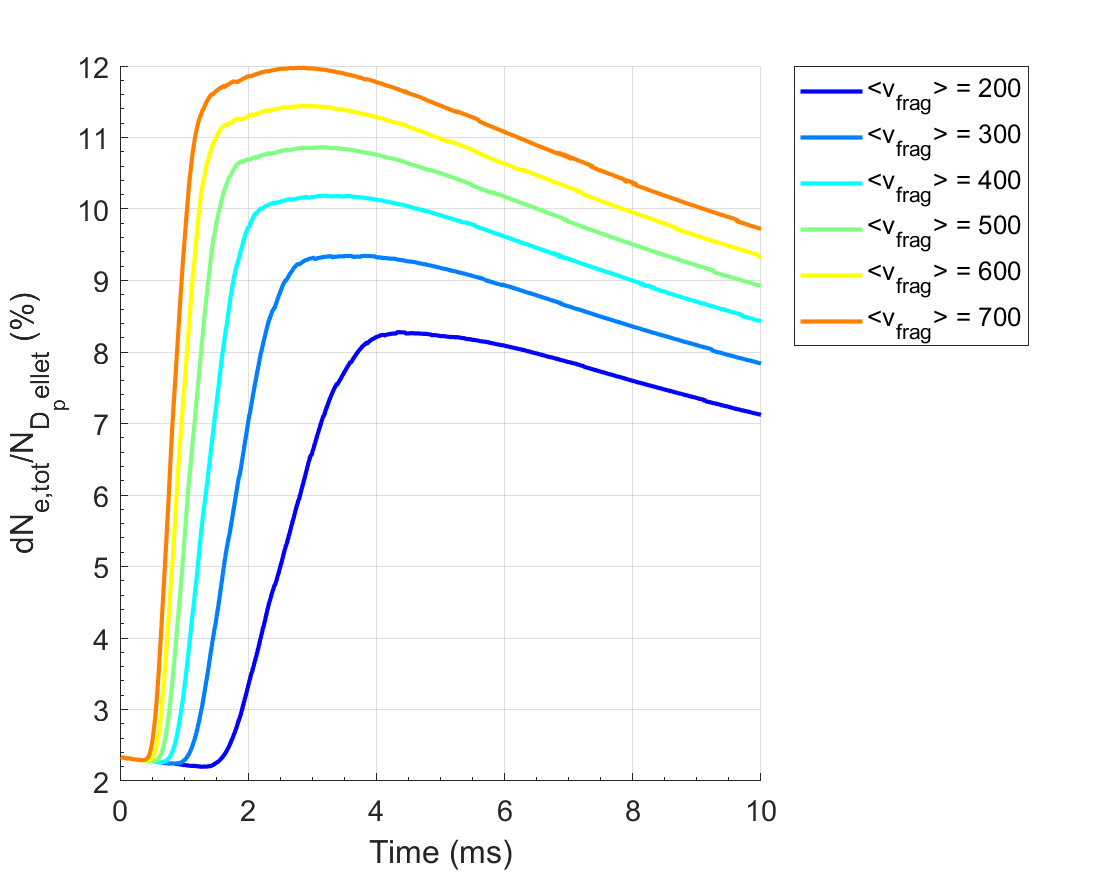}
         \caption{Fragment speeds.} 
         \label{fig:deuterium_speed_var_absolute assim} 
     \end{subfigure}
\caption{Simulated absolute electron assimilation fraction for varying fragment sizes and speeds.}
\label{fig:deuterium_size_speed_var_absolute assim}
\end{figure}

For assimilation comparison with the simulations, the absolute assimilation for the set of pure deuterium injection simulations carried out in \autoref{sssec:frag_size_results_pure_D} and \autoref{sssec:frag_speed_results_pure_D} can be referred to. While the trends shown in the aforementioned sections indicate larger and faster fragments being better for assimilation, a quantitative comparison of the assimilation fraction can be carried out. Hence, the assimilation fraction for the size and speed scans from the synthetic interferometer diagnostic is shown in \autoref{fig:deuterium_size_speed_var_absolute assim}. 

\begin{figure}[h]
    \centering
    \includegraphics[width = 0.4\linewidth]{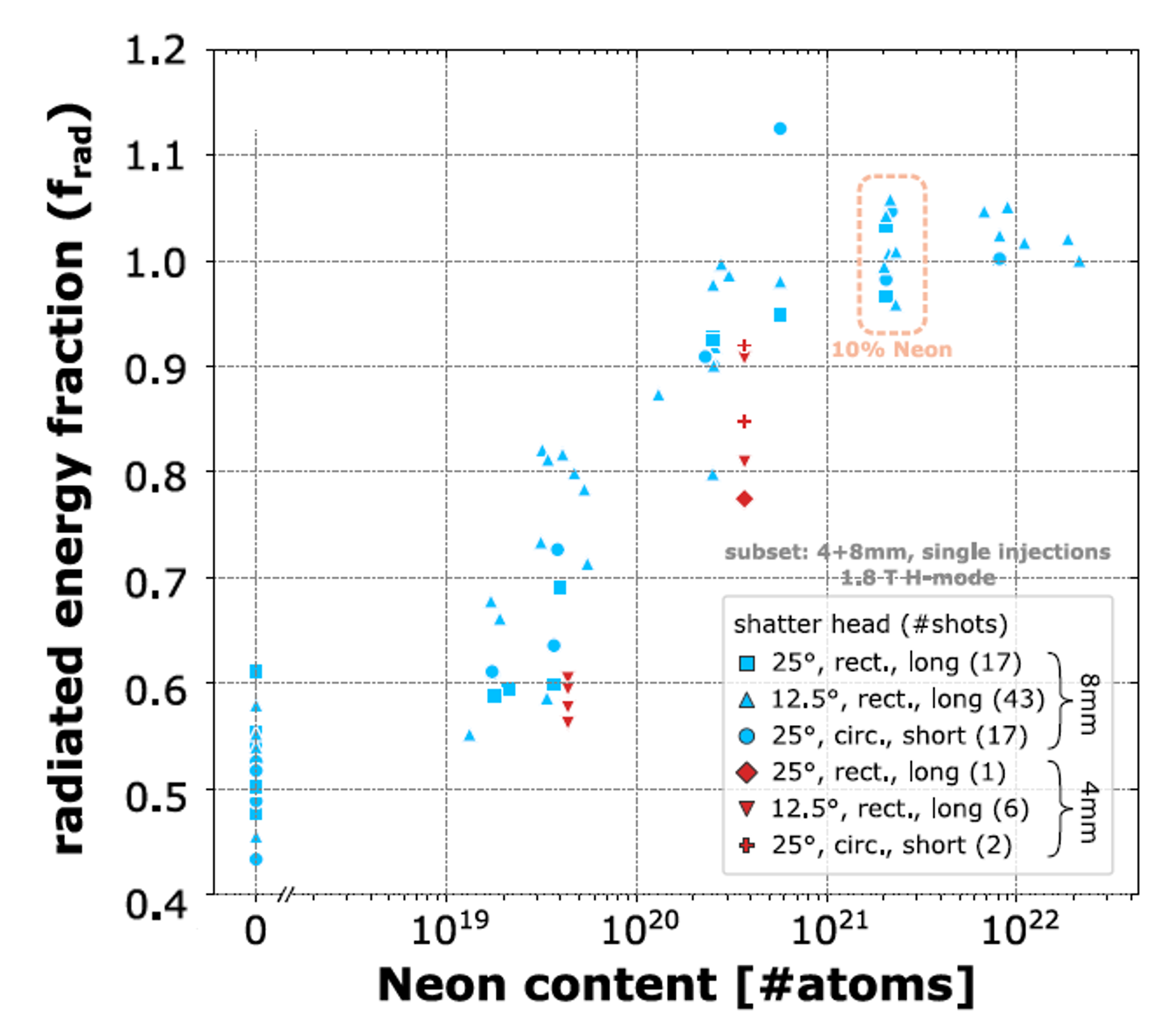}
    \caption{Radiated energy fraction ($f_\text{rad}$) as a function of the injected neon atoms. Different colors indicate the pellet diameter and different markers indicate the shattering tube used for the injection. $f_\text{rad}$ increases with increasing neon concentration. Figure taken from P. Heinrich et. al. \cite{paul_heinrich_analysis_2023}.} 
    \label{fig:Paulfrad}
\end{figure}

Assimilation measurements from the synthetic interferometer signals match the experimental trends of larger and faster fragments being better for higher assimilation. However, a striking difference in the absolute assimilation fraction between the simulations and the experiments is that the maximum experimental assimilation fractions lie between 25\% and 75\% while the simulated assimilation fraction lie between 6\% and 12\%. The main reason behind this significant discrepancy is that the experimental assimilation measurements take into account the peak density rise in the interferometer signal which has been preliminary observed to $\sim$ 2-8 times larger than the value after the peak rise (refer to \autoref{ssec:backavgTS}) although a more rigorous experimental analysis is being carried out by the AUG and ITER teams. The extent of density rise for fragments with different sizes or speeds might also be affected by this procedure. Regardless, the initial experimental analysis suggests larger and faster fragments being better for increased assimilation as also observed in the simulations. \\\\
For mixed D/Ne injections, comparison with qualitative assessments of neon assimilation can be carried out. S. Jachmich et. al. \cite{s_jachmich_shattered_2023} used the plasma current decay rate in the CQ phase to study the assimilated neon in the plasma. Higher injected neon amounts were reported to lead to faster current decay indicating a higher amount of assimilated neon which was also observed in simulations (see also \autoref{fig:xmol_var_absolute_neon_scatter}). First results on the radiated energy fraction ($f_\text{rad}$) for SPI induced disruptions had been reported for AUG \cite{paul_heinrich_analysis_2023}. The radiated energy fraction was defined as
\begin{equation}
    \mathrm{f}_{\text {rad }}=\frac{\mathrm{W}_{\text {rad }}}{\mathrm{W}_{\text {mag }}+\mathrm{W}_{\text {th }}-\mathrm{W}_{\text {coupled }}+\mathrm{W}_{\text {heating }}}
\end{equation}
and should ideally be close to 1 for an efficient mitigation scheme. Here, $W_\text{rad}$ is the radiated energy, $W_\text{mag}$ is the magnetic energy, $W_\text{th}$ is the thermal energy, $W_\text{coupled}$ is the magnetic energy coupled with the surrounding structures and $W_{heating}$ accounts for external heating sources. 
The results from the 2022 AUG experimental campaign  are shown in \autoref{fig:Paulfrad} and indicate that the radiation fraction increases with increasing amount of injected neon, however, for $n_\text{atoms,Ne}> 10^{21}$, most of the plasma thermal energy is radiated and the curve saturates \cite{paul_heinrich_analysis_2023}. A similar trend of simulated neon assimilation in the pre-TQ phase can be observed in \autoref{fig:xmol_var_absolute_neon_scatter}. It should be noted that the plasma dynamics during the TQ and CQ phase might influence the neon assimilation and hence the radiated energy fraction. By carrying out full disruption simulations in INDEX, a more direct comparison can be made. 

\subsection{Pre-TQ duration}
\begin{figure}[H]
    \centering
    \includegraphics[width = 0.7\linewidth]{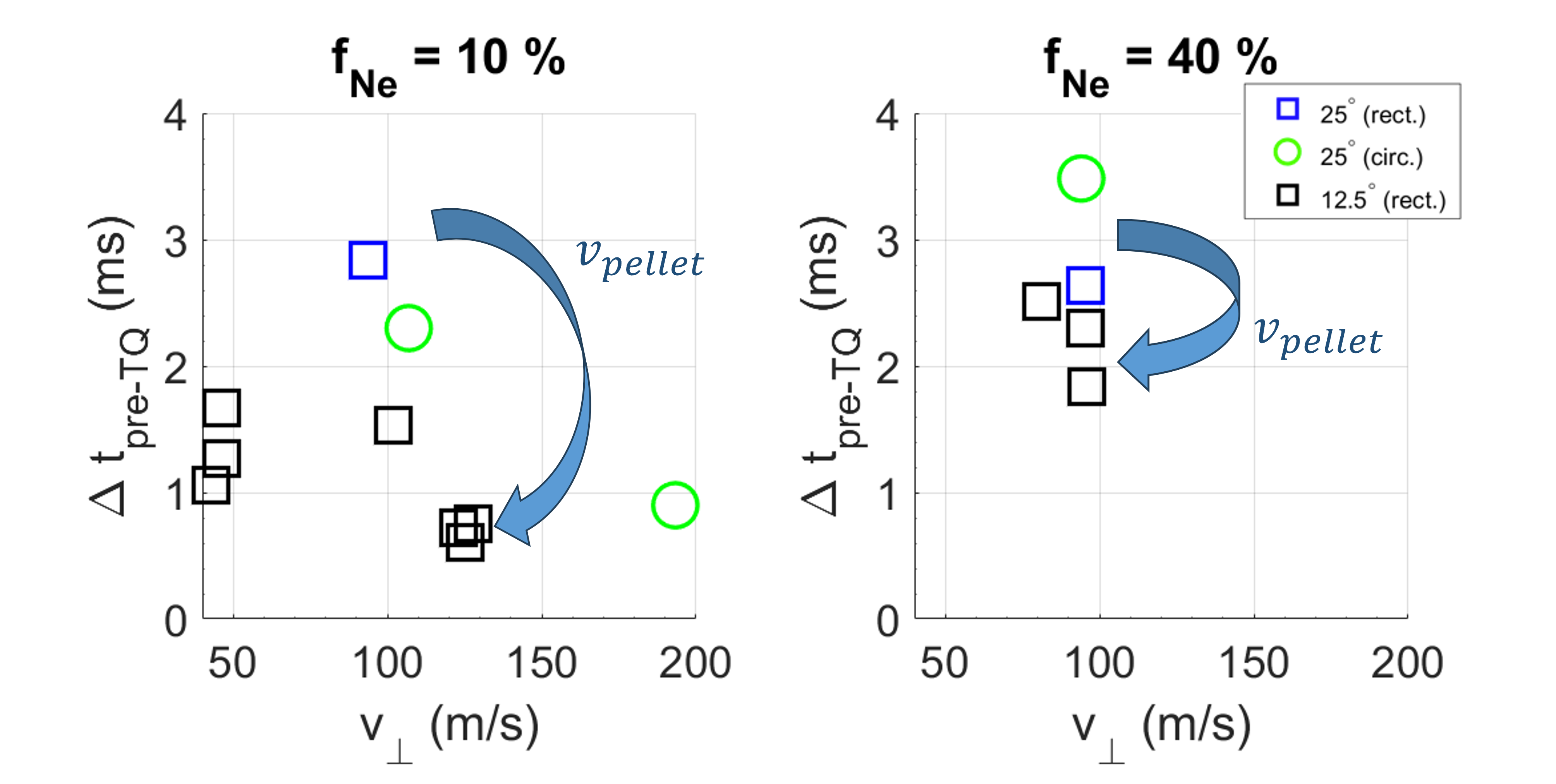}
    \caption{Experimental pre-TQ duration for varying perpendicular velocity for 10\% neon and 40\% neon pellets. $v_{pellet}$ is the pre-shattered pellet speed. Larger pellet speed ($\sim$ mean fragment speed) leads to a shorter pre-TQ duration.}
    \label{fig:v_normal_vs_Pre_TQ_all_neon_subplots}
\end{figure}

The pre-TQ duration is an important parameter for the DMS as a long pre-TQ duration can enable multiple injections to meet the material assimilation requirements at ITER. Hence, the pre-TQ duration for AUG SPI discharges was compared to the simulated pre-TQ duration for different fragment sizes and speeds. For the simulated discharges, the pre-TQ duration is defined in \autoref{ssec:exampleSim} and is related to the cold front reaching the $q=2$ surface. The formation of a cold front in simulations is only found in simulated mixed D/Ne injections and therefore, the pre-TQ duration can only compared for these injections. While qualitative trends in pre-TQ duration are compared, it should be noted that quantitative comparison might require MHD simulations to take into account the delay from MHD mode onset (refer to \autoref{sec:theoreticalback}) to grow enough to lead to a TQ. \\\\
For the mixed Ne/D injections, I have plotted the pre-TQ duration in  \autoref{fig:v_normal_vs_Pre_TQ_all_neon_subplots} for different fragment speeds and similar fragment sizes that were available for 10\% and 40\% mixed D/Ne injections. At similar impact velocity, higher pre-shattering pellet speeds ($\sim$ higher fragment speeds) lead to a shorter pre-TQ duration. A shorter pre-TQ duration for faster fragments has been observed in simulations in \autoref{ssec:frag_size_results_mixed_large_neon}, particularly in \autoref{fig:frag_speed_var_pre_TQ_duration}. The absolute value of the pre-TQ duration in the experiments differ from those found in the simulations however an exact comparison is difficult because of (a) the specific TQ onset condition used in the simulations, (b) differences in the modelled and experimental fragment size and speed distributions \cite{peherstorfer_fragmentation_2022} that modify the time taken by the cold front to reach the $q=2$ surface, (c) time taken by MHD activity growth for the TQ onset. Nevertheless, the parametric scans in this thesis can reproduce the shorter pre-TQ trend for faster fragments.

\begin{figure}[H]
    \centering
    \includegraphics[width = 0.5\linewidth]{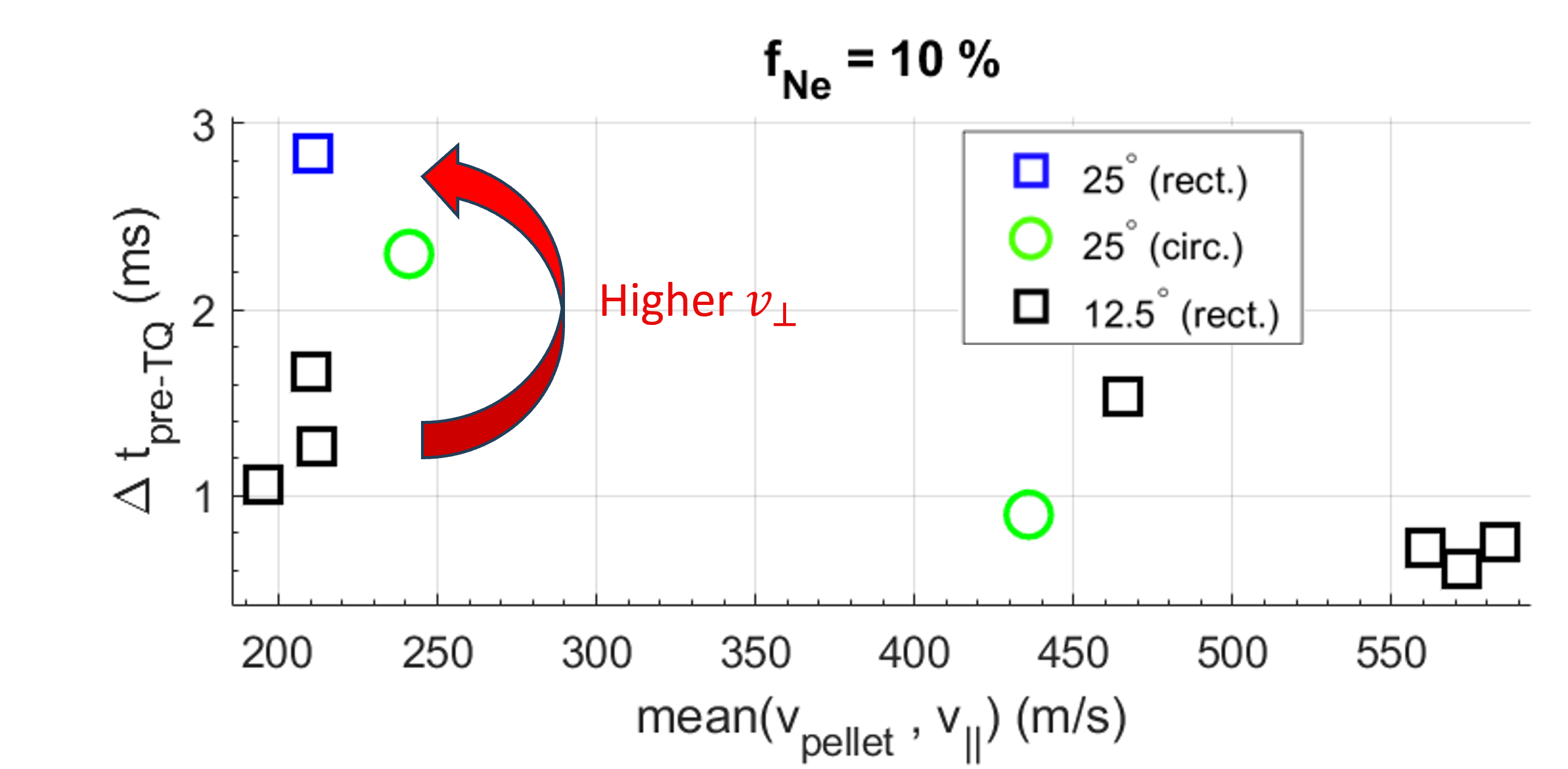}
    \caption{Experimental pre-TQ duration for varying mean fragment velocity for 10\% neon pellets. Larger impact velocity ($\sim$ smaller fragments) lead to a longer pre-TQ duration.}
    \label{fig:v_para_vs_Pre_TQ_all_neon_subplots}
\end{figure}

To assess the effect of fragment sizes, I have plotted the pre-TQ duration against the mean fragment velocity in \autoref{fig:v_para_vs_Pre_TQ_all_neon_subplots}. Comparable injections with similar fragment sizes but different fragment speeds were available for 10\% neon mixed deuterium/neon injections. 
The mean fragment velocity is assumed to be the mean of the pellet velocity and the parallel component of the pellet velocity with respect to the shattering tube as per the discussion in \autoref{ssec:AUGSPI}. For the same mean fragment velocity, a higher impact velocity $v_\perp$ i.e. smaller fragments lead to a longer pre-TQ duration. This is in contrast to the simulated pre-TQ duration for varying fragment sizes as shown in \autoref{fig:frag_size_var_preTQ_duration} where the pre-TQ duration is longer for larger fragments. Two possible reasons can be attributed to the contrasting trends of pre-TQ duration. First, shattering parameters leading to large fragment sizes in experiments (lower $v_\perp$) were often accompanied with a significantly higher number of small fragments as compared to the theoretical fragmentation model \cite{peherstorfer_fragmentation_2022} used in present simulations. In the presence of smaller fragments, which ablate quicker, the edge cooling might be accelerated leading to modification in the pre-TQ duration. Hence, the fragmentation model itself might have limited validity in the case of low impact velocities. Second, larger fragments might perturb the plasma more significantly due to more localized cooling, leading to stronger MHD activity and causing a disruption quicker in the experiments \cite{hu_collisional-radiative_2023}. To assess the impact of both these parameters, additional inputs are required from pellet shattering experimental analysis and 3D MHD modelling activities. \\\\
The comparisons above were carried out with the experimental data points at 200 m/s due to the simulations carried out in \autoref{section:frag_speed_var_mix_pellets} having similar mean fragment velocities. However, at $\sim$ 450 m/s in \autoref{fig:v_para_vs_Pre_TQ_all_neon_subplots}, a set of data points (green circle, black square) indicate that larger fragments lead to a longer pre-TQ, in line with the simulations. The possibility of these points being valid data points (and not statistical outliers) have to be considered motivating the requirement of more experimental data points.

\section{Summary and outlook}
I employed the 1.5D INDEX code in this thesis to model the plasma response before the thermal quench in the ASDEX Upgrade tokamak during shattered pellet injections. I studied the influence of three critical SPI parameters: fragment sizes, velocities, and pellet composition, with a particular focus on their impact on the penetration and assimilation of the injected material within the plasma. To simulate the effects of plasmoid drift during pure deuterium injections, a back-averaging model was used. I carried out interpretive simulations to determine the back-averaging parameter by matching assimilation measurements from experimental and synthetic Thomson scattering measurements. \\\\
In the context of mixed deuterium/neon injections, it is observed that larger and faster fragments result in increased penetration depth and higher material assimilation for an equal quantity of injected material. By applying a semi-empirical onset condition for thermal quench (TQ), the pre-TQ duration can be estimated. For varying fragment sizes, the pre-TQ duration is longer with increasing fragment sizes. Conversely, when investigating varying fragment speeds, faster fragments correspond to a shorter pre-TQ duration. The phenomenon of larger fragments leading to a longer pre-TQ phase can be attributed to their greater surface-area-to-volume ratio, enabling them to traverse deeper into the plasma and deposit more material before the plasma cools down, resulting in higher assimilation and an extended pre-TQ duration. On the other hand, faster fragments can achieve deeper penetration and higher assimilation compared to their slower counterparts, even for a shorter pre-TQ duration. While experimental data of the pre-TQ duration aligns with simulation results for variations in fragment speeds, there is an inconsistency for variations in fragment sizes. A possible reason for this mismatch may stem from disparities between the experimental and modelled fragment size distributions or MHD effects. Quantitative comparisons of the pre-TQ duration might require MHD modelling to take into account the time taken by destabilized MHD modes to grow and lead to the TQ.  
\\\\
For pure deuterium injections, I carried out simulations assuming the same back-averaging parameter for all fragment sizes and speeds. In line with the mixed deuterium/neon injections, larger and faster fragments lead to higher penetration and assimilation although the material assimilation is initially limited to the plasma edge. The core plasma density increases later on a diffusive time scale after most of the material has assimilated. Experimental comparisons of the material penetration in pure deuterium injections are in line with the simulated trends. Trends of experimental material assimilation also align qualitatively, with larger and faster fragments leading to higher assimilation. However, the absolute experimental assimilation values appear roughly 3-6 times higher than the simulated assimilation. The main source of this discrepancy can arise from the limitation in the experimental assimilation values. More accurate matching of assimilation trends requires further analysis of the experimental data. \\\\
I also carried out simulations with variations in the pellet composition.  Neon assimilation increased for increasing amounts of injected neon content. However, a self-regulating saturation of the neon assimilation was observed for neon content larger than $10^{21}$ atoms. The underlying cause of this effect is higher amounts of neon leading to accelerated cooling of the plasma, which, in turn, imposes limitations on further neon assimilation. Comparisons between simulated neon assimilation and indirect experimental measurements of neon assimilation were carried out. Experimental CQ decay rates increased with increasing neon content indicating higher neon assimilation as seen in the simulations. I noticed a similar trend for the experimental radiated energy fraction, which also  increased with higher neon content. The pre-TQ duration monotonically decreased with increasing neon concentration. Trace neon injections ($<10^{21}$ atoms) lead to an inside out temperature collapse in contrast with higher neon concentration injections where the plasma edge cooled down first. As a result, the highest amount of plasma energy was radiated before the TQ onset for trace neon injections. An important limitation for the trace neon injections is that the plasmoid drift can limit the material assimilation to the edge, however it wasn't included for these simulations and can affect the inside-out temperature collapse phenomenon. This limitation also motivates the use of improved plasmoid drift models. \\\\

As mentioned throughout the thesis, there are various other additional activities that will improve the simulations and matching with experiments to quantify the effect of different fragment sizes and speeds. These effects are summarised and further elaborated upon below:

\begin{itemize}
    \item \textbf{Experimental vs modelled fragment size distribution}\\
    As pointed out in \autoref{ssec:sim_exp_comp}, disparities between the fragment sizes in experimental and modelled distributions have been found \cite{peherstorfer_fragmentation_2022}. In the case of pellets with low impact velocities, the modelled fragment size distribution underestimates the amount of small fragments. As smaller fragments ablate faster as discussed in \autoref{ssec:frag_size_results_mixed_large_neon}, an underestimation of smaller sized fragments can affect the experiments differently than simulations leading to changes in assimilation and pre-TQ duration. 
    \item \textbf{Gas production}\\
    During a SPI, there are two possible avenues of gas entering the plasma. If the injector uses propellant gas to accelerate the pellet, as is the case in AUG, it is possible that the propellant gas can enter the vessel before the pellet. Additionally, gas can also be produced during the shattering process. It has already been shown in DIII-D SPI experiments \cite{raman_shattered_2020} that the gas entering the plasma can reduce the plasma energy content and increase the plasma density before the fragments arrive in the plasma. The effect of gas production can be taken into account in the simulations to further improve matching with experimental pre-TQ duration. 
    \item \textbf{Plasmoid drift}\\
    INDEX uses the back-averaging model to simulate the plasmoid drift affecting pure deuterium injections. A fixed back-averaging parameter $\beta$ has been used for all fragment sizes and speeds in this work. Plasma density measurements during pellet assimilation or inputs from 3D MHD simulations can assist in determining $\beta$ for different fragment sizes and speeds. Improved plasmoid drift models can also lead to more accurate estimates of assimilation for pure deuterium and trace neon injections. 
    \item \textbf{Assimilation measurements}\\
    As discussed in \autoref{ssec:backavgTS}, the present assimilation estimates have been obtained using the peak interferometer signal values. If the peak value corresponds to the line of sight measurement passing through a dense plasmoid which eventually dissipates in the background plasma, then the final assimilation measurements might be more reflective of the assimilated material. To quantitatively validate the assimilation measurements of INDEX, the final experimental density should be estimated. 
    \item \textbf{Current Quench simulations}\\
    To quantify the assimilated neon content in the experiments, CQ simulations of the AUG discharges can be carried out. These simulations can be initialized with a post-TQ plasma with assimilated deuterium and neon. As the plasma current decay rate in the CQ phase is strongly dependent on the impurity radiation rate, the amount of assimilated neon determines the current decay rate. By carrying out a scan of assimilated neon, the experimental current decay rate can be matched with the simulated current decay rate. This process can be carried out for injections with different amounts of assimilated neon to get an estimate of the assimilated neon amount. While some preliminary simulation setup work was carried out in the duration of this thesis in collaboration with Dr. Akinobu Matsuyama, additional setup of plasma control in the CQ needs to be implemented.     
\end{itemize}

The present simulations suggest that larger and faster fragments are beneficial for deeper penetration and higher assimilation for both pure deuterium and mixed deuterium/neon injections. However, for better validation of the INDEX code, quantitative comparisons of material assimilation for different fragment sizes and speeds should be carried out. This will require improved experimental assimilation measurements. Furthermore, CQ simulations can help in assessing the quantitative neon assimilation. Further simulations and experiments are also required to confirm the effect of different fragment sizes in determining the trends of pre-TQ duration for mixed deuterium/neon injections. 
After validating the INDEX code using AUG results on the effect of different fragment sizes, speeds and pellet composition, further simulations of different SPI parameters can be carried out for ITER plasma discharge scenarios. However, quantitative extrapolations of the aforementioned parameters to ITER depends on various other parameters, some of which are mentioned below:

\begin{itemize}
    \item \textbf{Thermal energy content} \\
    The assimilation of material in the plasma depends strongly on the thermal energy content of the plasma due to the dependence of the ablation rate on the plasma temperature. It has been reported in AUG SPI discharges as well \cite{s_jachmich_shattered_2023} that a higher thermal energy content leads to an increased assimilation. Hence, optimal fragment fragment parameters might depend on the target plasma. For this purpose, the INDEX code should be validated on different tokamaks with different thermal energy content. 
    \item \textbf{Pre-existing MHD activity} \\
    In the simulations carried out in this thesis, SPI was applied to an H-mode "healthy plasma" without any MHD activity due to a higher availability of experimental data to compare to. However, in practice, the DMS might be triggered when the plasma already has MHD activity and/or might have transitioned to L-mode \cite{hender_chapter_2007}. Wieschollek et. al. \cite{wieschollek_role_2022} studied the role of exiting MHD activity on plasma dynamics during SPI in AUG using JOREK and showed that pre-existing large magnetic islands might affect the pre-TQ duration. 
\end{itemize}
\newpage
\addcontentsline{toc}{section}{Bibliography}
\printbibliography

@article{vallhagen_drift_2023,
	title = {Drift of ablated material after pellet injection in a tokamak},
	volume = {89},
	url = {https://doi.org/10.1017/S0022377823000466},
	doi = {10.1017/S0022377823000466},
	language = {en},
	number = {3},
	urldate = {2023-10-15},
	journal = {Journal of Plasma Physics},
	author = {Vallhagen, O. and Pusztai, I. and Helander, P. and Newton, S. L. and Fülöp, T.},
	month = jun,
	year = {2023},
	note = {Publisher: Cambridge University Press},
	pages = {905890306},
}

@article{stroth_progress_2022,
	title = {Progress from {ASDEX} {Upgrade} experiments in preparing the physics basis of {ITER} operation and {DEMO} scenario development},
	volume = {62},
	issn = {0029-5515},
	url = {https://dx.doi.org/10.1088/1741-4326/ac207f},
	doi = {10.1088/1741-4326/ac207f},
	language = {en},
	number = {4},
	urldate = {2023-10-19},
	journal = {Nuclear Fusion},
	author = {Stroth, U. and Aguiam, D. and Alessi, E. and Angioni, C. and Arden, N. and Parra, R. Arredondo and Artigues, V. and Asunta, O. and Balden, M. and Bandaru, V. and Banon-Navarro, A. and Behler, K. and Bergmann, A. and Bergmann, M. and Bernardo, J. and Bernert, M. and Biancalani, A. and Bielajew, R. and Bilato, R. and Birkenmeier, G. and Blanken, T. and Bobkov, V. and Bock, A. and Body, T. and Bolzonella, T. and Bonanomi, N. and Bortolon, A. and Böswirth, B. and Bottereau, C. and Bottino, A. and Brand, H. van den and Brenzke, M. and Brezinsek, S. and Brida, D. and Brochard, F. and Bruhn, C. and Buchanan, J. and Buhler, A. and Burckhart, A. and Camenen, Y. and Cannas, B. and Megias, P. Cano and Carlton, D. and Carr, M. and Carvalho, P. and Castaldo, C. and Cavedon, M. and Cazzaniga, C. and Challis, C. and Chankin, A. and Cianfarani, C. and Clairet, F. and Coda, S. and Coelho, R. and Coenen, J. W. and Colas, L. and Conway, G. and Costea, S. and Coster, D. and Cote, T. and Creely, A. J. and Croci, G. and Zabala, D. J. Cruz and Cseh, G. and Czarnecka, A. and Cziegler, I. and D’Arcangelo, O. and Molin, A. Dal and David, P. and Day, C. and Baar, M. de and Marné, P. de and Delogu, R. and Denk, S. and Denner, P. and Siena, A. Di and Durán, J. J. Dominguez Palacios and Dunai, D. and Drenik, A. and Dreval, M. and Drube, R. and Dunne, M. and Duval, B. P. and Dux, R. and Eich, T. and Elgeti, S. and Encheva, A. and Engelhardt, K. and Erdös, B. and Erofeev, I. and Esposito, B. and Fable, E. and Faitsch, M. and Fantz, U. and Farnik, M. and Faugel, H. and Felici, F. and Ficker, O. and Fietz, S. and Figueredo, A. and Fischer, R. and Ford, O. and Frassinetti, L. and Fröschle, M. and Fuchert, G. and Fuchs, J. C. and Fünfgelder, H. and Futatani, S. and Galazka, K. and Galdon-Quiroga, J. and Escolà, D. Gallart and Gallo, A. and Gao, Y. and Garavaglia, S. and Muñoz, M. Garcia and Geiger, B. and Giannone, L. and Gibson, S. and Gil, L. and Giovannozzi, E. and Glöggler, S. and Gobbin, M. and Martin, J. Gonzalez and Goodman, T. and Gorini, G. and Görler, T. and Gradic, D. and Granucci, G. and Gräter, A. and Greuner, H. and Griener, M. and Groth, M. and Gude, A. and Guimarais, L. and Günter, S. and Haas, G. and Hakola, A. H. and Ham, C. and Happel, T. and Harder, N. den and Harrer, G. and Harrison, J. and Hauer, V. and Hayward-Schneider, T. and Heinemann, B. and Hellsten, T. and Henderson, S. and Hennequin, P. and Herrmann, A. and Heyn, E. and Hitzler, F. and Hobirk, J. and Höfler, K. and Holm, J. H. and Hölzl, M. and Hopf, C. and Horvath, L. and Höschen, T. and Houben, A. and Hubbard, A. and Huber, A. and Hunger, K. and Igochine, V. and Iliasova, M. and Ilkei, T. and Björk, K. Insulander and Ionita-Schrittwieser, C. and Ivanova-Stanik, I. and Jacob, W. and Jaksic, N. and Janky, F. and Vuuren, A. Jansen van and Jardin, A. and Jaulmes, F. and Jenko, F. and Jensen, T. and Joffrin, E. and Kallenbach, A. and Kálvin, S. and Kantor, M. and Kappatou, A. and Kardaun, O. and Karhunen, J. and Käsemann, C.-P. and Kasilov, S. and Kendl, A. and Kernbichler, W. and Khilkevitch, E. and Kirk, A. and Hansen, S. Kjer and Klevarova, V. and Kocsis, G. and Koleva, M. and Komm, M. and Kong, M. and Krämer-Flecken, A. and Krieger, K. and Krivska, A. and Kudlacek, O. and Kurki-Suonio, T. and Kurzan, B. and Labit, B. and Lackner, K. and Laggner, F. and Lahtinen, A. and Lang, P. T. and Lauber, P. and Leuthold, N. and Li, L. and Likonen, J. and Linder, O. and Lipschultz, B. and Liu, Y. and Lohs, A. and Lu, Z. and Cortemiglia, T. Luda di and Luhmann, N. C. and Lunt, T. and Lyssoivan, A. and Maceina, T. and Madsen, J. and Magnanimo, A. and Maier, H. and Mailloux, J. and Maingi, R. and Maj, O. and Maljaars, E. and Manas, P. and Mancini, A. and Manhard, A. and Mantica, P. and Mantsinen, M. and Manz, P. and Maraschek, M. and Marchetto, C. and Marrelli, L. and Martin, P. and Martitsch, A. and Matos, F. and Mayer, M. and Mayoral, M.-L. and Mazon, D. and McCarthy, P. J. and McDermott, R. and Merkel, R. and Merle, A. and Meshcheriakov, D. and Meyer, H. and Milanesio, D. and Cabrera, P. Molina and Monaco, F. and Muraca, M. and Nabais, F. and Naulin, V. and Nazikian, R. and Nem, R. D. and Nemes-Czopf, A. and Neu, G. and Neu, R. and Nielsen, A. H. and Nielsen, S. K. and Nishizawa, T. and Nocente, M. and Noterdaeme, J.-M. and Novikau, I. and Nowak, S. and Oberkofler, M. and Ochoukov, R. and Olsen, J. and Orain, F. and Palermo, F. and Pan, O. and Papp, G. and Perez, I. Paradela and Pau, A. and Pautasso, G. and Paz-Soldan, C. and Petersson, P. and Piovesan, P. and Piron, C. and Plank, U. and Plaum, B. and Plöck, B. and Plyusnin, V. and Pokol, G. and Poli, E. and Porte, L. and Pütterich, T. and Ramisch, M. and Rasmussen, J. and Ratta, G. and Ratynskaia, S. and Raupp, G. and Réfy, D. and Reich, M. and Reimold, F. and Reiser, D. and Reisner, M. and Reiter, D. and Ribeiro, T. and Riedl, R. and Riesch, J. and Rittich, D. and Rodriguez, J. F. Rivero and Rocchi, G. and Rodriguez-Fernandez, P. and Rodriguez-Ramos, M. and Rohde, V. and Ronchi, G. and Ross, A. and Rott, M. and Rubel, M. and Ryan, D. A. and Ryter, F. and Saarelma, S. and Salewski, M. and Salmi, A. and Samoylov, O. and Sanchez, L. Sanchis and Santos, J. and Sauter, O. and Schall, G. and Schlüter, K. and Schmid, K. and Schmitz, O. and Schneider, P. A. and Schrittwieser, R. and Schubert, M. and Schuster, C. and Schwarz-Selinger, T. and Schweinzer, J. and Seliunin, E. and Shabbir, A. and Shalpegin, A. and Sharapov, S. and Sheikh, U. and Shevelev, A. and Sias, G. and Siccinio, M. and Sieglin, B. and Sigalov, A. and Silva, A. and Silva, C. and Silvagni, D. and Simpson, J. and Sipilä, S. and Smigelskis, E. and Snicker, A. and Solano, E. and Sommariva, C. and Sozzi, C. and Spizzo, G. and Spolaore, M. and Stegmeir, A. and Stejner, M. and Stober, J. and Strumberge, E. and Lopez, G. Suarez and Sun, H.-J. and Suttrop, W. and Sytova, E. and Szepesi, T. and Tál, B. and Tala, T. and Tardini, G. and Tardocchi, M. and Terranova, D. and Teschke, M. and Thorén, E. and Tierens, W. and Told, D. and Treutterer, W. and Trevisan, G. and Trier, E. and Tripský, M. and Usoltceva, M. and Valisa, M. and Valovic, M. and Zeeland, M. van and Vannini, F. and Vanovac, B. and Varela, P. and Varoutis, S. and Vianello, N. and Vicente, J. and Verdoolaege, G. and Vierle, T. and Viezzer, E. and Voitsekhovitch, I. and Toussaint, U. von and Wagner, D. and Wang, X. and Weiland, M. and White, A. E. and Willensdorfer, M. and Wiringer, B. and Wischmeier, M. and Wolf, R. and Wolfrum, E. and Yang, Q. and Yu, Q. and Zagórski, R. and Zammuto, I. and Zehetbauer, T. and Zhang, W. and Zholobenko, W. and Zilker, M. and Zito, A. and Zohm, H. and Zoletnik, S. and Team, the EUROfusion MST1},
	month = mar,
	year = {2022},
	note = {Publisher: IOP Publishing},
	pages = {042006},
}

@article{vallhagen_effect_2022,
	title = {Effect of two-stage shattered pellet injection on tokamak disruptions},
	volume = {62},
	issn = {0029-5515},
	url = {https://dx.doi.org/10.1088/1741-4326/ac667e},
	doi = {10.1088/1741-4326/ac667e},
	language = {en},
	number = {11},
	urldate = {2023-10-15},
	journal = {Nuclear Fusion},
	author = {Vallhagen, O. and Pusztai, I. and Hoppe, M. and Newton, S. L. and Fülöp, T.},
	month = sep,
	year = {2022},
	note = {Publisher: IOP Publishing},
	pages = {112004},
}

@phdthesis{vallhagen_disruption_2023,
	type = {Licentiate thesis},
	title = {Disruption mitigation in tokamaks with massive material injection},
	url = {https://research.chalmers.se/publication/535522},
	school = {Chalmers University of Technology},
	author = {Vallhagen, Oskar},
	year = {2023},
}

@inproceedings{lehnen_iter_2021,
	address = {PPPL Workshop on Theory \& Simulations of Disruptions},
	title = {The {ITER} disruption mitigation system—{Design} progress and design validation},
	author = {Lehnen, M},
	year = {2021},
	pages = {19--23},
}

@article{kocsis_fast_2004,
	title = {A fast framing camera system for observation of acceleration and ablation of cryogenic hydrogen pellet in {ASDEX} {Upgrade} plasmas},
	volume = {75},
	issn = {0034-6748},
	url = {https://doi.org/10.1063/1.1808897},
	doi = {10.1063/1.1808897},
	number = {11},
	urldate = {2023-10-15},
	journal = {Review of Scientific Instruments},
	author = {Kocsis, G. and Kálvin, S. and Veres, G. and Cierpka, P. and Lang, P. T. and Neuhauser, J. and Wittman, C. and Team, ASDEX Upgrade},
	month = nov,
	year = {2004},
	pages = {4754--4762},
}

@article{summers_ionization_2006,
	title = {Ionization state, excited populations and emission of impurities in dynamic finite density plasmas: {I}. {The} generalized collisional–radiative model for light elements},
	volume = {48},
	issn = {0741-3335},
	shorttitle = {Ionization state, excited populations and emission of impurities in dynamic finite density plasmas},
	url = {https://dx.doi.org/10.1088/0741-3335/48/2/007},
	doi = {10.1088/0741-3335/48/2/007},
	language = {en},
	number = {2},
	urldate = {2023-10-15},
	journal = {Plasma Physics and Controlled Fusion},
	author = {Summers, H. P. and Dickson, W. J. and O'Mullane, M. G. and Badnell, N. R. and Whiteford, A. D. and Brooks, D. H. and Lang, J. and Loch, S. D. and Griffin, D. C.},
	month = jan,
	year = {2006},
	pages = {263},
}

@article{gruber_vertical_1993,
	title = {Vertical displacement events and halo currents},
	volume = {35},
	issn = {0741-3335},
	url = {https://dx.doi.org/10.1088/0741-3335/35/SB/015},
	doi = {10.1088/0741-3335/35/SB/015},
	language = {en},
	number = {SB},
	urldate = {2023-10-15},
	journal = {Plasma Physics and Controlled Fusion},
	author = {Gruber, O. and Lackner, K. and Pautasso, G. and Seidel, U. and Streibl, B.},
	month = dec,
	year = {1993},
	pages = {B191},
}

@inproceedings{shiraki_particle_2020,
	address = {IAEA Technical Meeting on Plasma Disruptions and their Mitigation (2020)},
	title = {Particle assimilation during shattered pellet injection},
	url = {https://conferences.iaea.org/event/217/contributions/16713/attachments/9356/12887/Shiraki_SPI_Assimilation.pdf},
	author = {Shiraki, Daisuke and Herfindal, J and Baylor, LR and Hollmann, EM and Lasnier, C and Bykov, I and Eidietis, N and Raman, R and Sweeney, R and Sheikh, U},
	month = jul,
	year = {2020},
}

@article{muller_high-_1999,
	title = {High- β {Plasmoid} {Drift} during {Pellet} {Injection} into {Tokamaks}},
	volume = {83},
	url = {https://link.aps.org/doi/10.1103/PhysRevLett.83.2199},
	doi = {10.1103/PhysRevLett.83.2199},
	number = {11},
	urldate = {2023-06-05},
	journal = {Physical Review Letters},
	author = {Müller, H. W. and Büchl, K. and Kaufmann, M. and Lang, P. T. and Lang, R. S. and Lorenz, A. and Maraschek, M. and Mertens, V. and Neuhauser, J. and {ASDEX Upgrade Team}},
	month = sep,
	year = {1999},
	note = {Publisher: American Physical Society},
	pages = {2199--2202},
}

@article{muller_high_2002,
	title = {High β plasmoid formation, drift and striations during pellet ablation in {ASDEX} {Upgrade}},
	volume = {42},
	issn = {0029-5515, 1741-4326},
	url = {https://iopscience.iop.org/article/10.1088/0029-5515/42/3/311},
	doi = {10.1088/0029-5515/42/3/311},
	language = {en},
	number = {3},
	urldate = {2022-12-23},
	journal = {Nuclear Fusion},
	author = {Müller, H.W and Dux, R and Kaufmann, M and Lang, P.T and Lorenz, A and Maraschek, M and Mertens, V and Neuhauser, J and Team, ASDEX Upgrade},
	month = mar,
	year = {2002},
	pages = {301--309},
}

@article{dibon_design_2023,
	title = {Design of the shattered pellet injection system for {ASDEX} {Upgrade}},
	volume = {94},
	issn = {0034-6748},
	url = {https://doi.org/10.1063/5.0141799},
	doi = {10.1063/5.0141799},
	number = {4},
	urldate = {2023-10-14},
	journal = {Review of Scientific Instruments},
	author = {Dibon, M. and de Marne, P. and Papp, G. and Vinyar, I. and Lukin, A. and Jachmich, S. and Kruezi, U. and Muir, A. and Rohde, V. and Lehnen, M. and Heinrich, P. and Peherstorfer, T. and Podymskii, D. and {ASDEX Upgrade Team}},
	month = apr,
	year = {2023},
	pages = {043504},
}

@inproceedings{papp_asdex_2020,
	address = {International Atomic Energy Agency (IAEA)},
	title = {{ASDEX} {Upgrade} {SPI}: design, status and plans},
	url = {http://inis.iaea.org/search/search.aspx?orig_q=RN:52097937},
	author = {Papp, Gergely and Dibon, Mathias and Bernert, Matthias and Eberl, Thomas and Lunt, Tilmann and Pautasso, Gabriella and Hoelzl, Matthias and Herrmann, Albrecht and Lehnen, Michael and Jachmich, Stefan and Kruezi, Uron},
	year = {2020},
	note = {INIS-XA--21M2166},
	pages = {2--3},
}

@inproceedings{peter_halldestam_modeling_2023,
	address = {Max Planck Institute for Plasma Physics, Garching, Germany},
	title = {Modeling of {SPI}-mitigated disruptions in {ASDEX} {Upgrade}},
	booktitle = {10th {Runaway} {Electron} {Modelling} ({REM}) meeting},
	author = {{Peter Halldestam}},
	month = jun,
	year = {2023},
}

@article{rozhansky_evolution_1995,
	title = {Evolution and stratification of a plasma cloud surrounding a pellet},
	volume = {37},
	issn = {0741-3335},
	url = {https://dx.doi.org/10.1088/0741-3335/37/4/003},
	doi = {10.1088/0741-3335/37/4/003},
	language = {en},
	number = {4},
	urldate = {2023-10-14},
	journal = {Plasma Physics and Controlled Fusion},
	author = {Rozhansky, V. and Veselova, I. and Voskoboynikov, S.},
	month = apr,
	year = {1995},
	pages = {399},
}

@article{nardon_origin_2023,
	title = {On the origin of the plasma current spike during a tokamak disruption and its relation with magnetic stochasticity},
	volume = {63},
	issn = {0029-5515},
	url = {https://dx.doi.org/10.1088/1741-4326/acc417},
	doi = {10.1088/1741-4326/acc417},
	language = {en},
	number = {5},
	urldate = {2023-10-14},
	journal = {Nuclear Fusion},
	author = {Nardon, E. and Särkimäki, K. and Artola, F. J. and Sadouni, S. and team, the JOREK and Contributors, J. E. T.},
	month = mar,
	year = {2023},
	note = {Publisher: IOP Publishing},
	pages = {056011},
}

@article{koppendorfer_asdex_1986,
	title = {The {ASDEX} {Upgrade} toroidal field magnet and poloidal divertor field coil system adapted to reactor requirements},
	volume = {3},
	issn = {0167-899X},
	url = {https://www.sciencedirect.com/science/article/pii/S0167899X86800179},
	doi = {10.1016/S0167-899X(86)80017-9},
	number = {3},
	urldate = {2023-10-12},
	journal = {Nuclear Engineering and Design. Fusion},
	author = {Köppendörfer, W. and Blaumoser, M. and Ennen, K. and Gruber, J. and Gruber, O. and Jandl, O. and Kaufmann, M. and Kollotzek, H. and Kotzlowski, H. and Lackner, E. and Lackner, K. and Von Larcher, T. and Noterdaeme, J. M. and Pillsticker, M. and Pöhlchen, R. and Preis, H. and Schneider, H. and Seidel, U. and Sombach, B. and Speth, E. and Streibl, B. and Vernickel, H. and Werner, F. and Wesner, F. and Wieczorek, A.},
	month = jan,
	year = {1986},
	pages = {265--272},
}

@article{lvovskiy_evolution_2022,
	title = {Evolution of {Density} and {Temperature} {Full} {Profiles} after {Pure} {Ne} and {D} 2 {Shattered} {Pellet} {Injections} on {DIII}-{D}},
	journal = {Bulletin of the American Physical Society},
	author = {Lvovskiy, Andrey and Eidietis, Nicholas and O'Gorman, Thomas and Shiraki, Daisuke and Matsuyama, Akinobu and Hollmann, Eric and Herfindal, Jeffery and LEHNEN, Michael and Boivin, Rejean},
	year = {2022},
	note = {Publisher: APS},
}

@article{miyamoto_inter-code_2014,
	title = {Inter-code comparison benchmark between {DINA} and {TSC} for {ITER} disruption modelling},
	volume = {54},
	issn = {0029-5515},
	url = {https://dx.doi.org/10.1088/0029-5515/54/8/083002},
	doi = {10.1088/0029-5515/54/8/083002},
	language = {en},
	number = {8},
	urldate = {2023-10-12},
	journal = {Nuclear Fusion},
	author = {Miyamoto, S. and Isayama, A. and Bandyopadhyay, I. and Jardin, S. C. and Khayrutdinov, R. R. and Lukash, V. E. and Kusama, Y. and Sugihara, M.},
	month = may,
	year = {2014},
	note = {Publisher: IOP Publishing},
	pages = {083002},
}

@article{khayrutdinov_studies_1993,
	title = {Studies of {Plasma} {Equilibrium} and {Transport} in a {Tokamak} {Fusion} {Device} with the {Inverse}-{Variable} {Technique}},
	volume = {109},
	issn = {0021-9991},
	url = {https://www.sciencedirect.com/science/article/pii/S0021999183712118},
	doi = {10.1006/jcph.1993.1211},
	number = {2},
	urldate = {2023-10-12},
	journal = {Journal of Computational Physics},
	author = {Khayrutdinov, R. R. and Lukash, V. E.},
	month = dec,
	year = {1993},
	pages = {193--201},
}

@inproceedings{w_tang_non-linear_2023,
	address = {London, United Kingdom},
	title = {Non-linear shattered pellet injection simulations based on {ASDEX} {Upgrade} experiments},
	booktitle = {{IAEA}-{CN}-316/2430},
	author = {{W, Tang} and {M, Hoelzl} and {M, Lehnen} and {P, Halldestam} and {P, Heinrich} and {G, Papp} and {D, Hu} and {FJ, Artola} and {E, Nardon} and {S, Jachmich} and {ASDEX Upgrade Team} and {JOREK Team}},
	year = {2023},
}

@article{fable_transport_2016,
	title = {Transport simulations of the pre–thermal–quench phase in {ASDEX} {Upgrade} massive gas injection experiments},
	volume = {56},
	issn = {0029-5515},
	url = {https://dx.doi.org/10.1088/0029-5515/56/2/026012},
	doi = {10.1088/0029-5515/56/2/026012},
	language = {en},
	number = {2},
	urldate = {2023-10-11},
	journal = {Nuclear Fusion},
	author = {Fable, E. and Pautasso, G. and Lehnen, M. and Dux, R. and Bernert, M. and Mlynek, A. and Team, the ASDEX Upgrade},
	month = jan,
	year = {2016},
	note = {Publisher: IOP Publishing},
	pages = {026012},
}

@article{reux_experimental_2010,
	title = {Experimental study of disruption mitigation using massive injection of noble gases on {Tore} {Supra}},
	volume = {50},
	issn = {0029-5515},
	url = {https://dx.doi.org/10.1088/0029-5515/50/9/095006},
	doi = {10.1088/0029-5515/50/9/095006},
	language = {en},
	number = {9},
	urldate = {2023-10-11},
	journal = {Nuclear Fusion},
	author = {Reux, C. and Bucalossi, J. and Saint-Laurent, F. and Gil, C. and Moreau, P. and Maget, P.},
	month = jul,
	year = {2010},
	pages = {095006},
}

@article{pautasso_disruption_2009,
	title = {Disruption studies in {ASDEX} {Upgrade} in view of {ITER}},
	volume = {51},
	issn = {0741-3335},
	url = {https://dx.doi.org/10.1088/0741-3335/51/12/124056},
	doi = {10.1088/0741-3335/51/12/124056},
	language = {en},
	number = {12},
	urldate = {2023-10-11},
	journal = {Plasma Physics and Controlled Fusion},
	author = {Pautasso, G. and Coster, D. and Eich, T. and Fuchs, J. C. and Gruber, O. and Gude, A. and Herrmann, A. and Igochine, V. and Konz, C. and Kurzan, B. and Lackner, K. and Lunt, T. and Marascheck, M. and Mlynek, A. and Reiter, B. and Rohde, V. and Zhang, Y. and Bonnin, X. and Beck, M. and Prausner, G. and Team, the ASDEX Upgrade},
	month = nov,
	year = {2009},
	pages = {124056},
}

@article{bozhenkov_generation_2008,
	title = {Generation and suppression of runaway electrons in disruption mitigation experiments in {TEXTOR}},
	volume = {50},
	issn = {0741-3335},
	url = {https://dx.doi.org/10.1088/0741-3335/50/10/105007},
	doi = {10.1088/0741-3335/50/10/105007},
	language = {en},
	number = {10},
	urldate = {2023-10-11},
	journal = {Plasma Physics and Controlled Fusion},
	author = {Bozhenkov, S. A. and Lehnen, M. and Finken, K. H. and Jakubowski, M. W. and Wolf, R. C. and Jaspers, R. and Kantor, M. and Marchuk, O. V. and Uzgel, E. and Wassenhove, G. Van and Zimmermann, O. and Reiter, D. and team, the TEXTOR},
	month = aug,
	year = {2008},
	pages = {105007},
}

@article{granetz_gas_2007,
	title = {Gas jet disruption mitigation studies on {Alcator} {C}-{Mod} and {DIII}-{D}},
	volume = {47},
	issn = {0029-5515},
	url = {https://dx.doi.org/10.1088/0029-5515/47/9/003},
	doi = {10.1088/0029-5515/47/9/003},
	language = {en},
	number = {9},
	urldate = {2023-10-11},
	journal = {Nuclear Fusion},
	author = {Granetz, R. S. and Hollmann, E. M. and Whyte, D. G. and Izzo, V. A. and Antar, G. Y. and Bader, A. and Bakhtiari, M. and Biewer, T. and Boedo, J. A. and Evans, T. E. and Hutchinson, I. H. and Jernigan, T. C. and Gray, D. S. and Groth, M. and Humphreys, D. A. and Lasnier, C. J. and Moyer, R. A. and Parks, P. B. and Reinke, M. L. and Rudakov, D. L. and Strait, E. J. and Terry, J. L. and Wesley, J. and West, W. P. and Wurden, G. and Yu, J.},
	month = aug,
	year = {2007},
	pages = {1086},
}

@article{hollmann_measurements_2005,
	title = {Measurements of impurity and heat dynamics during noble gas jet-initiated fast plasma shutdown for disruption mitigation in {DIII}-{D}},
	volume = {45},
	issn = {0029-5515},
	url = {https://dx.doi.org/10.1088/0029-5515/45/9/003},
	doi = {10.1088/0029-5515/45/9/003},
	language = {en},
	number = {9},
	urldate = {2023-10-11},
	journal = {Nuclear Fusion},
	author = {Hollmann, E. M. and Jernigan, T. C. and Groth, M. and Whyte, D. G. and Gray, D. S. and Austin, M. E. and Bray, B. D. and Brennan, D. P. and Brooks, N. H. and Evans, T. E. and Humphreys, D. A. and Lasnier, C. J. and Moyer, R. A. and McLean, A. G. and Parks, P. B. and Rozhansky, V. and Rudakov, D. L. and Strait, E. J. and West, W. P.},
	month = aug,
	year = {2005},
	pages = {1046},
}

@article{matsuyama_neutral_2022,
	title = {Neutral gas and plasma shielding ({NGPS}) model and cross-field motion of ablated material for hydrogen–neon mixed pellet injection},
	volume = {29},
	issn = {1070-664X},
	url = {https://doi.org/10.1063/5.0084586},
	doi = {10.1063/5.0084586},
	number = {4},
	urldate = {2023-10-11},
	journal = {Physics of Plasmas},
	author = {Matsuyama, Akinobu},
	month = apr,
	year = {2022},
	pages = {042501},
}

@article{pusztai_runaway_2023,
	title = {Runaway electron dynamics in shattered pellet mitigated {ITER} disruptions},
	journal = {Bulletin of the American Physical Society},
	author = {Pusztai, Istvan and Vallhagen, Oskar and Hanebring, Lise and Artola, Javier and Lehnen, Michael and Ekmark, Ida and Fulop, Tunde},
	year = {2023},
	note = {Publisher: APS},
}

@article{pusztai_runaway_2022,
	title = {Runaway dynamics in tokamak disruptions with current relaxation},
	volume = {88},
	issn = {0022-3778, 1469-7807},
	url = {https://www.cambridge.org/core/journals/journal-of-plasma-physics/article/runaway-dynamics-in-tokamak-disruptions-with-current-relaxation/C971C399150498D6A2011D906F21F04B},
	doi = {10.1017/S0022377822000733},
	language = {en},
	number = {4},
	urldate = {2023-10-11},
	journal = {Journal of Plasma Physics},
	author = {Pusztai, István and Hoppe, Mathias and Vallhagen, Oskar},
	month = aug,
	year = {2022},
	note = {Publisher: Cambridge University Press},
	pages = {905880409},
}

@article{ongena_magnetic-confinement_2016,
	title = {Magnetic-confinement fusion},
	volume = {12},
	copyright = {2016 Springer Nature Limited},
	issn = {1745-2481},
	url = {https://www.nature.com/articles/nphys3745},
	doi = {10.1038/nphys3745},
	language = {en},
	number = {5},
	urldate = {2023-10-11},
	journal = {Nature Physics},
	author = {Ongena, J. and Koch, R. and Wolf, R. and Zohm, H.},
	month = may,
	year = {2016},
	note = {Number: 5
Publisher: Nature Publishing Group},
	pages = {398--410},
}

@article{editors_chapter_1999,
	title = {Chapter 1: {Overview} and summary},
	volume = {39},
	issn = {0029-5515},
	shorttitle = {Chapter 1},
	url = {https://dx.doi.org/10.1088/0029-5515/39/12/301},
	doi = {10.1088/0029-5515/39/12/301},
	language = {en},
	number = {12},
	urldate = {2023-10-11},
	journal = {Nuclear Fusion},
	author = {Editors, ITER Physics Basis and Chairs, ITER Physics Expert Group and {Co-Chairs} and Team, ITER Joint Central and Unit, Physics Integration},
	month = dec,
	year = {1999},
	pages = {2137},
}

@book{harms_principles_2002,
	title = {Principles of fusion energy},
	isbn = {81-7764-233-2},
	publisher = {Allied Publishers},
	author = {Harms, Archie A},
	year = {2002},
}

@book{freidberg_plasma_2008,
	title = {Plasma physics and fusion energy},
	isbn = {1-139-46215-6},
	publisher = {Cambridge university press},
	author = {Freidberg, Jeffrey P},
	year = {2008},
}

@book{wesson_tokamaks_2011,
	title = {Tokamaks},
	volume = {149},
	isbn = {0-19-959223-3},
	publisher = {Oxford university press},
	author = {Wesson, John and Campbell, David J},
	year = {2011},
}

@inproceedings{illerhaus_machine_2022,
	title = {Machine {Learning} {Applications} in {Control} at {ASDEX} {Upgrade}},
	publisher = {DPG},
	author = {Illerhaus, J and Treutterer, W and Bock, A and Fischer, R and Heinrich, P and Jenko, F and Kudlacek, O and Papp, G and Peherstorfer, T and Sieglin, B},
	year = {2022},
}

@article{wieschollek_role_2022,
	title = {On the role of preexisting {MHD} activity for the plasma response to massive deuterium injection},
	volume = {29},
	issn = {1070-664X},
	url = {https://doi.org/10.1063/5.0075473},
	doi = {10.1063/5.0075473},
	number = {3},
	urldate = {2023-10-09},
	journal = {Physics of Plasmas},
	author = {Wieschollek, F. and Hoelzl, M. and Nardon, E. and {JOREK Team} and {ASDEX Upgrade Team} and {EUROfusion MST1 Team}},
	month = mar,
	year = {2022},
	pages = {032509},
}

@article{hu_radiation_2021,
	title = {Radiation asymmetry and {MHD} destabilization during the thermal quench after impurity shattered pellet injection},
	volume = {61},
	issn = {0029-5515},
	url = {https://dx.doi.org/10.1088/1741-4326/abcbcb},
	doi = {10.1088/1741-4326/abcbcb},
	language = {en},
	number = {2},
	urldate = {2023-10-04},
	journal = {Nuclear Fusion},
	author = {Hu, D. and Nardon, E. and Hoelzl, M. and Wieschollek, F. and Lehnen, M. and Huijsmans, G. T. A. and Vugt, D. C. van and Kim, S.-H. and contributors, J. E. T. and team, JOREK},
	month = jan,
	year = {2021},
	note = {Publisher: IOP Publishing},
	pages = {026015},
}

@article{kim_shattered_2019,
	title = {Shattered pellet injection simulations with {NIMROD}},
	volume = {26},
	issn = {1070-664X},
	url = {https://doi.org/10.1063/1.5088814},
	doi = {10.1063/1.5088814},
	number = {4},
	urldate = {2023-10-04},
	journal = {Physics of Plasmas},
	author = {Kim, Charlson C. and Liu, Yueqiang and Parks, Paul B. and Lao, Lang L. and Lehnen, Michael and Loarte, Alberto},
	month = apr,
	year = {2019},
	pages = {042510},
}

@article{hu_3d_2018,
	title = {{3D} non-linear {MHD} simulation of the {MHD} response and density increase as a result of shattered pellet injection},
	volume = {58},
	issn = {0029-5515},
	url = {https://dx.doi.org/10.1088/1741-4326/aae614},
	doi = {10.1088/1741-4326/aae614},
	language = {en},
	number = {12},
	urldate = {2023-10-04},
	journal = {Nuclear Fusion},
	author = {Hu, D. and Nardon, E. and Lehnen, M. and Huijsmans, G. T. A. and Vugt, D. C. van and Contributors, J. E. T.},
	month = oct,
	year = {2018},
	note = {Publisher: IOP Publishing},
	pages = {126025},
}

@book{glasstone_controlled_1960,
	address = {New York},
	title = {Controlled thermonuclear reactions: an introduction to theory and experiment},
	publisher = {Robert E. Krieger Publishing},
	author = {Glasstone, Samuel and Lovberg, Ralph Harvey},
	year = {1960},
}

@article{pegourie_review_2007,
	title = {Review: {Pellet} injection experiments and modelling},
	volume = {49},
	issn = {0741-3335},
	shorttitle = {Review},
	url = {https://dx.doi.org/10.1088/0741-3335/49/8/R01},
	doi = {10.1088/0741-3335/49/8/R01},
	language = {en},
	number = {8},
	urldate = {2023-10-04},
	journal = {Plasma Physics and Controlled Fusion},
	author = {Pégourié, B.},
	month = jul,
	year = {2007},
	pages = {R87},
}

@article{milora_pellet_1995,
	title = {Pellet fuelling},
	volume = {35},
	issn = {0029-5515},
	url = {https://dx.doi.org/10.1088/0029-5515/35/6/I04},
	doi = {10.1088/0029-5515/35/6/I04},
	language = {en},
	number = {6},
	urldate = {2023-10-04},
	journal = {Nuclear Fusion},
	author = {Milora, S. L. and Houlberg, W. A. and Lengyel, L. L. and Mertens, V.},
	month = jun,
	year = {1995},
	pages = {657},
}

@inproceedings{parks_theoretical_2017,
	title = {A theoretical model for the penetration of a shattered-pellet debris plume},
	url = {https://tsdw.pppl.gov/Talks/2017/Lexar/Paul%20PPPL%20talk.pdf},
	booktitle = {Theory and {Simulation} of {Disruptions} {Workshop}},
	author = {Parks, P.},
	year = {2017},
	pages = {17--19},
}

@inproceedings{paul_heinrich_analysis_2023,
	address = {Bordeaux, France},
	title = {Analysis of shattered pellet injection experiments at {ASDEX} {Upgrade}},
	booktitle = {Tu\_MCF54, 49th {European} {Conference} on {Plasma} {Physics}},
	author = {{Paul Heinrich} and {G,  Papp} and {M,  Bernert} and {M,  Dibon} and {P,  de Marné} and {S,  Jachmich} and {M,  Lehnen} and {T,  Peherstorfer} and {N,  Schwarz} and {U,  Sheikh} and {J,  Svoboda} and {ASDEX Upgrade Team}},
	month = jul,
	year = {2023},
}

@inproceedings{s_jachmich_shattered_2023,
	address = {Bordeaux, France},
	title = {Shattered {Pellet} {Injection} experiments at {ASDEX}-{Upgrade} for design optimisation of the {ITER} {Disruption} {Mitigation} {System}},
	booktitle = {O2.103, 49th {European} {Conference} on {Plasma} {Physics}},
	author = {{S, Jachmich} and {M, Lehnen} and {G, Papp} and {M, Dibon} and {P, Heinrich} and {U, Kruezi1} and {P, de Marné} and {M, Bernert} and {A, Bock} and {J, Cerovsky} and {O, Ficker} and {D, Fiorucci} and {G, Kocsis} and {B, Kurzan} and {A, Lukin} and {T, Peherstorfer} and {U, Sheikh} and {I Vinyar} and {T, Eich} and {J, Hobirk} and {ASDEX Upgrade Team}},
	month = jul,
	year = {2023},
}

@article{hirshman_neoclassical_1977,
	title = {Neoclassical conductivity of a tokamak plasma},
	volume = {17},
	issn = {0029-5515},
	url = {https://dx.doi.org/10.1088/0029-5515/17/3/016},
	doi = {10.1088/0029-5515/17/3/016},
	language = {en},
	number = {3},
	urldate = {2023-09-26},
	journal = {Nuclear Fusion},
	author = {Hirshman, S. P. and Hawryluk, R. J. and Birge, B.},
	month = jun,
	year = {1977},
	pages = {611},
}

@article{greenwald_new_1988,
	title = {A new look at density limits in tokamaks},
	volume = {28},
	issn = {0029-5515},
	url = {https://dx.doi.org/10.1088/0029-5515/28/12/009},
	doi = {10.1088/0029-5515/28/12/009},
	language = {en},
	number = {12},
	urldate = {2023-09-21},
	journal = {Nuclear Fusion},
	author = {Greenwald, M. and Terry, J. L. and Wolfe, S. M. and Ejima, S. and Bell, M. G. and Kaye, S. M. and Neilson, G. H.},
	month = dec,
	year = {1988},
	pages = {2199},
}

@techreport{ansh_patel_internship_2023,
	address = {France},
	title = {Internship report: {Analysis} of camera data from experiments with {Shattered} {Pellet} {Injection} at {ASDEX} {Upgrade} tokamak},
	url = {https://user.iter.org/?uid=8WXECM},
	institution = {ITER Organization},
	author = {{Ansh Patel}},
	month = mar,
	year = {2023},
}

@article{jardin_fast_2000,
	title = {A fast shutdown technique for large tokamaks},
	volume = {40},
	issn = {0029-5515},
	url = {https://dx.doi.org/10.1088/0029-5515/40/5/305},
	doi = {10.1088/0029-5515/40/5/305},
	language = {en},
	number = {5},
	urldate = {2023-09-14},
	journal = {Nuclear Fusion},
	author = {Jardin, S. C. and Schmidt, G. L. and Fredrickson, E. D. and Hill, K. W. and Hyun, J. and Merrill, B. J. and Sayer, R.},
	month = may,
	year = {2000},
	pages = {923},
}

@article{herfindal_injection_2019,
	title = {Injection of multiple shattered pellets for disruption mitigation in {DIII}-{D}},
	volume = {59},
	issn = {0029-5515},
	url = {https://dx.doi.org/10.1088/1741-4326/ab3693},
	doi = {10.1088/1741-4326/ab3693},
	language = {en},
	number = {10},
	urldate = {2023-06-10},
	journal = {Nuclear Fusion},
	author = {Herfindal, J. L. and Shiraki, D. and Baylor, L. R. and Eidietis, N. W. and Hollmann, E. M. and Lasnier, C. J. and Moyer, R. A.},
	month = sep,
	year = {2019},
	note = {Publisher: IOP Publishing},
	pages = {106034},
}

@article{baylor_design_2021,
	title = {Design and performance of shattered pellet injection systems for {JET} and {KSTAR} disruption mitigation research in support of {ITER}*},
	volume = {61},
	issn = {0029-5515},
	url = {https://dx.doi.org/10.1088/1741-4326/ac1bc3},
	doi = {10.1088/1741-4326/ac1bc3},
	language = {en},
	number = {10},
	urldate = {2023-06-10},
	journal = {Nuclear Fusion},
	author = {Baylor, L. R. and Meitner, S. J. and Gebhart, T. E. and Caughman, J. B. O. and Shiraki, D. and Wilson, J. R. and Craven, D. and Fortune, M. and Silburn, S. and Muir, A. and Peacock, A. T. and Park, S. H. and Kim, K. P. and Kim, J. H. and Lee, K. S. and Ellwood, G. and Jachmich, S. and Kruezi, U. and Lehnen, M. and Contributors, J. E. T.},
	month = aug,
	year = {2021},
	note = {Publisher: IOP Publishing},
	pages = {106001},
}

@article{bosviel_near-field_2021,
	title = {Near-field models and simulations of pellet ablation in tokamaks},
	volume = {28},
	issn = {1070-664X},
	url = {https://doi.org/10.1063/5.0029721},
	doi = {10.1063/5.0029721},
	number = {1},
	urldate = {2023-06-07},
	journal = {Physics of Plasmas},
	author = {Bosviel, Nicolas and Parks, Paul and Samulyak, Roman},
	month = jan,
	year = {2021},
	pages = {012506},
}

@inproceedings{matsuyama_requirements_2020,
	title = {Requirements for {Runaway} {Electron} {Avoidance} in {ITER} {Disruption} {Mitigation} {Scenario} by {Shattered} {Pellet} {Injection}},
	url = {https://nucleus.iaea.org/sites/fusionportal/Shared%20Documents/FEC%202020/fec2020-preprints/preprint0817.pdf},
	author = {Matsuyama, A and Nardon, E and Honda, M and Shiroto, T and Lehnen, M},
	year = {2020},
}

@article{hu_collisional-radiative_2023,
	title = {Collisional-radiative simulation of impurity assimilation, radiative collapse and {MHD} dynamics after {ITER} shattered pellet injection},
	volume = {63},
	issn = {0029-5515},
	url = {https://dx.doi.org/10.1088/1741-4326/acc8e9},
	doi = {10.1088/1741-4326/acc8e9},
	language = {en},
	number = {6},
	urldate = {2023-06-05},
	journal = {Nuclear Fusion},
	author = {Hu, D. and Nardon, E. and Artola, F. J. and Lehnen, M. and Bonfiglio, D. and Hoelzl, M. and Huijsmans, G. T. A. and Lee, S.-J. and Team, the JOREK},
	month = apr,
	year = {2023},
	note = {Publisher: IOP Publishing},
	pages = {066008},
}

@article{parks_effect_1978,
	title = {Effect of transonic flow in the ablation cloud on the lifetime of a solid hydrogen pellet in a plasma},
	volume = {21},
	issn = {0031-9171},
	url = {https://aip.scitation.org/doi/abs/10.1063/1.862088},
	doi = {10.1063/1.862088},
	number = {10},
	urldate = {2023-03-22},
	journal = {The Physics of Fluids},
	author = {Parks, P. B. and Turnbull, R. J.},
	month = oct,
	year = {1978},
	note = {Publisher: American Institute of Physics},
	pages = {1735--1741},
}

@inproceedings{nardon_theory_2021,
	title = {Theory and modelling activities in support of the {ITER} disruption mitigation system},
	url = {https://hal-cea.archives-ouvertes.fr/cea-03253650},
	language = {en},
	urldate = {2023-03-22},
	author = {Nardon, E. and Huijsmans, G. and Peysson, Y. and Reux, C. and Matsuyama, A. and Lehnen, M. and Aleynikov, P. and Artola, F. J. and Bandaru, V. and Hoelzl, M. and Papp, G. and Bardsley, O. and Kong, M. and Beidler, M. and Del-Castillo-Negrete, D. and Spong, D. and Bonfiglio, Daniele and Boozer, A. and Paz-Soldan, C. and Breizman, B. and Kiramov, D. and Brennan, D. and Ferraro, N. and Jardin, S. and Liu, C. and Garland, N. and Tang, X. and Decker, J. and Sommariva, C. and Embreus, O. and Harvey, R. and Hu, D. and Izzo, V. and Kim, C. and Konovalov, S. and Lao, L. and Liu, Y. and Lyons, B. and Mcclenaghan, J. and Parks, P. and Lee, S. J. and Martín-Solís, J. R. and Mcdevitt, C. and Samulyak, R. and Strauss, H.},
	month = may,
	year = {2021},
}

@inproceedings{lehnen_iter_2020,
	title = {The {ITER} disruption mitigation strategy},
	url = {https://conferences.iaea.org/event/217/contributions/17867/attachments/9322/12801/Lehnen_IAEA_TM2020_final.pdf},
	author = {Lehnen, M and Jachmich, S and Kruezi, U},
	year = {2020},
	pages = {20--23},
}

@article{birol_world_nodate,
	title = {World {Energy} {Outlook} 2022},
	url = {https://www.iea.org/reports/world-energy-outlook-2022},
	language = {en},
	author = {Birol, Dr Fatih},
}

@article{matsuyama_enhanced_2022,
	title = {Enhanced {Material} {Assimilation} in a {Toroidal} {Plasma} {Using} {Mixed} {H} 2 + {Ne} {Pellet} {Injection} and {Implications} to {ITER}},
	volume = {129},
	issn = {0031-9007, 1079-7114},
	url = {https://link.aps.org/doi/10.1103/PhysRevLett.129.255001},
	doi = {10.1103/PhysRevLett.129.255001},
	language = {en},
	number = {25},
	urldate = {2023-02-25},
	journal = {Physical Review Letters},
	author = {Matsuyama, A. and Sakamoto, R. and Yasuhara, R. and Funaba, H. and Uehara, H. and Yamada, I. and Kawate, T. and Goto, M.},
	month = dec,
	year = {2022},
	pages = {255001},
}

@article{hender_chapter_2007,
	title = {Chapter 3: {MHD} stability, operational limits and disruptions},
	volume = {47},
	issn = {0029-5515},
	shorttitle = {Chapter 3},
	url = {https://dx.doi.org/10.1088/0029-5515/47/6/S03},
	doi = {10.1088/0029-5515/47/6/S03},
	language = {en},
	number = {6},
	urldate = {2023-02-25},
	journal = {Nuclear Fusion},
	author = {Hender, T. C. and Wesley, J. C. and Bialek, J. and Bondeson, A. and Boozer, A. H. and Buttery, R. J. and Garofalo, A. and Goodman, T. P. and Granetz, R. S. and Gribov, Y. and Gruber, O. and Gryaznevich, M. and Giruzzi, G. and Günter, S. and Hayashi, N. and Helander, P. and Hegna, C. C. and Howell, D. F. and Humphreys, D. A. and Huysmans, G. T. A. and Hyatt, A. W. and Isayama, A. and Jardin, S. C. and Kawano, Y. and Kellman, A. and Kessel, C. and Koslowski, H. R. and Haye, R. J. La and Lazzaro, E. and Liu, Y. Q. and Lukash, V. and Manickam, J. and Medvedev, S. and Mertens, V. and Mirnov, S. V. and Nakamura, Y. and Navratil, G. and Okabayashi, M. and Ozeki, T. and Paccagnella, R. and Pautasso, G. and Porcelli, F. and Pustovitov, V. D. and Riccardo, V. and Sato, M. and Sauter, O. and Schaffer, M. J. and Shimada, M. and Sonato, P. and Strait, E. J. and Sugihara, M. and Takechi, M. and Turnbull, A. D. and Westerhof, E. and Whyte, D. G. and Yoshino, R. and Zohm, H. and the ITPA MHD, Disruption and Group, Magnetic Control Topical},
	month = jun,
	year = {2007},
	pages = {S128},
}

@article{murmann_thomson_1992,
	title = {The {Thomson} scattering systems of the {ASDEX} upgrade tokamak},
	volume = {63},
	issn = {0034-6748, 1089-7623},
	url = {http://aip.scitation.org/doi/10.1063/1.1143504},
	doi = {10.1063/1.1143504},
	language = {en},
	number = {10},
	urldate = {2023-02-23},
	journal = {Review of Scientific Instruments},
	author = {Murmann, H. and Götsch, S. and Röhr, H. and Salzmann, H. and Steuer, K. H.},
	month = oct,
	year = {1992},
	pages = {4941--4943},
}

@article{mlynek_infrared_2012,
	title = {Infrared {Interferometry} with {Submicrosecond} {Time} {Resolution} in {Massive} {Gas} {Injection} {Experiments} on {ASDEX} {Upgrade}},
	volume = {61},
	issn = {1536-1055},
	url = {https://doi.org/10.13182/FST12-A13582},
	doi = {10.13182/FST12-A13582},
	number = {4},
	urldate = {2023-02-23},
	journal = {Fusion Science and Technology},
	author = {Mlynek, Alexander and Pautasso, Gabriella and Maraschek, Marc and Eixenberger, Horst},
	month = may,
	year = {2012},
	note = {Publisher: Taylor \& Francis
\_eprint: https://doi.org/10.13182/FST12-A13582},
	pages = {290--300},
}

@article{mlynek_simple_2017,
	title = {A simple and versatile phase detector for heterodyne interferometers},
	volume = {88},
	issn = {0034-6748, 1089-7623},
	url = {http://aip.scitation.org/doi/10.1063/1.4975992},
	doi = {10.1063/1.4975992},
	language = {en},
	number = {2},
	urldate = {2023-02-23},
	journal = {Review of Scientific Instruments},
	author = {Mlynek, A. and Faugel, H. and Eixenberger, H. and Pautasso, G. and Sellmair, G. and {ASDEX Upgrade Team}},
	month = feb,
	year = {2017},
	pages = {023504},
}

@article{jachmich_shattered_2022,
	title = {Shattered pellet injection experiments at {JET} in support of the {ITER} disruption mitigation system design},
	volume = {62},
	issn = {0029-5515, 1741-4326},
	url = {https://iopscience.iop.org/article/10.1088/1741-4326/ac3c86},
	doi = {10.1088/1741-4326/ac3c86},
	language = {en},
	number = {2},
	urldate = {2023-02-21},
	journal = {Nuclear Fusion},
	author = {Jachmich, S. and Kruezi, U. and Lehnen, M. and Baruzzo, M. and Baylor, L.R. and Carnevale, D. and Craven, D. and Eidietis, N.W. and Ficker, O. and Gebhart, T.E. and Gerasimov, S. and Herfindal, J.L. and Hollmann, E. and Huber, A. and Lomas, P. and Lovell, J. and Manzanares, A. and Maslov, M. and Mlynar, J. and Pautasso, G. and Paz-Soldan, C. and Peacock, A. and Piron, L. and Plyusnin, V. and Reinke, M. and Reux, C. and Rimini, F. and Sheikh, U. and Shiraki, D. and Silburn, S. and Sweeney, R. and Wilson, J. and Carvalho, P. and JET Contributors, the},
	month = feb,
	year = {2022},
	pages = {026012},
}

@inproceedings{heinrich_characterization_2022,
	title = {Characterization of the {ASDEX} {Upgrade} shattered pellet injector},
	publisher = {2nd Technical Meeting on Plasma Disruptions and their Mitigation},
	author = {Heinrich, P and Papp, G and Dibon, M and Marne, P de and Peherstorfer, T and Jachmich, S and Lehnen, M and Bock, A and {ASDEX Upgrade Team}},
	year = {2022},
}

@article{sheikh_disruption_2021,
	title = {Disruption thermal load mitigation with shattered pellet injection on the {Joint} {European} {Torus} ({JET})},
	volume = {61},
	issn = {0029-5515, 1741-4326},
	url = {https://iopscience.iop.org/article/10.1088/1741-4326/ac3191},
	doi = {10.1088/1741-4326/ac3191},
	language = {en},
	number = {12},
	urldate = {2023-02-16},
	journal = {Nuclear Fusion},
	author = {Sheikh, U.A. and Shiraki, D. and Sweeney, R. and Carvalho, P. and Jachmich, S. and Joffrin, E. and Lehnen, M. and Lovell, J. and Nardon, E. and Silburn, S. and {JET Contributors}},
	month = dec,
	year = {2021},
	pages = {126043},
}

@inproceedings{luce_progress_2021,
	title = {Progress on the {ITER} {DMS} design and integration 28th {IAEA} {Fusion} {Energy} {Conf}},
	publisher = {TECH/1-4Ra](https://conference. iaea. org/event/214/contributions/)},
	author = {Luce, T. C.},
	year = {2021},
}

@article{riccardo_jet_2010,
	title = {{JET} disruption studies in support of {ITER}},
	volume = {52},
	issn = {0741-3335, 1361-6587},
	url = {https://iopscience.iop.org/article/10.1088/0741-3335/52/12/124018},
	doi = {10.1088/0741-3335/52/12/124018},
	language = {en},
	number = {12},
	urldate = {2023-02-14},
	journal = {Plasma Physics and Controlled Fusion},
	author = {Riccardo, V and Arnoux, G and Cahyna, P and Hender, T C and Huber, A and Jachmich, S and Kiptily, V and Koslowski, R and Krlin, L and Lehnen, M and Loarte, A and Nardon, E and Paprok, R and Tskhakaya, D and {JET-EFDA contributors}},
	month = dec,
	year = {2010},
	pages = {124018},
}

@article{loarte_transient_2007,
	title = {Transient heat loads in current fusion experiments, extrapolation to {ITER} and consequences for its operation},
	volume = {T128},
	issn = {0031-8949, 1402-4896},
	url = {https://iopscience.iop.org/article/10.1088/0031-8949/2007/T128/043},
	doi = {10.1088/0031-8949/2007/T128/043},
	language = {en},
	urldate = {2023-02-14},
	journal = {Physica Scripta},
	author = {Loarte, A and Saibene, G and Sartori, R and Riccardo, V and Andrew, P and Paley, J and Fundamenski, W and Eich, T and Herrmann, A and Pautasso, G and Kirk, A and Counsell, G and Federici, G and Strohmayer, G and Whyte, D and Leonard, A and Pitts, R A and Landman, I and Bazylev, B and Pestchanyi, S},
	month = mar,
	year = {2007},
	pages = {222--228},
}

@article{riccardo_timescale_2005,
	title = {Timescale and magnitude of plasma thermal energy loss before and during disruptions in {JET}},
	volume = {45},
	issn = {0029-5515},
	url = {https://dx.doi.org/10.1088/0029-5515/45/11/025},
	doi = {10.1088/0029-5515/45/11/025},
	language = {en},
	number = {11},
	urldate = {2023-02-14},
	journal = {Nuclear Fusion},
	author = {Riccardo, V. and Loarte, A. and Contributors, the JET EFDA},
	month = oct,
	year = {2005},
	pages = {1427},
}

@article{nardon_fast_2020,
	title = {Fast plasma dilution in {ITER} with pure deuterium shattered pellet injection},
	volume = {60},
	issn = {0029-5515},
	url = {https://dx.doi.org/10.1088/1741-4326/abb749},
	doi = {10.1088/1741-4326/abb749},
	language = {en},
	number = {12},
	urldate = {2023-02-09},
	journal = {Nuclear Fusion},
	author = {Nardon, E. and Hu, D. and Hoelzl, M. and Bonfiglio, D. and team, the JOREK},
	month = oct,
	year = {2020},
	note = {Publisher: IOP Publishing},
	pages = {126040},
}

@techreport{lehnen_rd_2018,
	address = {International Atomic Energy Agency (IAEA)},
	title = {R\&{D} for {Reliable} {Disruption} {Mitigation} in {ITER}},
	url = {http://inis.iaea.org/search/search.aspx?orig_q=RN:50052430},
	author = {Lehnen, M. and Campbell, D.J. and Hu, D. and Kruezi, U. and Luce, T.C. and Maruyama, S. and Snipes, J.A. and Sweeney, R.},
	month = oct,
	year = {2018},
	note = {IAEA-CN--258},
	pages = {368},
}

@article{hollmann_status_2014,
	title = {Status of research toward the {ITER} disruption mitigation system},
	volume = {22},
	copyright = {© 2014 AIP Publishing LLC.},
	issn = {1070-664X},
	url = {https://aip.scitation.org/doi/abs/10.1063/1.4901251},
	doi = {10.1063/1.4901251},
	language = {en},
	number = {2},
	urldate = {2023-02-08},
	journal = {Physics of Plasmas},
	author = {Hollmann, E. M. and Aleynikov, P. B. and Fülöp, T. and Humphreys, D. A. and Izzo, V. A. and Lehnen, M. and Lukash, V. E. and Papp, G. and Pautasso, G. and Saint-Laurent, F. and Snipes, J. A.},
	month = nov,
	year = {2014},
	note = {Publisher: AIP Publishing LLCAIP Publishing},
	pages = {021802},
}

@article{matsuyama_transport_2022,
	title = {Transport simulations of pre-thermal quench shattered pellet injection in {ITER}: code verification and assessment of key trends},
	volume = {64},
	issn = {0741-3335, 1361-6587},
	shorttitle = {Transport simulations of pre-thermal quench shattered pellet injection in {ITER}},
	url = {https://iopscience.iop.org/article/10.1088/1361-6587/ac89b2},
	doi = {10.1088/1361-6587/ac89b2},
	language = {en},
	number = {10},
	urldate = {2022-12-23},
	journal = {Plasma Physics and Controlled Fusion},
	author = {Matsuyama, A and Hu, D and Lehnen, M and Nardon, E and Artola, J},
	month = oct,
	year = {2022},
	pages = {105018},
}

@article{breizman_physics_2019,
	title = {Physics of runaway electrons in tokamaks},
	volume = {59},
	issn = {0029-5515, 1741-4326},
	url = {https://iopscience.iop.org/article/10.1088/1741-4326/ab1822},
	doi = {10.1088/1741-4326/ab1822},
	language = {en},
	number = {8},
	urldate = {2021-03-15},
	journal = {Nuclear Fusion},
	author = {Breizman, Boris N. and Aleynikov, Pavel and Hollmann, Eric M. and Lehnen, Michael},
	month = aug,
	year = {2019},
	pages = {083001},
}

@article{kurzan_edge_2011,
	title = {Edge and core {Thomson} scattering systems and their calibration on the {ASDEX} {Upgrade} tokamak},
	volume = {82},
	issn = {0034-6748},
	url = {https://aip.scitation.org/doi/full/10.1063/1.3643771},
	doi = {10.1063/1.3643771},
	number = {10},
	urldate = {2023-01-24},
	journal = {Review of Scientific Instruments},
	author = {Kurzan, B. and Murmann, H. D.},
	month = oct,
	year = {2011},
	note = {Publisher: American Institute of Physics},
	pages = {103501},
}

@article{jachmich_shattered_2021,
	title = {Shattered pellet injection experiments at {JET} in support of the {ITER} disruption mitigation system design},
	volume = {62},
	issn = {0029-5515},
	url = {https://doi.org/10.1088/1741-4326/ac3c86},
	doi = {10.1088/1741-4326/ac3c86},
	language = {en},
	number = {2},
	urldate = {2022-09-05},
	journal = {Nuclear Fusion},
	author = {Jachmich, S. and Kruezi, U. and Lehnen, M. and Baruzzo, M. and Baylor, L. R. and Carnevale, D. and Craven, D. and Eidietis, N. W. and Ficker, O. and Gebhart, T. E. and Gerasimov, S. and Herfindal, J. L. and Hollmann, E. and Huber, A. and Lomas, P. and Lovell, J. and Manzanares, A. and Maslov, M. and Mlynar, J. and Pautasso, G. and Paz-Soldan, C. and Peacock, A. and Piron, L. and Plyusnin, V. and Reinke, M. and Reux, C. and Rimini, F. and Sheikh, U. and Shiraki, D. and Silburn, S. and Sweeney, R. and Wilson, J. and Carvalho, P. and Contributors, the JET},
	month = dec,
	year = {2021},
	pages = {026012},
}

@article{bandyopadhyay_summary_2021,
	title = {Summary of the {IAEA} technical meeting on plasma disruptions and their mitigation},
	volume = {61},
	issn = {0029-5515},
	url = {https://doi.org/10.1088/1741-4326/abfe76},
	doi = {10.1088/1741-4326/abfe76},
	language = {en},
	number = {7},
	urldate = {2022-09-06},
	journal = {Nuclear Fusion},
	author = {Bandyopadhyay, Indranil and Barbarino, Matteo and Bhattacharjee, Amitava and Eidietis, Nicholas and Huber, Alexander and Isayama, Akihiko and Kim, Jayhyun and Konovalov, Sergey and Lehnen, Michael and Nardon, Eric and Pautasso, Gabriella and Rea, Cristina and Sozzi, Carlo and Villone, Fabio and Zeng, Long},
	month = jun,
	year = {2021},
	pages = {077001},
}

@article{lehnen_disruptions_2015,
	title = {Disruptions in {ITER} and strategies for their control and mitigation},
	volume = {463},
	issn = {00223115},
	url = {https://linkinghub.elsevier.com/retrieve/pii/S0022311514007594},
	doi = {10.1016/j.jnucmat.2014.10.075},
	language = {en},
	urldate = {2021-03-15},
	journal = {Journal of Nuclear Materials},
	author = {Lehnen, M. and Aleynikova, K. and Aleynikov, P.B. and Campbell, D.J. and Drewelow, P. and Eidietis, N.W. and Gasparyan, Yu. and Granetz, R.S. and Gribov, Y. and Hartmann, N. and Hollmann, E.M. and Izzo, V.A. and Jachmich, S. and Kim, S.-H. and Kočan, M. and Koslowski, H.R. and Kovalenko, D. and Kruezi, U. and Loarte, A. and Maruyama, S. and Matthews, G.F. and Parks, P.B. and Pautasso, G. and Pitts, R.A. and Reux, C. and Riccardo, V. and Roccella, R. and Snipes, J.A. and Thornton, A.J. and de Vries, P.C.},
	month = aug,
	year = {2015},
	pages = {39--48},
}

@misc{peherstorfer_fragmentation_2022,
	title = {Fragmentation {Analysis} of {Cryogenic} {Pellets} for {Disruption} {Mitigation}},
	url = {http://arxiv.org/abs/2209.01024},
	language = {en},
	urldate = {2022-12-23},
	publisher = {arXiv},
	author = {Peherstorfer, Tobias},
	month = sep,
	year = {2022},
	note = {arXiv:2209.01024 [physics]},
}

@article{park_deployment_2020,
	title = {Deployment of multiple shattered pellet injection systems in {KSTAR}},
	volume = {154},
	issn = {0920-3796},
	url = {https://www.sciencedirect.com/science/article/pii/S0920379620300831},
	doi = {10.1016/j.fusengdes.2020.111535},
	language = {en},
	urldate = {2022-11-21},
	journal = {Fusion Engineering and Design},
	author = {Park, SooHwan and Lee, KunSu and Baylor, Larry R. and Meitner, Steven J. and Lee, HyunMyung and Song, JaeIn and Gebhart, Trey E. and Yun, SangWon and Kim, Jayhyun and Kim, KwangPyo and Park, KapRai and Yoon, SiWoo},
	month = may,
	year = {2020},
	pages = {111535},
}

@article{raman_shattered_2020,
	title = {Shattered pellet penetration in low and high energy plasmas on {DIII}-{D}},
	volume = {60},
	issn = {0029-5515, 1741-4326},
	url = {https://iopscience.iop.org/article/10.1088/1741-4326/ab686f},
	doi = {10.1088/1741-4326/ab686f},
	language = {en},
	number = {3},
	urldate = {2021-05-10},
	journal = {Nuclear Fusion},
	author = {Raman, R. and Sweeney, R. and Moyer, R.A. and Eidietis, N.W. and Shiraki, D. and Herfindal, J.L. and Sachdev, J. and Hollmann, E.M. and Jardin, S.C. and Baylor, L.R. and Wilcox, R. and Carlstrom, T. and Osborne, T. and Eldon, D. and Menard, J.E. and Lunsford, R. and Grierson, B.},
	month = mar,
	year = {2020},
	pages = {036014},
}
\end{document}